\documentclass[aps,pre,longbibliography,notitlepage,floatfix,nofootinbib]{revtex4-1}

\usepackage{bm,amsmath,amssymb,graphicx,stmaryrd,float,subcaption,upgreek,sidecap,MnSymbol,mathtools,lipsum}
\usepackage{epstopdf}
\usepackage[toc]{appendix}
\usepackage[justification=raggedright]{caption}
\usepackage[abs]{overpic}
\usepackage{comment,footnote}
\usepackage{enumitem}
\usepackage{tikz}
\usepackage{hyperref}
\usepackage{wasysym}
\usepackage{relsize}

\usepackage{xcolor}

\newcommand{\blackdiamond}{\text{\smaller[1.0] $\!\!\! \blacklozenge$}}
\newcommand{\blackstar}{\text{\smaller[3.0]$\bigstar$}}
\newcommand{\lefttriangle}{\text{\larger[2.0] $\!\!\! \blacktriangleleft \!$}}
\newcommand{\righttriangle}{\text{\larger[2.0] $\!\!\! \blacktriangleright$}}

\newcommand{\triangleup}{\text{\smaller[1.0] $\!\!\! \blacktriangle \!$}}

\usepackage{hyperref}

\usepackage{fullpage}

\graphicspath{{Figures/}}

\begin{document}
	\title{Bistability and equilibria of creased annular sheets and strips}
	\author{Tian Yu}
	\email{tiany@princeton.edu}
	\affiliation{Department of Civil and Environmental Engineering, Princeton University, Princeton, NJ 08544}
	\date{\today}
	
	\begin{abstract}
		A creased thin disk is generally bistable since the crease could be pushed through to form a stable cone-like inverted state with an elastic singularity corresponding to the vertex of the conical surface. In a recent study, we found that this bistability could be destroyed by removing the singularity through cutting a hole around the vertex, depending on the size and shape of the hole. Particularly, to maintain the bistability, a circular hole normally cannot exceed approximately $20 \%$ of the disk size.  
		This paper extends our recent work and is based on the following observations in tabletop models of creased disks with circular holes: (i) reducing the circumference of the creased disk by removing an annular sector could increase the hole size to be as large as the disk without destroying the bistability, (ii) with a single crease, the circular hole could be as large as the disk without loss of the bistability, and (iii) a family of stable inverted states can be obtained by inverting the disk almost anywhere along the crease. 
		An inextensible strip model is implemented to investigate these phenomena. We formulate a minimal facet of the creased disk as a two-point boundary value problem with the creases modeled as nonlinear hinges, and use numerical continuation to conduct parametric studies. Specifically, we focus on geometric parameters which include an \emph{angle deficit} that determines the circumference of the disk, the rest crease angle, the number of evenly distributed creases, and an \emph{eccentricity} that determines the position of the hole on the crease. Our numerical results confirm the qualitative observations in (i)-(iii) and further reveal unexpected results caused by the coupling between these geometric parameters.
		Our results demonstrate that by varying the geometry of a simply creased disk, surprisingly rich nonlinear behaviors can be obtained, which shed new light on the mechanics and design of origami, kirigami, and morphable structures.     
	\end{abstract}
	

	\maketitle

	\section{Introduction}

	Creases and vertices often occur together in the extreme deformation of thin sheets \cite{witten09spontaneous, walsh2011weakening,korte11triangular,nasto2014localized,chopin16disclinations}, such as squeezing a soda can and crumpling a piece of paper \cite{walsh2011weakening,blair2005geometry}, in which deformations are highly localized around the creases and vertices with the rest surface remaining relatively flat. For engineering applications, discrete crease patterns have been introduced to both thin and thick plates to achieve different functions and forms, 
	such as the foldability and free-form surfaces in rigid and curved origami \cite{miura1985method,schenk2013geometry,chen2015origami,dang2020inverse,feng2020designs,duncan1982folded,demaine2011curved,dias2012geometric,feng2021concentrated,callens2018flat} and sheet metals \cite{hu2002mechanics}, 	
	energy absorption in crash tubes \cite{gattas2015behaviour,garrett2016curved,song2012axial}, and the redistribution of bending stiffness \cite{woodruff2020curved}.
	 Introducing flexibility to the facets of creased thin sheets leads to the creation of new equilibria, which extend the configuration space of the traditional rigid origami \cite{badger2019normalized,silverberg15origami,silverberg2014using,liu2018topological,hanna2014waterbomb}.

	It is the competition between the mechanics of creases and the flexibility of the facets that determines the mechanics of creased thin structures \cite{dias2014non,dias2012geometric,badger2019normalized}. Thin sheets prefer to bend rather than to stretch due to the large ratio of stretching to bending stiffness. Various continuum theories have been employed to study the mechanics of thin sheets and strips, e.g., F{\"o}ppl-von K{\'a}rm{\'a}n theory \cite{lechenault15generic}, 1-director Cosserat plate theory \cite{kumar2020investigation}, small-deflection inextensible plate theory \cite{mansfield1955inextensional,ashwell57equilibrium,mansfield59large,mansfield71analogy}, and geometrically exact inextensible strip model \cite{dias2014non,badger2019normalized,starostin2007shape}. Under the inextensible theory, a flat sheet will be deformed into a developable surface.   
	Sadowsky \cite{hinz2015translation} and Wunderlich \cite{todres2015translation} derived the energy functional for inextensible strips with infinitesimal width and finite width respectively. Based on Wunderlich's functional, Starostin and van der Heijden first derived the Euler-Lagrange equations of the inextensible strip model, which has been employed to study the shapes of Mobius bands \cite{starostin2007shape}, the triangular buckling patterns of twisted ribbons \cite{korte11triangular}, the cascade unlooping of helical ribbons \cite{starostin2008tension}, and the mechanics of elastic annuli \cite{heijdenannular}.

	The inextensible strip model is known to have singular behaviors for geometries where the local stretch of the surface would be preferred to be incorporated \cite{starostin2015equilibrium,yu2019bifurcations,borum2018manipulation,freddi2016corrected,MooreHealey18,audoly2021one,neukirch2021convenient}. On the other hand, it works well to capture the mechanical behaviors of thin sheets with singularity-free and stretching-free geometries \cite{starostin2008tension,dias2014non,audoly2015buckling,badger2019normalized}.
	With the proper choice of materials and lighting, it is possible to approximately see the generators (i.e., unbent lines), and potential singularities (where generators intersect with each other on the material surface) and stretching areas in deformed thin sheets, which help determine if the inextensible strip model could be applied to the whole geometry or only part of it \cite{witten09spontaneous,korte11triangular}. Our choice of the inextensible strip model in this study is based on the observation that our deformed geometries are singularity free and could be parameterized by a family of straight lines.

Creases play a key role in the mechanics of creased thin sheets \cite{lechenault2014mechanical,korte11triangular,lechenault15generic,dias2014non,bende2018overcurvature,duffy2021shape,mowitz2020finite}. A single crease unfolds quickly at first and then slowly in terms of a progressive relaxation \cite{thiria2011relaxation}. The origami length \cite{lechenault2014mechanical} and a similar hinge index \cite{francis2013origami} are able to quantify the competition between the deformations of the crease and the facets. Creases are normally modeled as rotational hinges with a finite stiffness that balances the bending moments from the thin sheets \cite{lechenault2014mechanical,walker2020mechanics,barbieri2019curvature,dharmadasa2018characterizing,dias2014non}. Creases could also be modeled as continuous structures, where the local tangent makes a rapid turn within a short material length \cite{walker2019flexural,jules19local,hernandez2016modeling}. Accurate prediction of the mechanical responses of creased thin sheets requires incorporating both the mechanics of thin sheets and creases. In flexible origami, thin sheets have been modeled as inextensible strips with the creases modeled as elastic hinges \cite{dias2014non,yu2021cutting}. Various discrete models are also developed to study nonrigid origami, such as the bar and hinge model \cite{gillman2018truss,liu2017nonlinear,filipov2017bar}, triangular mesh model \cite{dias2012geometric,kleiman2016influence}, and the hinge and facet model \cite{walker18shape}.

A thin sheet with a single crease is generally bistable with a second stable state obtained by locally inverting the crease, which results in a conical shape with a singularity corresponding to the vertex of the cone \cite{lechenault15generic,walker18shape,andrade2019foldable}. Elastic singularities play important roles in the mechanics of thin sheets \cite{witten09spontaneous,moshe2019nonlinear} and are used to generate concentrated Gaussian curvatures \cite{feng2020evolving,feng2021concentrated,guven2013dipoles}. The bistable behavior in a simply creased sheet is generally insensitive to the constituent materials and the shape of the sheet \cite{lechenault15generic,walker18shape}. Excising the singularity by making a hole around the vertex could reduce the forces needed to invert the crease \cite{walker18shape}. It is found that when indented at the center of a creased metal disk, a localized dimple first forms surrounding the center and then propagates towards the disk edge before the structure snaps to the conical shape \cite{walker2020mechanics}. In a recent work, the author and collaborators demonstrated with both experiments and numerical continuation of an inextensible strip model that a creased thin disk could lose its bistability if the vertex of the inverted shape is cut by making a large enough hole, with the critical size dependent on the shape of the hole \cite{yu2021cutting}. For example, we found that the critical size of a circular hole should be less than approximately $20 \%$ of the the disk for the purpose of retaining bistability.

This paper extends our recent work \cite{yu2021cutting} and investigates several additional factors that affect the mechanics of creased annular sheets and strips. The rest of the paper is organized as follows. Section \ref{se:tabletop} introduces the geometric parameters and novel mechanics phenomena of creased sheets and strips through tabletop models.
In Section \ref{se:stripmodel}, we use an inextensible strip model to describe a minimal facet of the creased annular strip with the creases modeled as nonlinear hinges whose angle-moment relationship follows a sinusoidal form. 
From Sections \ref{se:angledeficit} to \ref{se:eccentrichole}, we present numerical results obtained through numerical continuation of the inextensible strip model. Specifically, Section \ref{se:angledeficit} reports the influence of the angle deficit $\alpha$ on the bistability of creased thin disks with two creases. Section \ref{se:restangle} presents the effect of the rest crease angle $\gamma_{0}$ on the mechanics of creased thin disks with two creases. In Section \ref{se:numbercreases}, we solve both the folded and inverted state of creased thin disks with different number of evenly space creases. Section \ref{se:eccentrichole} introduces an eccentricity to the position of the hole with $N_c=2$ and studies its effect on the bistability. We give a summary and further discussion in Section \ref{se:conclusiondiscussion}. In Appendix \ref{appse:BVPcontinuation}, we document the details of formulating a creased annular strip as a two-point boundary value problem and the procedures of solving it with numerical continuation. Appendix \ref{appse:linearcrease} gives an example ($N_c=2$) with the crease following a linear angle-moment relationship. Appendix \ref{appse:3Dprofile2Dprojection} displays the 3D profile and corresponding 2D projections of the outer and inner circumferences of some renderings shown in Sections \ref{se:angledeficit}-\ref{se:eccentrichole}. Additional renderings of the folded and inverted state with different eccentricities, hole sizes, and number of creases are documented in Appendix \ref{appse:morerenderings} for the interest of the reader.

\section{Tabletop Demonstrations and definition of the geometries}\label{se:tabletop}

\begin{figure}[h!]
		\centering
		\includegraphics[width=0.95\textwidth]{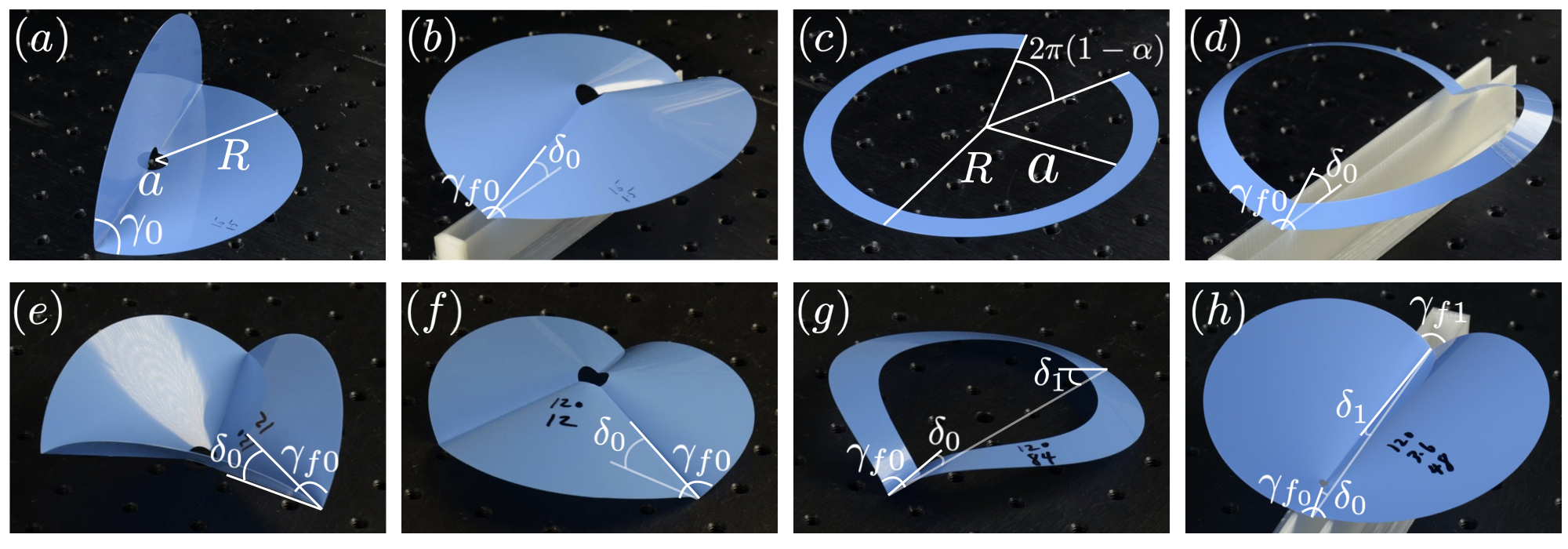}
		\caption{Photographs of creased annular sheets and strips. $(a)$ An energy-free folded state with two evenly spaced creases, a rest crease angle $\gamma_0$, and a circular hole with radius $a$ that is concentric to the disk with radius $R$. $(b)$ The inverted state of $(a)$. $\gamma_{f0}$ represents the final crease angle at one end and $\delta_0$ measures the inclined angle between the crease and the horizontal plane. (c) Cutting an annular sector $2 \pi (1-\alpha)$ could significantly increase the hole size without loss of the inverted state, shown in (d). (c-d) have $(\alpha,a/R)=(0.75,0.85)$. $(e)$ A folded state with three evenly spaced creases contains bent facets. One of the creases is characterized by $\gamma_{f0}$ and $\delta_0$. $(f)$ The inverted state of $(e)$. ($g$) The inverted state with a single crease could admit a hole as large as the disk. ($h$) A continuous family of stable inverted states can be obtained by inverting almost anywhere along the crease. Shown is the inverted state of a creased disk with a small hole that is nonconcentric to the disk. The nonvanishing eccentricity results in different final crease angles $\gamma_{f0}$ and $\gamma_{f1}$ and different inclined angles $\delta_0$ and $\delta_1$ at the two creases. } \label{fig:exptmodel}
	\end{figure}

The tabletop models in Figure \ref{fig:exptmodel} include disks with radius $R=75$ mm (Figures \ref{fig:exptmodel}(a-d)) and 60 mm (Figures \ref{fig:exptmodel}(e-h)), thickness $t=0.127$ mm, and different hole size $a$. They are cut from polyester shim stock (Artus Corp., Englewood, NJ) by a Silhouette Cameo 3 cutter, and subsequently creased using a vise. In this study, we did not attempt to obtain creases with precise rest crease angles considering their complex relaxation mechanisms \cite{thiria2011relaxation}. These models are used only to demonstrate the qualitative behaviors of creased thin disks with different geometries.

We refer to the stable creased configuration in Figure \ref{fig:exptmodel}(a) as the \emph{folded state} and the stable inverted configuration in Figure \ref{fig:exptmodel}(b) as the \emph{inverted state}; between these two stable states exists an unstable \emph{energy barrier}, which is captured by numerical modeling with an inextensible strip model (Section \ref{se:angledeficit}). In addition, numerical results predict \emph{flipped states} with the crease being inverted to bend in the other direction, due to our choice of a sinusoidal constitutive law for the crease (Sections \ref{se:stripmodel} and \ref{se:angledeficit}). Figure \ref{fig:exptmodel} summarizes some tabletop models whose mechanical features are influenced by several geometric parameters, which include the hole size $a/R$ and the rest crease angle $\gamma_0$ (Figure \ref{fig:exptmodel}(a)), an \emph{angle deficit} $\alpha$ that determines the circumference of the annular strip (Figures \ref{fig:exptmodel}(c-d)), the number of evenly distributed creases (Figures \ref{fig:exptmodel}(e-g)), and an \emph{eccentricity} (Figure \ref{fig:exptmodel}(h)) that determines the position of the hole on the crease (see Figure \ref{fig:exptmodel}). In this paper, we will address the following points:

	\begin{itemize}
		\item Figures \ref{fig:exptmodel}(c-d) demonstrate that by cutting an annular sector $2 \pi (1-\alpha)$, the size of a circular hole could increase significantly without destroying the bistability. The model in Figure \ref{fig:exptmodel}(d) is sequentially made by joining the two ends of the open annulus in Figure \ref{fig:exptmodel}(c) with transparent tapes, making two evenly spaced creases to create the folded state (not shown), and inverting the folded state. 
		In our definition, $\alpha < 1$ corresponds to removing a sector, $\alpha > 1$ corresponds to inserting a sector, and $\alpha=1$ represents an annulus with an exact angle of $2 \pi$. The stable inverted state in Figure \ref{fig:exptmodel}(d) has $(\alpha,a/R)=(0.75,0.85)$. We are interested in the effect of the angle deficit $\alpha$ on the bistability of the creased thin disk.
		
		\item The bistability is created by decorating a thin disk with creases.
		 How does the crease angle and crease stiffness affect the mechanical behaviors? 
		
		\item How does the number of creases, $N_c$, affect the mechanics of creased annular sheets and strips? We focus on radial creases that are evenly spaced along the circumference. Figures \ref{fig:exptmodel}(e-f) show the folded state and inverted state with $N_c=3$ and 4, respectively. In addition, with a single crease $N_c=1$, the circular hole can be as large as the disk without loss of the bistability. Figure \ref{fig:exptmodel}(g) shows the stable inverted state with $(N_c,a/R)=(1,0.7)$ (its stable folded state is not included).

		\item A creased thin disk can be inverted about almost anywhere along the crease, resulting in a continuous family of inverted states. Figure \ref{fig:exptmodel}(h) shows an example of an inverted state, with the small hole corresponding to the singularity of the conical surface being nonconcentric to the disk. We will introduce an eccentricity to the position of the hole and study its influence on the mechanics.
	\end{itemize}

	We take advantage of the symmetry in the structure and use the inextensible strip model to study a minimal facet of the folded and inverted state, which are characterized by the final crease angle and the inclined angle between the crease and the horizontal plane. With $N_c \ge 2$, the folded and inverted state have $N_c$-fold mirror symmetries, and the structure could be characterized by the final crease angle $\gamma_{f0}$ and 
	and the inclined angle $\delta_0$ at one end of a minimal facet, shown in Figures \ref{fig:exptmodel}(b), \ref{fig:exptmodel}(d), and \ref{fig:exptmodel}(e-f). However, with $N_c=1$, the inverted state has one-fold mirror symmetry and we study half of the structure whose two ends have different inclined angles $\delta_0$ and $\delta_1$ (Figure \ref{fig:exptmodel}(g)). 
	This is also true for the case with $N_c=2$ and a nonvanishing eccentricity (Figure \ref{fig:exptmodel}(h)), which further results in two different final crease angles $\gamma_{f0}$ and $\gamma_{f1}$ at the two creases. In addition, the crease with a shorter length is observed to have a larger final crease angle, i.e., $\gamma_{f0} > \gamma_{f1}$ in Figure \ref{fig:exptmodel}(h).

	\section{an inextensible strip model}\label{se:stripmodel}

	We describe a creased annular strip as a developable surface decorated with creases that are modeled as nonlinear hinges. The equilibrium equations presented in this section have been derived in our recent work \cite{yu2021cutting}, which follows directly from Starostin and van der Heijden's, and Dias and Audoly's pioneering works on the mechanics of inextensible straight and curved strips \cite{korte11triangular,starostin2015equilibrium,dias2015wunderlich}. Here, we only include a brief discussion of the inextensible theory and focus on applying it to the current study.

\begin{figure}[h!]
	 	\centering
	 	\includegraphics[width=0.95\textwidth]{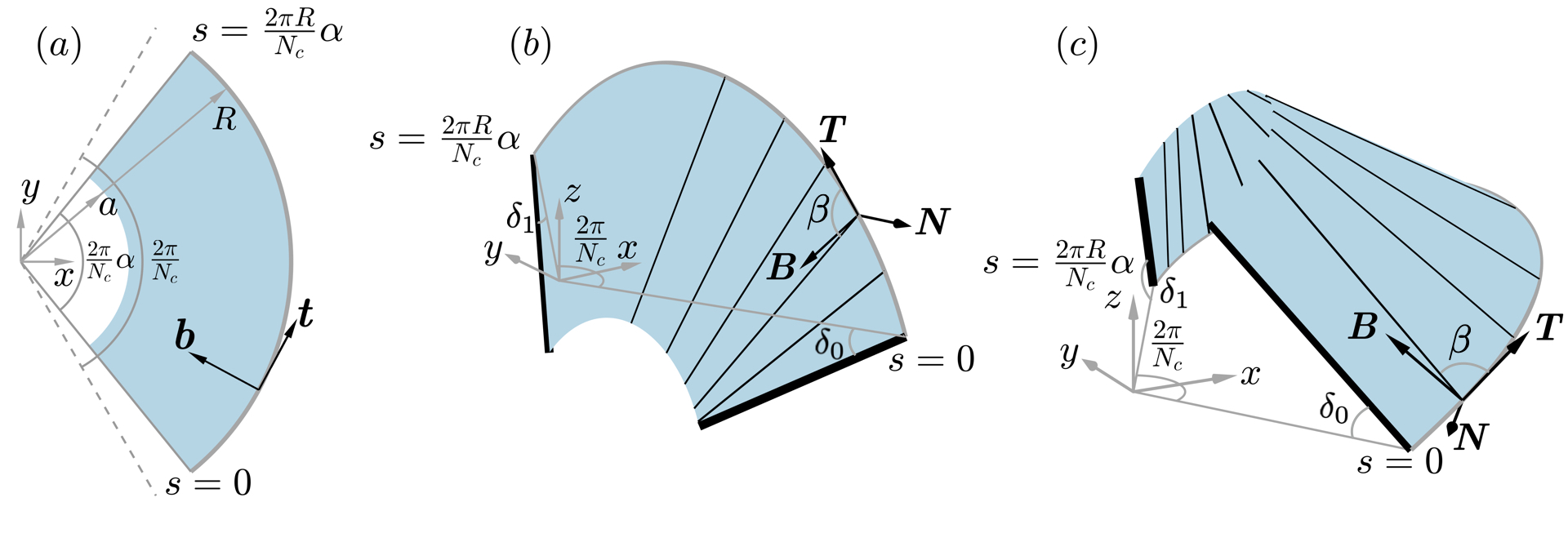}
	 	\caption{A minimal facet of a creased annular strip with $N_c$ evenly spaced creases and an angle deficit $\alpha$ is described by an inextensible strip model. (a) The flat configuration corresponds to an annular sector with a central angle $2 \pi \alpha /N_c$. The undeformed outer circle carries a right-handed orthonormal frame $(\bm{t}, \bm{n}, \bm{b})$ with $\bm{n}$ going into the plane. (b) Folded state. (c) Inverted state.}\label{fig:annularstripkine}
	 \end{figure}

	We take advantage of symmetries in the system and only solve a minimal facet. For example, with $N_c$ evenly spaced creases, we solve one piece bounded by two adjacent creases, as shown in Figure \ref{fig:annularstripkine}. 
	The description involves an orthonormal Darboux frame $(\bm{T},\bm{N}, \bm{B} )$ attached to the directrix $\bm{r}(s)$ of the deformed configurations, corresponding to the outer circle. Here $s$ is the arc length of the directrix. $\bm{T}$ represents the tangent of $\bm{r}(s)$, $\bm{N}$ represents the normal of the surface, and $\bm{B}=\bm{T} \times \bm{N}$.

	Figure \ref{fig:annularstripkine}(a) shows a flat annular sector with an angle deficit $\alpha$ ($s\in [0, 2 \pi R \alpha/N_c]$) and a right-handed orthonormal frame $(\bm{t},\bm{n},\bm{b} )$ attached to the undeformed directrix. The annular sector has an inner radius $a$ and outer radius $R$ and is positioned symmetrically about the $x-z$ plane of a Cartesian coordinate system. Figures \ref{fig:annularstripkine}(b-c) respectively correspond to its folded and inverted state, with the creases rendered as thick black lines.
	The two ends $s=0$ and $s=2 \pi R \alpha /N_c$ of the folded state (Figure \ref{fig:annularstripkine}(b)) and the inverted state (Figure \ref{fig:annularstripkine}(c)) are constrained in the $x-y$ plane to slide along the two rays $y=-\tan \tfrac{\pi}{N_c} x$ and $y=\tan \tfrac{\pi}{N_c} x$, respectively. In addition, the rotation axis of the full structure (which can be constructed by using symmetry properties) aligns with the $z$ axis.
	 $\delta_0$ and $\delta_1$ correspond to the inclined angle of the crease at the two ends. Because of the symmetry, in both the folded and inverted state, we have $\delta_0=\delta_1$. In our definition, $\delta_0 > 0$ for the inverted state and $\delta_0 < 0$ for the folded state. We assume that creases at the two ends remain straight as two generators.

	Moving of the the Darboux frame $(\bm{T}, \bm{N}, \bm{B})$ on the directrix can be described as $\bm{T}'= \kappa_n \bm{N}  - \kappa_g \bm{B}$, $\bm{N}'= -\kappa_n \bm{T} + \tau_g \bm{B}$, and $\bm{B}'= \kappa_g \bm{T} - \tau_g \bm{N}$, where a prime denotes an $s$-derivative, $\kappa_g$ is the preserved geodesic curvature under isometric deformation, and $\tau_g$ represents the geodesic torsion. In our notation, $\kappa_g=-1/R$. Generators (thin black lines) align with the vector $(\bm{B}+\eta \bm{T})$ and make a local angle $\beta$ with $\bm{T}$. Here $\eta$ is related to $\beta$ through $\eta=\cot \beta$. Adjacent generators intersect each other on a space curve called the edge of regression, whose analytical expression is included in Appendix \ref{appse:BVPcontinuation}.
	The developable annular sector in Figures \ref{fig:annularstripkine}(b-c) can be parameterized as 
	\begin{equation}\label{eq:developabledescription} 
	\begin{aligned}
	\bm{X}(s,v)=\bm{r}(s)+v (\bm{B}(s)+\eta(s) \bm{T}(s)) \,, \\
	\end{aligned}
	\end{equation}
	
with $v \in [0,V]$ and $s \in [0, 2 \pi R \alpha/N_c]$. $v$ is the coordinate along the generator whose length is $V \sqrt {1+ \eta ^2}$; $V$ can be determined by $\eta$, $s$ and the hole geometry through an implicit function $\chi (V, s, \eta,a,R) =0$. 
	 In Figure \ref{fig:annularstripkine}, for a thin disk with a concentric circular hole, $\chi$ can be written as \cite{yu2021cutting} 
	 
	 \begin{equation} \label{eq:circularkai} 
	 \begin{aligned}
	 \chi( \eta, V) = V^2 +(R^2-2V R -a^2) \sin^2 \beta \, ,
	 \end{aligned}
	 \end{equation}

	 where $V$ could be explicitly solved as a function of $\eta$ and the geometric parameters $a$ and $R$ \cite{dias2015wunderlich}. 
	 Later we will show that $\chi$ becomes complicated in the case with a nonvanishing eccentricity, where solving $V$ explicitly becomes nontrivial. Following \cite{yu2021cutting}, we treat $V$ as a variable and differentiate the algebraic constraint $\chi=0$ to obtain an additional differential equation. This technique makes it convenient for using the inextensible strip model to solve developable surfaces with any smooth boundaries, where the length of the generator may not be explicitly solved in terms of the geometry. Details are discussed in Appendix \ref{appse:BVPcontinuation}. 
	 
	 The mean curvature of the developable surface represented by Equation \eqref{eq:developabledescription} is $H=\tfrac{\kappa_n (1+\eta^2)}{2[1+v(\eta'+\kappa_g (1+\eta^2))]}$ and an area element can be written as $dA=[1+v(\eta' + \kappa_g (1+\eta^2))]dsdv$ \cite{yu2021cutting}.
	 The identical vanishing of the Gaussian curvature further requires $\eta=\tau_g / \kappa_n$.

	The total elastic energy $U$ of a creased annular strip includes elastic energy stored in the crease and the bending energy of the facets. We assume that the bending moment generated by the crease follows $K_c (R-a) \sin (\tilde{\gamma}_f - \gamma_0)$, where $K_c$ is the crease stiffness per unit length, $\gamma_0$ is the rest crease angle, and $\tilde{\gamma}_f$ measures the deformed crease angle \cite{brunck2016elastic}. We wish to emphasize that the mechanics of crease is complicated and a precise description of its angle-moment relationship does not exist \cite{thiria2011relaxation,lechenault2014mechanical}. In Appendix \ref{appse:linearcrease}, we demonstrate that with the crease following a linear angle-moment relationship, the major conclusions of this study (i.e., the qualitative observations in Figure \ref{fig:exptmodel}) are not affected and the numerical results only contain slight quantitative differences. 
	Note that in our crease model, the crease moment periodically vanishes at $\tilde{\gamma}_f=\gamma_{0} +i \pi$ ($i$ is an integer). With $\tilde{\gamma}_f \in [\gamma_0, \gamma_0 + \tfrac{\pi}{2}]$, the crease moment increases with the opening of the crease. On the other hand, with $\tilde{\gamma}_f \in [\gamma_0 + \tfrac{\pi}{2}, \gamma_0 + \pi]$, the crease enters a softening regime, where the crease moment decreases with the further opening of the crease. 
	Our numerical results in Section \ref{se:angledeficit} show that in certain parameter spaces, the crease could flip to reach a final crease angle around $\gamma_{0} +\pi$. We assume the thin sheet has a bending rigidity  $D=E t^3 /[12 (1 - \nu ^2)]$, where $E$ and $\nu$ are the material's Young's modulus and Poisson's ratio, respectively. The total elastic energy of the inverted and folded state of a thin disk with $N_c$ evenly spaced creases ($N_c \ge 2$) can be written as \cite{yu2021cutting}

	\begin{equation}\label{eq:totalenergygeneral}
	\begin{aligned}
	\frac{U}{N_c D}&=  \frac{K_{c}}{D}(R-a)  \int_{\gamma_0}^{\gamma_{f0}} \sin(\tilde\gamma_f-\gamma_0) d \tilde\gamma_f + \frac{1}{2}\int_0^{\tfrac{2 \pi R \alpha}{N_c}} \! \int_{0}^{V}{(2H)}^2 \,dA  \,, \\
	&= \frac{K_{c} R}{D} \left(1-\frac{a}{R}\right) \left[1-\cos (\gamma_{f0} - \gamma_0)\right] + \int_0^{\tfrac{2 \pi R \alpha}{N_c}} YW ds  \,,
	\end{aligned}
	\end{equation}

	with $Y=\frac{\kappa^2_n (1+\eta^2)^2}{2[\eta'+\kappa_g(1+\eta^2)]} $ and $W=\ln[1+V (\eta'+\kappa_g (1+\eta^2))]$. We have assumed that all the creases have the same length $(R-a)$ and the same final crease angle $\gamma_{f0}$. \emph{In this study, we assume the final crease angle is always constant along the crease length}.
	 Equation \eqref{eq:totalenergygeneral} needs minor modifications for a few cases in this paper. For example with $N_c=2$, introducing an eccentricity to the position of the hole results in creases with different lengths and different final crease angles (Figure \ref{fig:exptmodel}(h)). In the following sections, we will include the corresponding modifications of Equation \eqref{eq:totalenergygeneral} when necessary. 
	For thin sheets, the origami length $D/K_c$ is found to be proportional to the thickness of the material \cite{lechenault2014mechanical}. This makes the dimensionless crease stiffness $K_c R/D$ diverge as the material thickness goes to zero, resulting in a rigid crease that will not store any elastic energy.
	
	With our modeling of the creases as discrete hinges, the mechanics of the crease only balances the moments of the thin sheets at the boundaries, and does not appear in the Euler-Lagrange equations, given by \cite{starostin2007shape,dias2015wunderlich}
	
	\begin{align}
	\bm{F}' &= \bm{0} \,, \label{eq:stripgovern1} \\
	\bm{M}' + \bm{T} \times \bm{F} &=\bm{0} \,, \label{eq:stripgovern2} \\
	\partial_{\kappa_n} (YW) -\eta M_1 -M_3 &=0 \,, \label{eq:stripgovern3} \\ 
	\partial_\eta (YW) -(\partial_{\eta'} (YW))'-\kappa_n M_1 &=0 \label{eq:stripgovern4}\,,
	\end{align}
	where forces and moments, normalized by $D$, are resolved in the material frame through $\bm{F}=F_1 \bm{T}+F_2 \bm{N}+F_3 \bm{B}$ and  $\bm{M}=M_1 \bm{T}+M_2 \bm{N}+M_3 \bm{B}$. Equations (\ref{eq:stripgovern1}), (\ref{eq:stripgovern2}), and (\ref{eq:stripgovern3}-\ref{eq:stripgovern4}) represent the force balance, moment balance, and the constitutive laws, respectively. Together with a quaternion description of the rotations of the material frame and boundary conditions imposed at the two ends of a minimal facet, we obtain a two-point boundary value problem (TPBVP) and solve it with the continuation package AUTO 07P \cite{doedel2007auto}. To obtain a consistent prescription of boundary conditions for the quaternions, we follow Healey and Metha and introduce a dummy parameter \cite{healey2006straightforward,MooreHealey18}. The current implementation combined the merits of quaternions, which are free of polar singularity that Euler angles could suffer, and the merits of Euler angles, which are convenient for imposing boundary conditions explicitly containing ``rotation angles", such as the moment balance at the crease. 
	Detailed formulation of the TPBVP can be found in Appendix \ref{appse:BVPcontinuation}.

	Throughout the rest of this paper, the results from numerical continuation of the inextensible strip model are presented as solution curves, loci of the fold (which connect the inverted state and the energy barrier), and renderings corresponding to the symbols on the solution curves. All the numerical results have $R$ set to unity. 
	The solution curves measure the response of the creased disk through the angle $\delta_0$, the total elastic energy, and the change of the crease angle $(\gamma_{f0}-\gamma_{0})$ as certain parameter varies, e.g., the hole size $a/R$. In numerical continuation, we constrain the hole size $a/R$ in the range $[0.001,0.96]$. 
	The solution curves include the inverted state, the folded state, the energy barrier, and possibly a half-flipped state and a flipped state that exist only in certain parameter spaces. Numerical continuation could fail at a point (indicated by a cross) where the edge of regression contacts the material surface resulting in the blow-up of the local bending \cite{starostin2007shape,starostin2015equilibrium}.
	The renderings include the 3D deformed configurations and their developments on the 2D flat configurations. Both of them display the bending energy density (color maps of twice the squared mean curvature $2H^2$) on a minimal facet of the creased thin disk and the generators (black lines) on the rest facets, which are shown in grey. Only the edge of the regression of the facet to the left of the color map is included as red lines on the 3D renderings. In the flat developments of the configurations with $\alpha >1$ (i.e., with inserted sectors), the facets are slightly shifted outward to avoid overlapping.
	 We also examine some solutions in detail by reporting the distribution of the contact forces/moments, the curvature $\kappa_n$, the geodesic torsion $\tau_g$, and $\eta$.

	\section{Angle deficit} \label{se:angledeficit}
	
	With tabletop models, we observed that cutting a sector could allow for the increase in hole size of a creased disk significantly without destroying the bistability (Figures \ref{fig:exptmodel}(c-d)). 
	In this section, we study the effect of the angle deficit $\alpha$ on the bistability of a creased thin disk with two creases. $\alpha$ enters the two-point BVP through a scaling factor corresponding to the length of the directrix (see Appendix \ref{appse:BVPcontinuation}), which enables us to insert or remove materials by varying $\alpha$ directly. With $N_c=2$ and $\alpha=1$ , the folded state always contains two flat facets and is energy free. However with $N_c=2$ and $\alpha \neq 1$, the facets of the folded state could also be deformed. We will discuss the numerical results of folded state in Section \ref{se:numbercreases}.

		\begin{figure}[h!]
		\centering
		\includegraphics[width=0.95\textwidth]{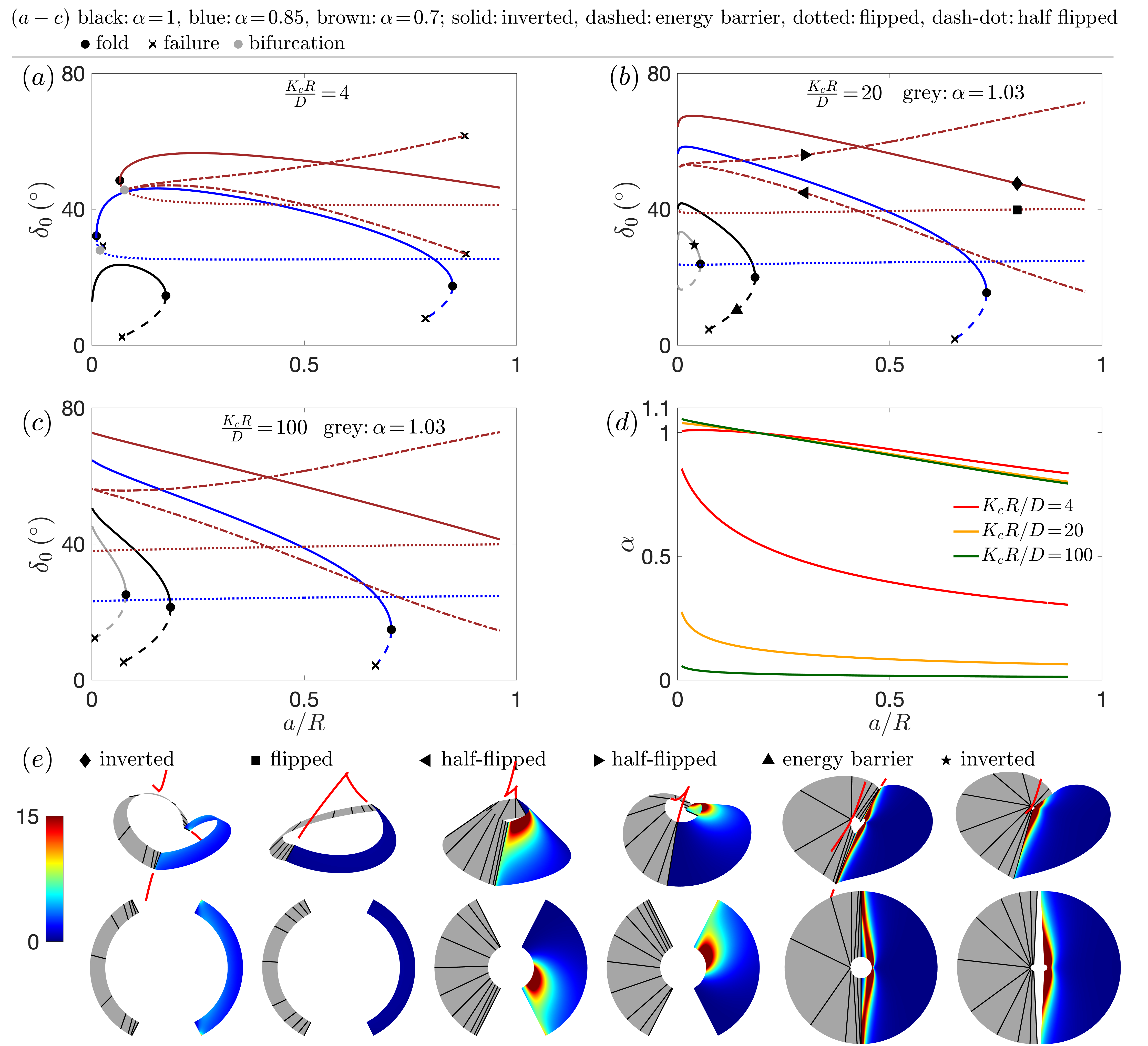}
		\caption{Solution curves ($\delta_0$ versus $a/R$) with different ($K_cR/D,\alpha$), and loci of folds ($\alpha$ versus $a/R$) with different $K_cR/D$. $(N_c,\gamma_0)$ is fixed to $(2,45^{\circ})$. (a) Solutions with a small crease stiffness $K_cR/D=4$. (b) Solutions with a moderate crease stiffness $K_cR/D=20$.
			(c) Solutions with a large crease stiffness $K_cR/D=100$. 
			(d) Loci of the fold. The area enclosed by the upper and lower boundary corresponds to the bistable region. (e) Renderings that correspond to the symbols in $(a-c)$.}
		\label{fig:alphaepsbifurphase}
	\end{figure}

	Figure \ref{fig:alphaepsbifurphase} reports solution curves with different dimensionless crease stiffness $K_cR/D$ and angle deficit $\alpha$ in panels (a-c), and the loci of the fold in panel (d). The rest crease angle $\gamma_0$ is fixed to $45^{\circ}$. Figure \ref{fig:alphaepsbifurphase}(e) shows several renderings corresponding to the symbols in Figure \ref{fig:alphaepsbifurphase}(b). 
	Figures \ref{fig:alphaepsbifurphase}(a-c) employ the hole size $a/R$ as the continuation parameter and the angle $\delta_0$ as the solution measure, and respectively correspond to a weak crease ($K_cR/D=4$), a crease with an intermediate stiffness ($K_cR/D=20$), and a strong crease ($K_cR/D=100$). Two elements are employed in Figures \ref{fig:alphaepsbifurphase}(a-c) to improve their readability. First, the solid lines, dashed lines, dotted lines, and dash-dot lines represent the stable inverted state, the unstable energy barrier, the flipped state, and the half flipped state, respectively. Second, colors are used for different $\alpha$ with black, blue, and brown corresponding to $\alpha=1.0$, 0.85, and 0.7, respectively. The grey curves in Figures \ref{fig:alphaepsbifurphase}(b-c) have $\alpha=1.03$. For example, the black solid and dashed lines in Figure \ref{fig:alphaepsbifurphase}(a) correspond to an inverted branch and an energy barrier branch with $\alpha=1$, respectively.

	Increasing the hole size $a/R$ could destroy the bistability through a fold (black circle), resulting in a critical hole size.
	Decreasing $\alpha$ generally leads to the increase of the critical hole size. With $\alpha=0.7$, the inverted branch could approach $a/R=1$ without a fold (we only report the portion up to 0.96), implying that the hole can be as large as the disk without loss of the bistability. \emph{This qualitatively matches our experimental observations demonstrated in Figures \ref{fig:exptmodel}c-\ref{fig:exptmodel}d}. On the other hand, increasing $\alpha$ (i.e., inserting a sector) reduces the critical hole size quickly. For example, see the grey curves in Figures \ref{fig:alphaepsbifurphase}(b-c), whose counterpart disappears in Figure \ref{fig:alphaepsbifurphase}(a) with a weak crease. We also notice that while the solution curves of the inverted branch (solid lines) in Figure \ref{fig:alphaepsbifurphase}(c) decline monotonically with the increase of $a/R$, they first rise a bit and then start declining in Figures \ref{fig:alphaepsbifurphase}(a-b). This is due to the fact that a small $a/R$ will generate a large bending moment from the conical shape at the two boundaries, which will open the crease angle a lot with a weak crease. Opening the crease angle generally flattens the inverted state and thus reduces its inclined angle $\delta_0$.

	A flipped state and a pair of half flipped states exist in certain parameter spaces.
	In Figure \ref{fig:alphaepsbifurphase}(a), with $\alpha=0.7$ and 0.85, decreasing $a/R$ could also lead to instability through a fold (black circle), which further connects to a flipped state and a pair of half-flipped states through a bifurcation point (grey circle). The pair of flipped states with $\alpha=0.85$ terminate soon after the bifurcation, where the edge of regression contacts the material surface and bending energy blows up locally. With a larger crease stiffness in Figures \ref{fig:alphaepsbifurphase}(b-c), the inverted state, the pair of half flipped states, and the flipped state are separated for $\alpha=0.7$; with $\alpha=0.85$, the pair of half flipped states cannot be obtained due to the local contact between the edge of regression and the material surface. The final crease angle of the flipped crease is about $(\pi+\gamma_0)$, at which the crease generates almost a vanishing moment but stores a finite elastic energy. This is different from the rest angle $\gamma_0$, at which both the crease moment and crease energy vanish. We remark that the additional fold and bifurcation at a small $a/R$ is due to our specific choice of the constitutive law for the crease, which follows a sinusoidal form and has a second fictitious rest angle of $(\pi+\gamma_0)$. In Appendix \ref{appse:linearcrease}, we give an example to show that with linear creases (i.e., a crease with linear angle-moment relationship), the additional fold and bifurcation at small $a/R$ disappear.

 Figure \ref{fig:alphaepsbifurphase}(d) reports the loci of the fold in the $\alpha$ versus $a/R$ plane with different crease stiffness $K_cR/D$. For each crease stiffness, we obtain an upper and a lower boundary, with the enclosed area corresponding to the bistable regime. The upper boundary corresponds to the fold that connects to the unstable energy barrier, and the lower boundary corresponds to the fold that connects to the flipped state. Starting with a bistable geometry, both increasing and decreasing $\alpha$ could cross the stability boundary and thus destroy the inverted state. 
 The curves at the upper left corner ($\alpha>1$) set the limit of the largest sector that can be inserted without destroying the inverted state. Actually only with small holes, $\alpha$ can slightly exceed unity and is always less than 1.1, which implies that only a small sector could be inserted without loss of the bistability. 
 In addition, the upper boundary slowly declines with the increase of $a/R$, following almost a linear relationship. This implies that by cutting a small sector, the critical hole size could be increased significantly. For example, with $(K_c R/D,\alpha)=(100,0.794)$, the critical hole size increases to $a/R=0.92$, corresponding to the right end of the green curve. The lower boundary sets the limit of the minimal material needed to preserve the inverted state. Although increasing the crease stiffness does not remarkably change the upper boundary, it shifts the lower boundary significantly downward, implying that the inverted state exists in a larger parameter space with a stronger crease. We remark that with the crease following a linear response, there is no such lower boundary. However, a similar upper boundary exists. Figure \ref{fig:alphaepsbifurphase}(e) displays several renderings, whose 3D profiles and their 2D projections of the outer and inner circumferences are documented in Figure \ref{appfig:alphaepsProjection} (Appendix \ref{appse:3Dprofile2Dprojection}).

\begin{figure}[h!]
	\centering
	\includegraphics[width=0.95\textwidth]{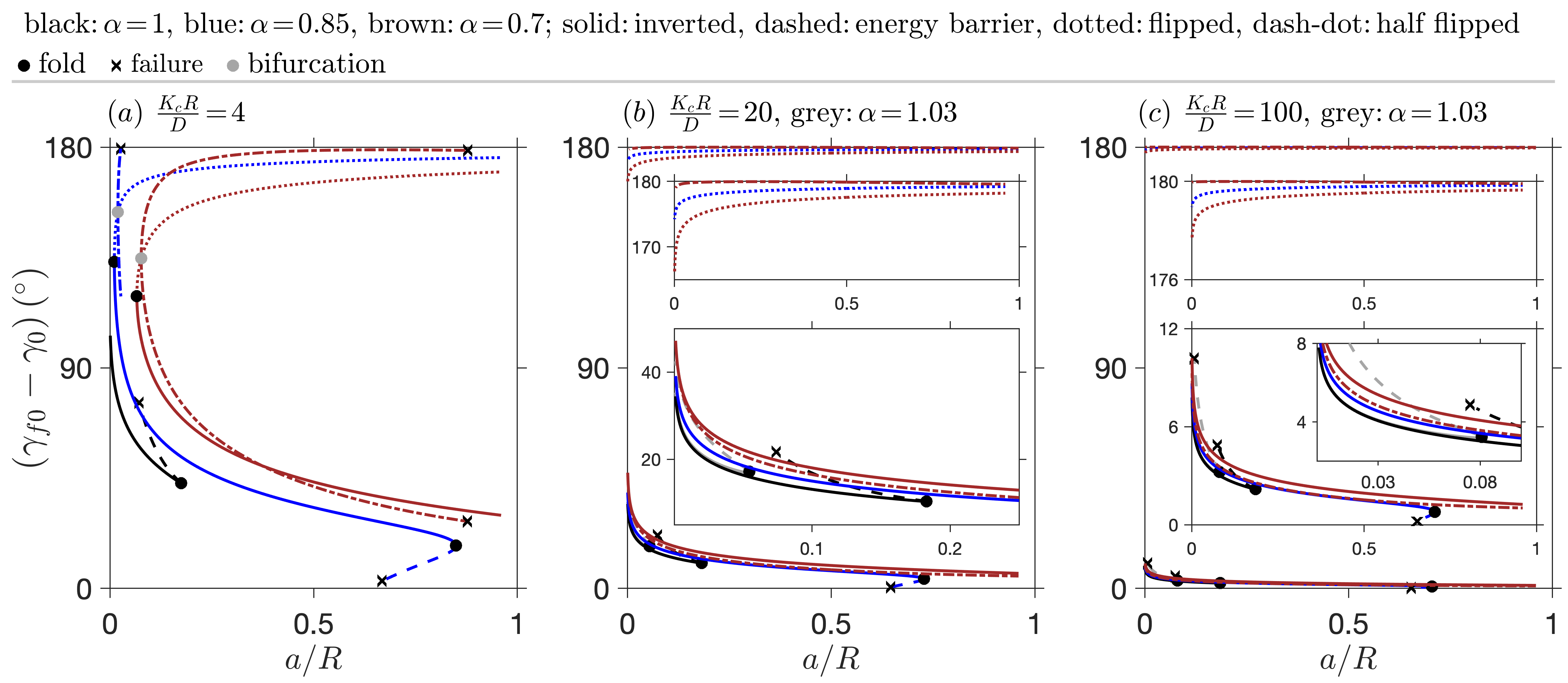}
	\caption{Changes of the crease angle at $s=0$, $(\gamma_{f0} - \gamma_0)$. (a), (b), and (c) correspond to the solutions in Figures \ref{fig:alphaepsbifurphase}(a), \ref{fig:alphaepsbifurphase}(b), and Figure \ref{fig:alphaepsbifurphase}(c), respectively.}
	\label{fig:alphaepsbiurgammaf0}
\end{figure}

 Figure \ref{fig:alphaepsbiurgammaf0} reports the deviation of the crease from the rest angle $(\gamma_{f0}-\gamma_{0})$ at $s=0$ for the solutions in Figures \ref{fig:alphaepsbifurphase}(a-c). Vertically, the curves are approximately divided into two groups: the top group is close to $180^{\circ}$, including the flipped state and one of the half flipped state with the crease at $s=0$ flipped; the bottom group contains the inverted state, the energy barrier, and the other half flipped state with the crease flipped at $s=\pi R \alpha$.
  With a weak crease $K_cR/D=4$, the top group is connected to the bottom group through folds and bifurcations. Increasing $K_cR/D$ to 20 and 100 separates the top and bottom groups and pushes the deviation $(\gamma_{f0}-\gamma_{0})$ to approach $180^{\circ}$ and $0^{\circ}$, respectively. Large $K_cR/D$ corresponds to a relatively rigid crease, which will force the final crease angle $\gamma_{f0}$ to be close to the moment-free crease angle $\gamma_0$ and $\gamma_0+ \pi$.
  In Figures \ref{fig:alphaepsbiurgammaf0}(b-c), the final crease angle $\gamma_{f0}$ of the bottom group approaches $\gamma_{0}$, while the final crease angle $\gamma_{f0}$ of the top group approaches $(\gamma_{0}+\pi)$, with the latter storing more crease energy.

\begin{figure}[h!]
	\centering
	\includegraphics[width=0.95\textwidth]{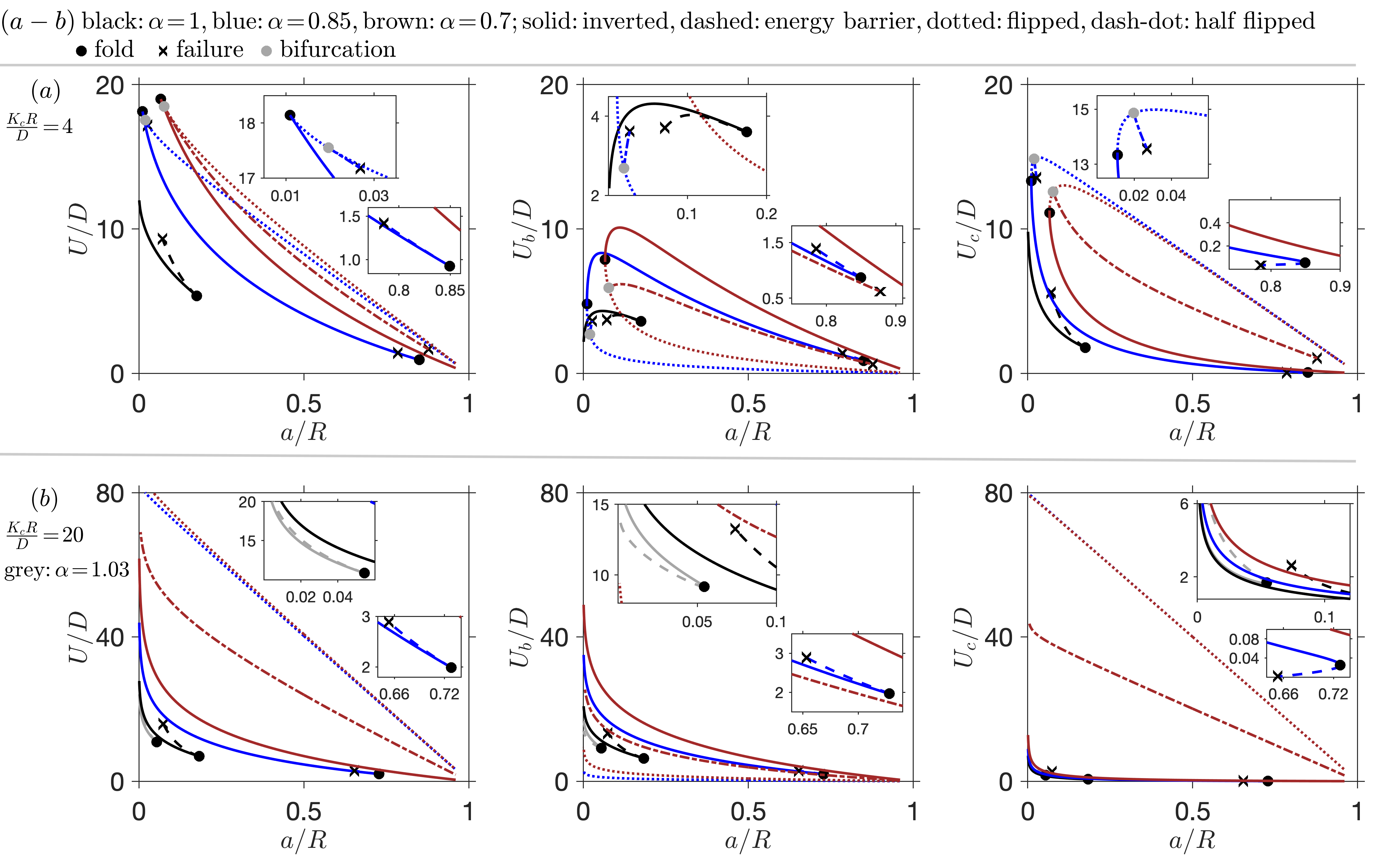}
	\caption{Normalized total energy $U/D$, bending energy $U_b/D$, and crease energy $U_c/D$. (a) and (b)  correspond to the solutions in Figures \ref{fig:alphaepsbifurphase}(a) and \ref{fig:alphaepsbifurphase}(b), respectively.}
	\label{fig:alphaepsN2Gamma45Energy}
\end{figure}

Figures \ref{fig:alphaepsN2Gamma45Energy}(a-b) report respectively the normalized elastic energies of the solutions in Figures \ref{fig:alphaepsbifurphase}(a-b), including the total energy $U/D$, the bending energy $U_b/D$, and the crease energy $U_c/D$. 
The total elastic energy $U/D$ of different states follows: stable inverted state $<$ energy barrier $<$ half-flipped state $<$ flipped state. Later we will show that the folded state generally contains much lower elastic energy than the inverted state. The bending energy $U_b/D$ follows: flipped state $<$ half-flipped state $<$ stable inverted state. However, the relationship between the bending energy of the inverted state and the energy barrier could reverse. For example, with $(K_cR/D, \alpha)=(4,1)$, energy barrier $<$ inverted state, while this is reversed for $(K_cR/D, \alpha)=(4,0.85)$. The crease energy $U_c/D$ follows: stable inverted state $<$ half-flipped state $<$ flipped state. The relationship between the crease energy of the inverted state and the energy barrier could reverse. For example, with $(K_cR/D, \alpha)=(4,1)$, inverted state $<$ energy barrier, while this is reversed for $(K_cR/D, \alpha)=(4,0.85)$. We conclude that flipping the crease generally reduces the bending energy, but increases the crease energy significantly, and thus increases the total energy.

With the hole size $a/R$ fixed, cutting material in the circumferential direction (i.e., decreasing $\alpha$) generally leads to an increase in both the bending energy and the crease energy of the inverted state. 
The former is similar to an exact cone made by joinning the two ends of a flat sector, in which a smaller sector generally results in a conical structure with a higher elastic energy.
Tabletop models show that for the inverted state, decreasing $\alpha$ generally leads to an increase in bending of the facet and the opening of the crease angle. Here our numerical results show that the intuitive decrease of the elastic energy due to the decrease of the bending area (caused by reducing $\alpha$) is exceeded by the increase of the bending energy density and the crease energy density. 
In addition, increasing $a/R$ (i.e., cutting materials in the radial direction) generally leads to a decrease in all the energies of all the states. An exception to this is the bending energy of disks with a weak crease ($K_c R/D=4$), in which the energy curves first rise a bit, and then decline with an increase in $a/R$. The reason this occurs also explains why the inclined angle $\delta_0$ first rises a bit and then declines in Figure \ref{fig:alphaepsbifurphase}(a): the large crease moment caused by a small hole will open a weak crease a lot and thus flattens the facet, which decreases the bending energy.

The ratio of the crease energy to the bending energy in the structure mainly depends on the dimensionless crease stiffness $K_c R/D$ and the hole size $a/R$. For the inverted state with a weak crease $K_c R/D=4$ and small $a/R$, the two creases contribute more energy than the bending of the facets. 
On the other hand, increasing $a/R$ will unload the crease and increase the bending deformation of the facet, which reverses the energy contribution. In other words, with $K_c R/D=4$ and large $a/R$, the bending of the facets contributes more energy than the two creases. However, for the flipped and half flipped branch, the crease energy contributes more than the bending energy.
Increasing $K_c R/D$ to 20 significantly increases the total elastic energy in the system. With $K_c R/D=20$, the portion from the crease energy reduces significantly for the inverted state and the energy barrier, in which most of the elastic energy comes from the bending of the facets. However, for the half flipped and particularly the flipped states, the contribution from the bending energy is small and most of the energy comes from the flipped crease.

Symmetries in the structure could facilitate our understanding of the contact force and moment on the directrix $\bm{r}(s)$. The inverted state, the flipped state and the energy barrier have two-fold mirror symmetries, with the plane spanned by the two creases and the $x-z$ plane (see Figure \ref{fig:annularstripkine}) being planes of symmetry. These mirror symmetries vanish the contact force $\bm{F}$ identically and result in a constant contact moment in the $z$ direction. The pair of half flipped states is only symmetric about the plane spanned by the two creases (i.e. the $y-z$ plane), resulting in a constant contact force in the $x$ direction and a nonconstant contact moment in the $y-z$ plane.

\begin{figure}[h!]
		\centering
		\includegraphics[width=0.95\textwidth]{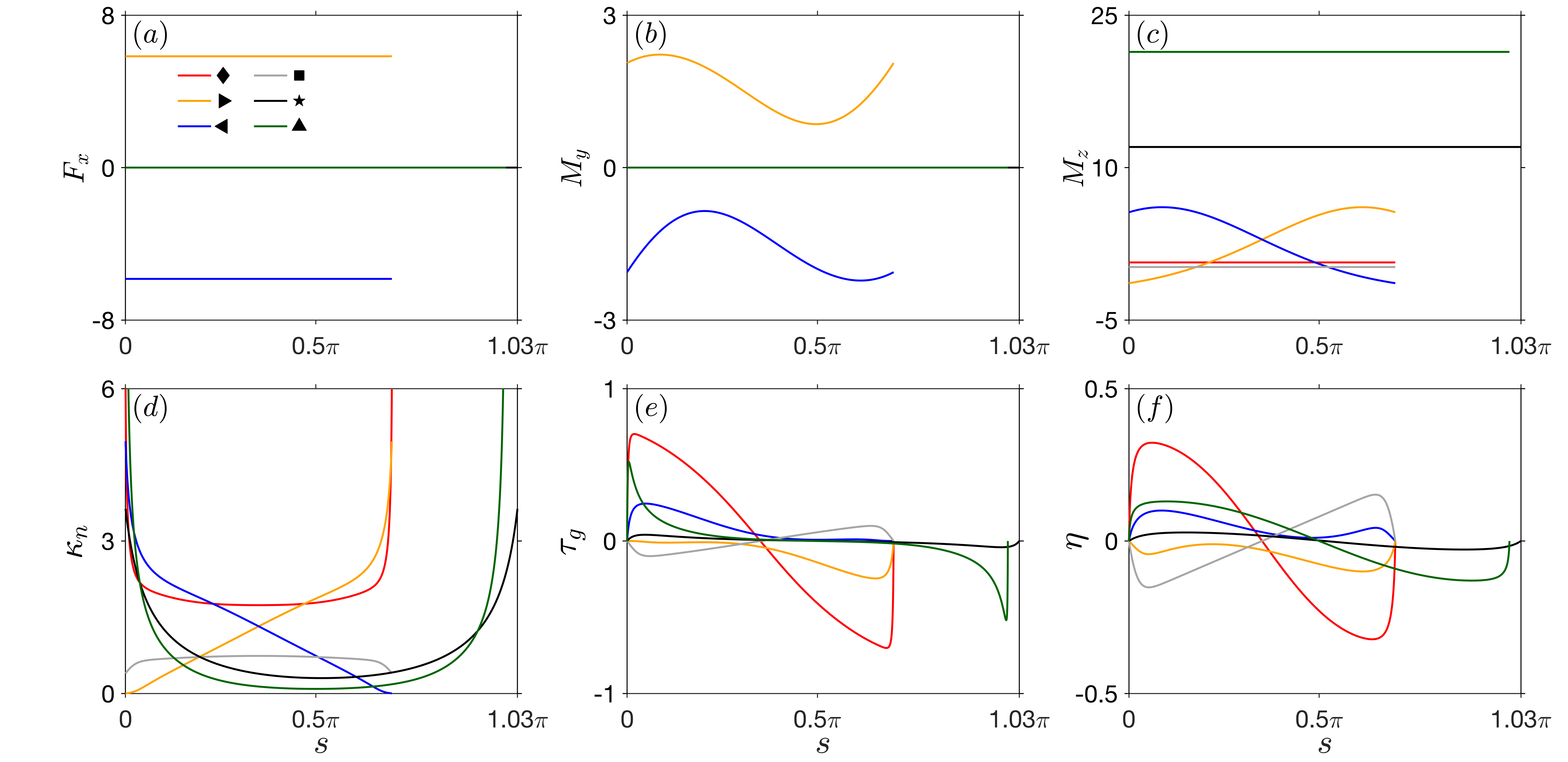}
		\caption{The Cartesian component of contact force/moment and several geometric quantities, corresponding to the renderings in Figure \ref{fig:alphaepsbifurphase}(e). Because of the symmetry, only one facet is reported. (a) $F_x$. (b) $M_y$. (c) $M_z$. (d) Normal curvature $\kappa_n$. (e) Geodesic torsion $\tau_g$. (f) $\eta$.}
		\label{fig:alphaepsForcesCurvatures}
	\end{figure}

Figure \ref{fig:alphaepsForcesCurvatures} reports the Cartesian component of the contact force/moment and several geometric quantities for the renderings shown in Figure \ref{fig:alphaepsbifurphase}(e).
Figures \ref{fig:alphaepsForcesCurvatures}(a-c) report $F_x$, $M_y$, and $M_z$, respectively. The horizontal axis represents the arc length $s$, up to $1.03\pi$ for $\blackstar$ ($\alpha=1.03$).
Other Cartesian components $F_y$, $F_z$, and $M_x$ are set to zero through boundary conditions due to the mirror symmetry about the $y-z$ plane (Appendix \ref{appse:BVPcontinuation}). Only the pair of half flipped state ($\righttriangle$ and $\lefttriangle$) have nonvanishing $F_x$ (which is constant) and nonvanishing $M_y$ (which is nonconstant). $M_z$ in the pair of half flipped states keeps varying along the arc length, and is constant for the other renderings including the two inverted states ($\blackdiamond$ and $\blackstar$), a flipped state ($\blacksquare$), and an energy barrier ($\triangleup$). These predictions match with our symmetry analysis. In addition, $F_x$ of the two half flipped states have equal magnitude but opposite sign ($F_x<0$ for $\lefttriangle$ and $F_x>0$ for $\righttriangle$), which imply that the non-flipped end of the directrix is under compression, while the flipped end is under tension.     
Figures \ref{fig:alphaepsForcesCurvatures}(d-f) report the normal curvature $\kappa_n$, geodesic torsion $\tau_g$, and $\eta$, respectively. Notice that in the two half-flipped states, the normal curvature $\kappa_n$ approaches zero at the flipped crease, implying that a singularity (corresponding to $\kappa_n=0$) is about to form and could move inside the integral interval \cite{starostin2015equilibrium,yu2019bifurcations}.

\newpage
	
	\section{Rest crease angle}\label{se:restangle} 
	
	The bistability of a creased thin disk is created by introducing non-flat crease angles; decreasing $\gamma_0$ (i.e., folding the crease more heavily) generally makes the inverted state more stable. However, in certain parameter spaces, we observed that decreasing $\gamma_0$ could destroy the inverted state. For example, with the removal of a sector (i.e., $\alpha<1$), it is observed with tabletop models that the rest crease angle $\gamma_0$ must be large enough to stabilize the inverted state. In this section, we study the effect of the rest crease angle $\gamma_0$ on the bistability. We vary the rest crease angle $\gamma_0$ with different hole size and angle deficit. Other parameters $(N_c, K_cR/D)$ are fixed to $(2,20)$.

	\begin{figure}[h!]
	\centering
	\includegraphics[width=0.95\textwidth]{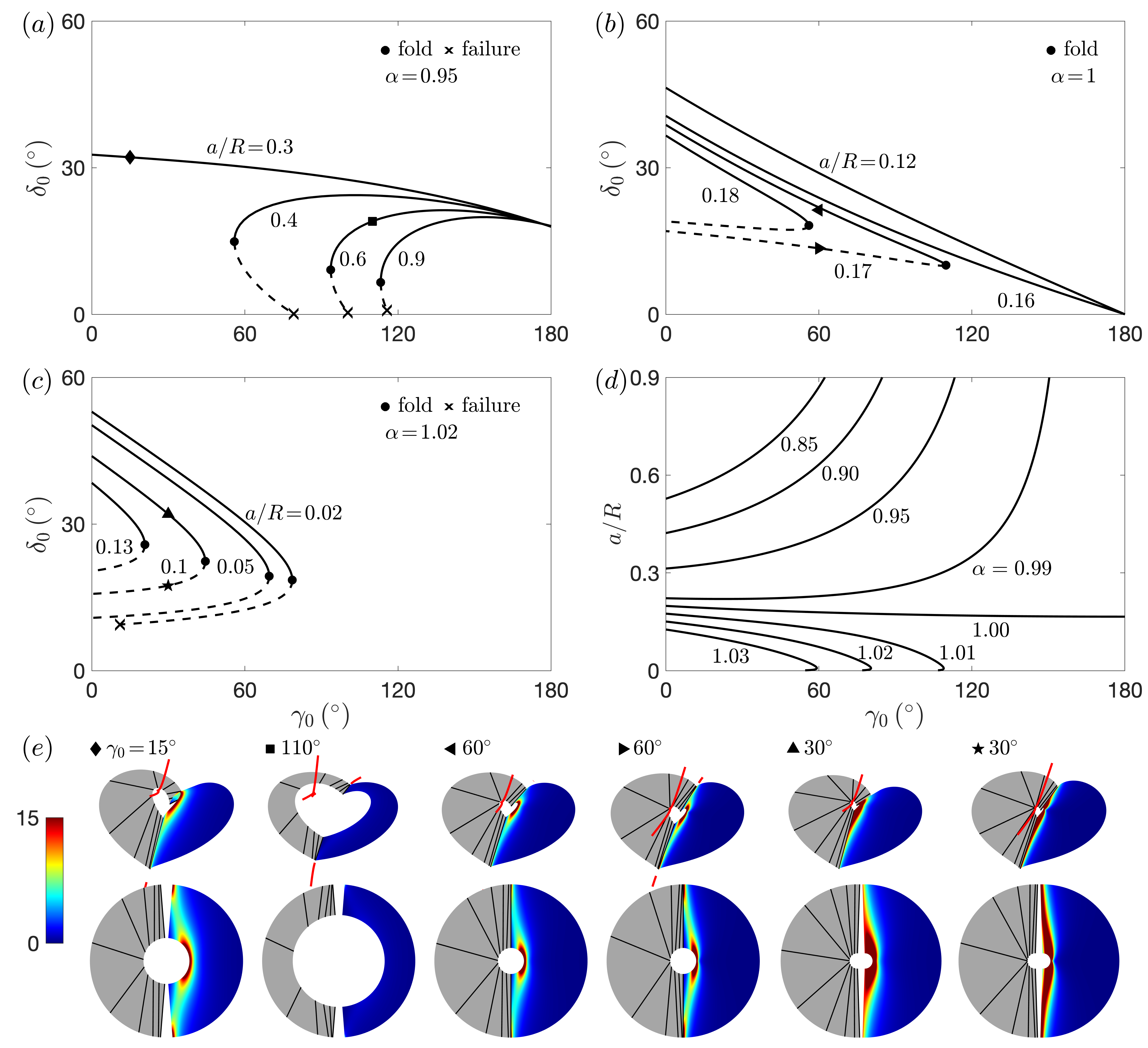}
	\caption{Solutions curves ($\delta_0$ versus $\gamma_0$) of the inverted state (solid lines) and energy barrier (dashed lines), and loci of folds ($a/R$ versus $\gamma_0$) with different $(\alpha,a/R)$. $(K_cR/D, N_c)$ is fixed to $(20, 2)$. (a) $\alpha=0.95$. (b) $\alpha=1.0$
		(c) $\alpha=1.02$. 
		(d) Loci of the fold with different $\alpha$. (e) Renderings that correspond to the symbols in (a-c). }\label{fig:diepsbifurphase}
\end{figure}

	Figure \ref{fig:diepsbifurphase} summarizes the numerical results from the inextensible strip model.  Figures \ref{fig:diepsbifurphase}(a-c) employ the rest angle $\gamma_{0}$ as the continuation parameter and the angle $\delta_0$ as the solution measure, and have respectively $\alpha=0.95$, $1.0$, and $1.02$. With $\alpha=0.95$ and large hole sizes $a/R=0.4$, 0.6, and 0.9, decreasing $\gamma_0$ could destroy the bistability through a fold; with a smaller hole such as $a/R=0.3$, the inverted state exists for the entire range $\gamma_0 \in [0^{\circ},180^{\circ}]$. 
	 At $\gamma_0=180^{\circ}$, the four solution curves merge approximately at the point $(\gamma_{0}, \delta_0) \approx (180^{\circ}, \cos^{-1} 0.95)$, which corresponds to a perfect cone. With a rigid crease, they will merge exactly at $(180^{\circ}, \cos^{-1} 0.95)$.

	  With $\alpha=1.0$ and 1.02 (Figures \ref{fig:diepsbifurphase}(b-c)), increasing $\gamma_0$ destroys the bistability through a fold. With $\alpha=1$, the solution curves could merge at $(\gamma_0, \delta_0) =(180^{\circ},0^{\circ})$ for small holes, corresponding to a flat annulus. With $\alpha=1.02$, the inverted state is destroyed far before the rest crease angle reaches $\pi$ for $a/R \ge 0.02$. Figure \ref{fig:diepsbifurphase}(d) shows the loci of the fold in the $\gamma_0$ versus $a/R$ plane with different $\alpha$. The area below the loci curve corresponds to the bistable region. Starting with a bistable geometry, increasing the hole size $a/R$ will generally cross the stability boundary from the bistable region to a monostable region, and thus destroys the bistability. With $\alpha \le 0.99$, decreasing $\gamma_0$ could destroy the bistability, while with $\alpha \geq 1$, increasing $\gamma_0$ destroys the bistability. With $\alpha =1$, the stability boundary is almost a horizontal line and the critical hole size is not sensitive to $\gamma_0$.
	 A tiny reentry exists at the bottom right of the curves with $\alpha=1.01, 1.02$ and 1.03. We did not explore the details of these structures in this paper. Figure \ref{fig:diepsbifurphase}(e) shows several renderings corresponding to the symbols in Figures \ref{fig:diepsbifurphase}(a-c). \emph{The numerical results presented in this section confirm that together with the angle deficit $\alpha$, both decreasing and increasing the rest crease angle $\gamma_{0}$ could stabilize or destabilize the inverted state}.

\section{Number of evenly distributed creases}\label{se:numbercreases}

A thin disk can be decorated with a pattern of creases. 	
In this section, we study the inverted and folded state with different number of evenly spaced creases. The facets of the folded state are generally deformed for $N_c \neq 2$. With $N_c = 2$ and $\alpha \neq 1$, the facets of the folded state could also be deformed. With a single crease $N_c = 1$, the inverted state has one mirror symmetry. We solve half of the structure and only impose the crease boundary condition at one end (Appendix \ref{appse:BVPcontinuation}). In this case, we have only one crease contributing to the elastic energy in Equation \ref{eq:totalenergygeneral}.

\subsection{The inverted state}\label{sse:invertedstate}

\begin{figure}[h!]
	\centering
	\includegraphics[width=0.95\textwidth]{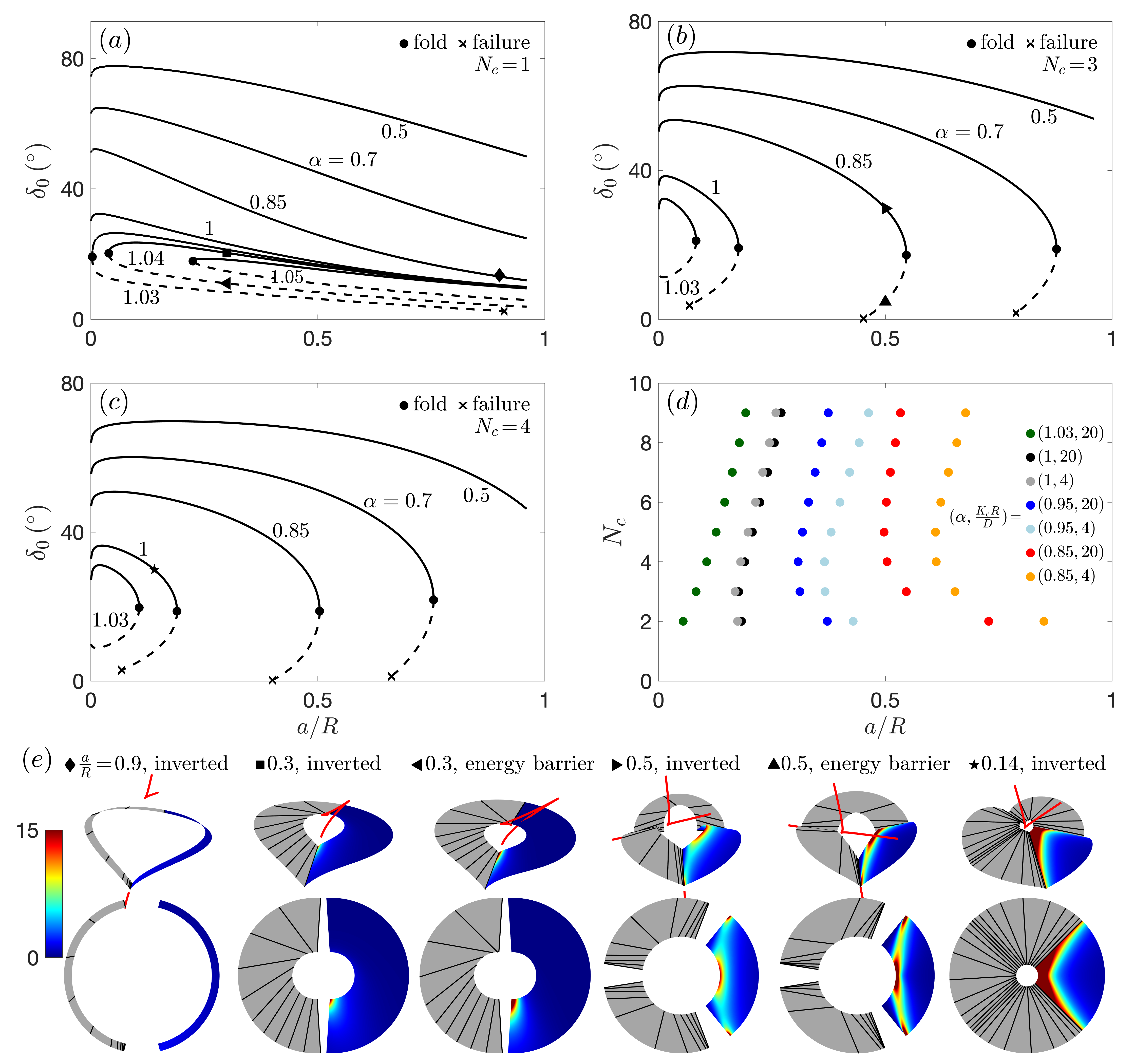}
	\caption{Solution curves ($\delta_0$ versus $a/R$) of the inverted state (solid lines) and energy barrier (dashed lines) with different $(N_c, \alpha)$, and loci of the fold ($a/R$ versus $N_c$) with different $(\alpha, K_c R/D)$. In $(a-c)$, $(K_cR/D, \gamma_{0})$ is fixed to $(20,45^{\circ})$. (a) $N_c=1$. 
		(b) $N_c=3$.  
		(c) $N_c=4$.
		(d) Loci of the fold with $\gamma_0$ fixed to $45^{\circ}$.
		(e) Renderings that correspond to the symbols in $(a-c)$.} \label{fig:nfepsbifurphaseinverted}
\end{figure}

In Figure \ref{fig:nfepsbifurphaseinverted}, the crease angle is fixed to $\gamma_0=45^{\circ}$. We employ $a/R$ as the continuation parameter and $\delta_0$ as the solution measure with different $(N_c,\alpha)$ in Figures \ref{fig:nfepsbifurphaseinverted}(a-c), which correspond to $N_c=1$, $3$, and $4$, respectively. Figure \ref{fig:nfepsbifurphaseinverted}(d) reports the loci of the fold in the $N_c$ versus $a/R$ plane with different $(\alpha, K_c R/D)$ through a series of discrete points.

Figure \ref{fig:nfepsbifurphaseinverted}(a) shows that with a single crease $N_c=1$ and $\alpha \le 1$, $a/R$ on the inverted state (solid lines) could be continued in the entire range $[0.001, 0.96]$ without any fold or bifurcation. \emph{This matches with our experimental observation that with a single crease, the hole could be as large as the disk without loss of the bistability (Figure \ref{fig:exptmodel}g)}. In the same diagram with $\alpha = 1.03$, 1.04, and 1.05, the hole could still be as large as the disk. However, decreasing the hole size could destroy the bistability through a fold, which connects to the energy barrier state. Increasing $\alpha$ quickly moves the fold toward the right limit $a/R=1$. Our results show that to preserve the inverted branch, $\alpha$ cannot exceed 1.1. In other words, with a single crease, only a small sector is allowed to be inserted without loss of the inverted state.
The solution curves in Figures \ref{fig:nfepsbifurphaseinverted}(b-c) with $N_c=3$ and 4 share several features: increasing hole size $a/R$ destroys the bistability through a fold that connects to the energy barrier; decreasing $\alpha$ from 1 to 0.7 leads to a significant increase of the critical hole size; the critical hole size quickly drops to zero as $\alpha$ is slightly larger than 1. Figure \ref{fig:nfepsbifurphaseinverted}(d) reports the loci of the fold in the $N_c$ versus $a/R$ plane (up to nine creases) with several $(\alpha, K_cR/D)$. While the critical hole size increases monotonically with the increase of $N_c$ for $\alpha=1.03$, it first decreases a bit and then reverses to increase for $\alpha \le 1$. The reverse effect becomes more pronounced with the decrease of $\alpha$.
In addition, a stronger crease generally leads to a larger critical hole size.
The mechanical behavior with $N_c=1$ is qualitatively different from $N_c \ge 2$ and is not included in Figure \ref{fig:nfepsbifurphaseinverted}(d). For example, with $\alpha \le 1$, there is no fold in the entire range $a/R \in [0.001, 0.96]$. Figure \ref{fig:nfepsbifurphaseinverted}(e) shows several renderings corresponding to the symbols in Figures \ref{fig:nfepsbifurphaseinverted}(a-c).

\begin{figure}[h!]
		\centering
		\includegraphics[width=0.95\textwidth]{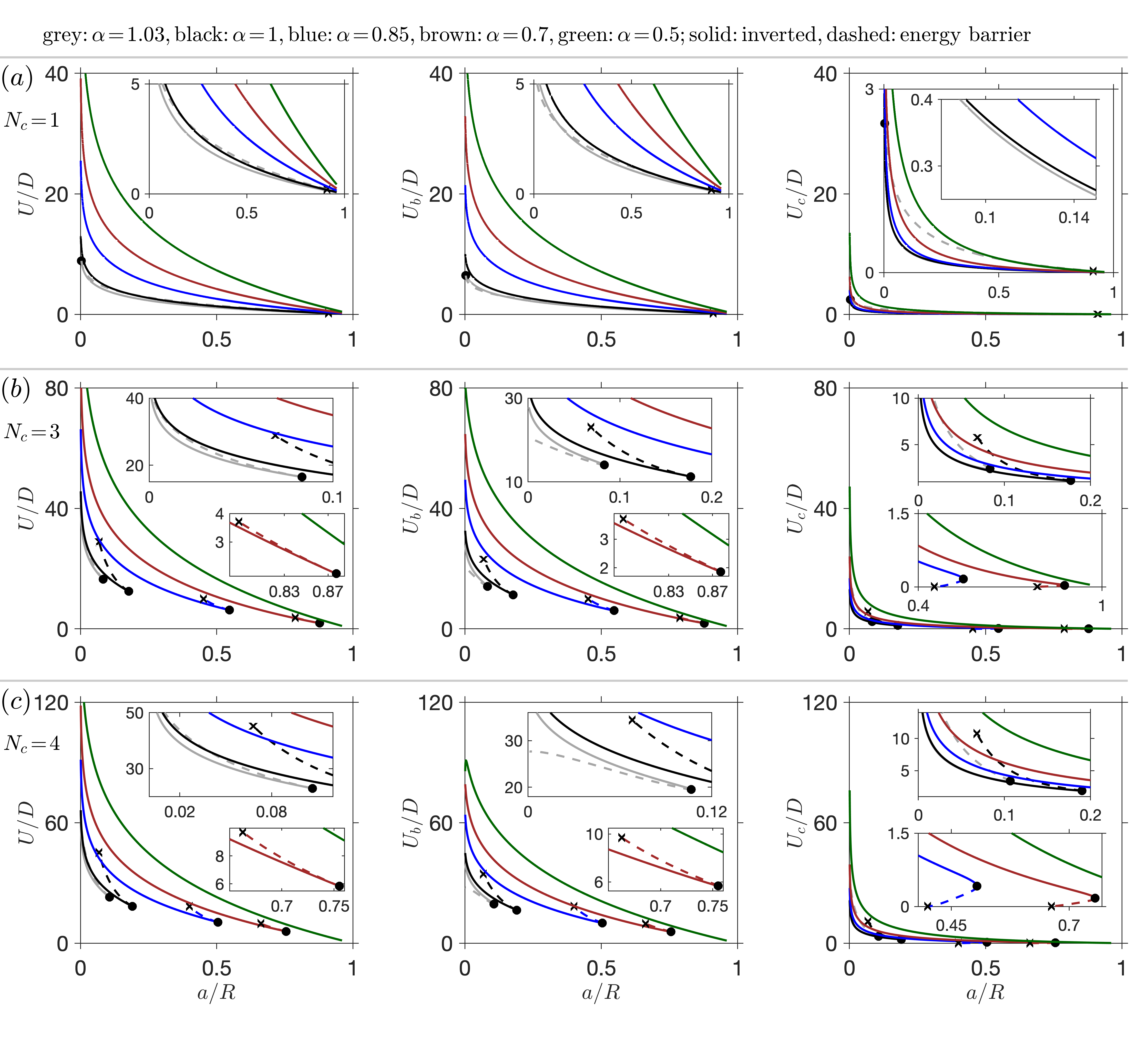}
		\caption{Normalized total energy $U/D$, bending energy $U_b/D$, and crease energy $U_c/D$. (a), (b), and (c) correspond to
the solutions in Figures \ref{fig:nfepsbifurphaseinverted}(a), \ref{fig:nfepsbifurphaseinverted}(b), and \ref{fig:nfepsbifurphaseinverted}(c), respectively.} \label{fig:nfepsN2Gamma45Energy}
\end{figure}

Figures \ref{fig:nfepsN2Gamma45Energy}(a-c) report the normalized elastic energies of the solutions in Figures \ref{fig:nfepsbifurphaseinverted}(a-c), respectively, including
the total energy $U/D$, the bending energy $U_b/D$, and the crease energy $U_c/D$. The solutions in Figure \ref{fig:nfepsbifurphaseinverted}(a) with $\alpha=1.04$ and 1.05 are not included for clarity. Similar to the results in Figure \ref{fig:alphaepsN2Gamma45Energy} with $N_c=2$, decreasing $\alpha$ generally leads to the increase of the various energies, and increasing $a/R$ generally reduces these energies. In addition, a thin disk decorated with more creases generally contains more bending energy, crease energy, and thus the total elastic energy.

\begin{figure}[htp!!!]
	\centering
	\captionsetup[subfigure]{labelfont=normalfont,textfont=normalfont}
	\centering
	\includegraphics[width=0.95\textwidth]{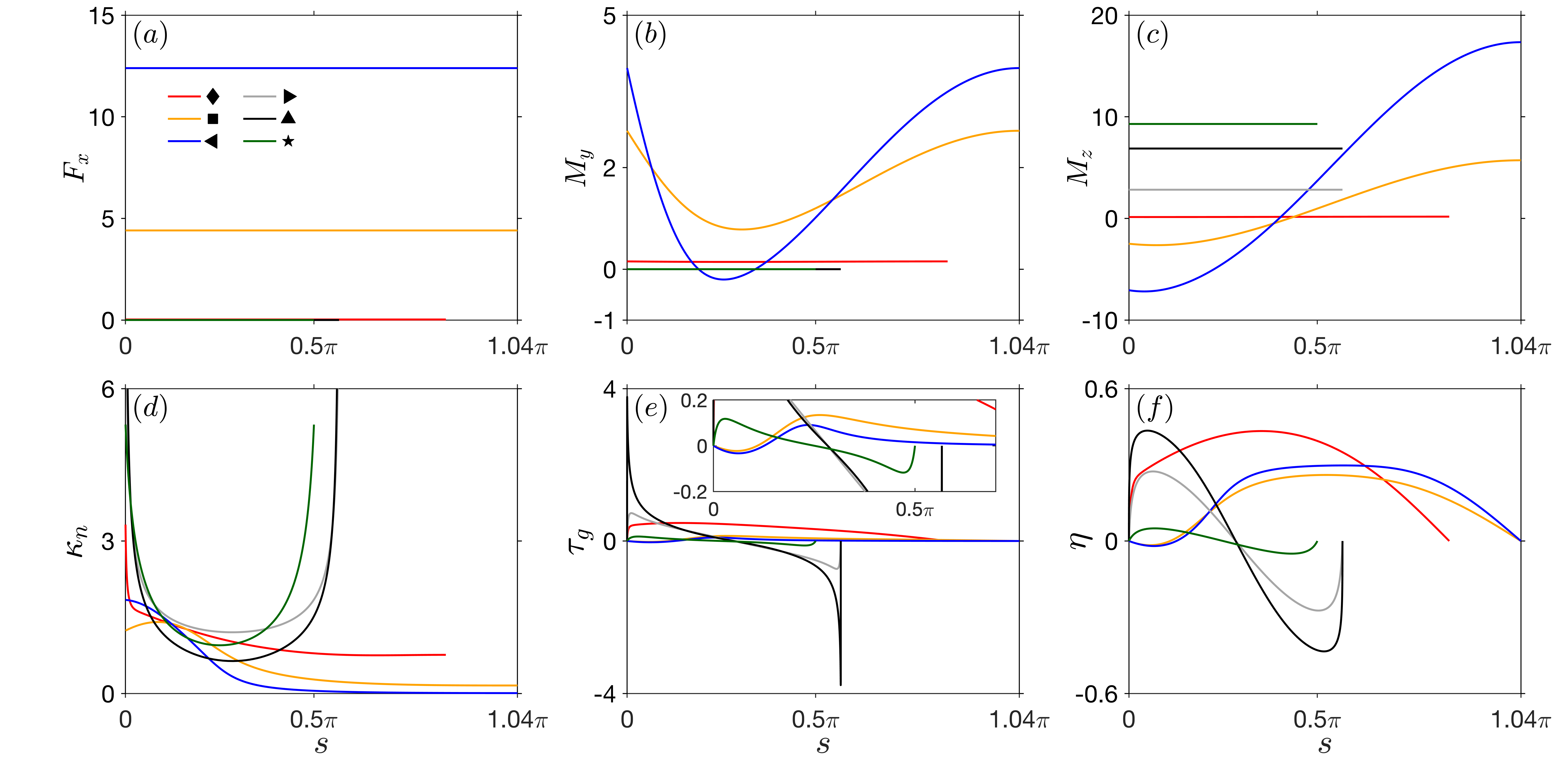}
	\caption{The Cartesian component of the contact force/moment and several geometric quantities of the renderings in Figure \ref{fig:nfepsbifurphaseinverted}(e). Because of the symmetry, only one facet is
reported. (a) $F_x$. (b) $M_y$. (c) $M_z$. (d) Normal curvature $\kappa_n$. (e) Geodesic torsion $\tau_g$. (f) $\eta$. }\label{fig:nfepsForcesCurvatures}
\end{figure}

With a single crease $N_c=1$, the system has one mirror symmetry; while with $N_c \ge 2$, the inverted state and energy barrier have $N_c$-fold mirror symmetries. As a result, with $N_c=1$, the contact force $\bm{F}$ is a constant vector in the $x$ direction and the contact moment is restricted in the $y-z$ plane. With $N_c=3$ and 4, the contact force vanishes identically, and the contact moment $\bm{M}$ is a constant vector in the $z$ direction. 
Figures \ref{fig:nfepsForcesCurvatures}(a-c) present the Cartesian component of the contact force and moment $F_x$, $M_y$, and $M_z$ of the renderings in Figure \ref{fig:nfepsbifurphaseinverted}(e), respectively. The numerical results match with our symmetry analysis.
With $(N_c, a/R)=(1,0.9)$ (i.e., $\blackdiamond$), $F_x$ is found to be a nonvanishing small constant 0.0252, and its corresponding $M_y$ and $M_z$ varies slowly along the arc length. $F_x>0$ for all the renderings with $N_c=1$ ($\blackdiamond$, $\blacksquare$, and $\lefttriangle$), which implies that the creased end of the directrix $\bm{r}(s)$ is in tension, while the non-creased end of the directrix is under compression. 
Figures \ref{fig:nfepsForcesCurvatures}(d-f) display the normal curvature $\kappa_n$, geodesic torsion $\tau_g$, and $\eta$ of the renderings in Figure \ref{fig:nfepsbifurphaseinverted}(e), respectively. Notice that in the energy barrier $\lefttriangle$, the normal curvature $\kappa_n$ approaches zero at the noncreased end at $s=\alpha \pi R$, implying that a singularity (corresponding to $\kappa_n=0$) is about to form and could move inside the integral interval.	

\clearpage

\subsection{The folded state}\label{sse:foldedstate}
In tabletop models, we observed that with a single crease $N_c=1$, an inflection point (i.e., $\kappa_n=0$) exists on the folded state, which could lead to local divergence of the bending energy and will bring significant difficulty in solving the inextensible strip model \cite{starostin2015equilibrium,yu2019bifurcations}.
We did not include the solutions of the folded state with $N_c=1$ in this study.

\begin{figure}[h!]
	\centering
	\includegraphics[width=0.95\textwidth]{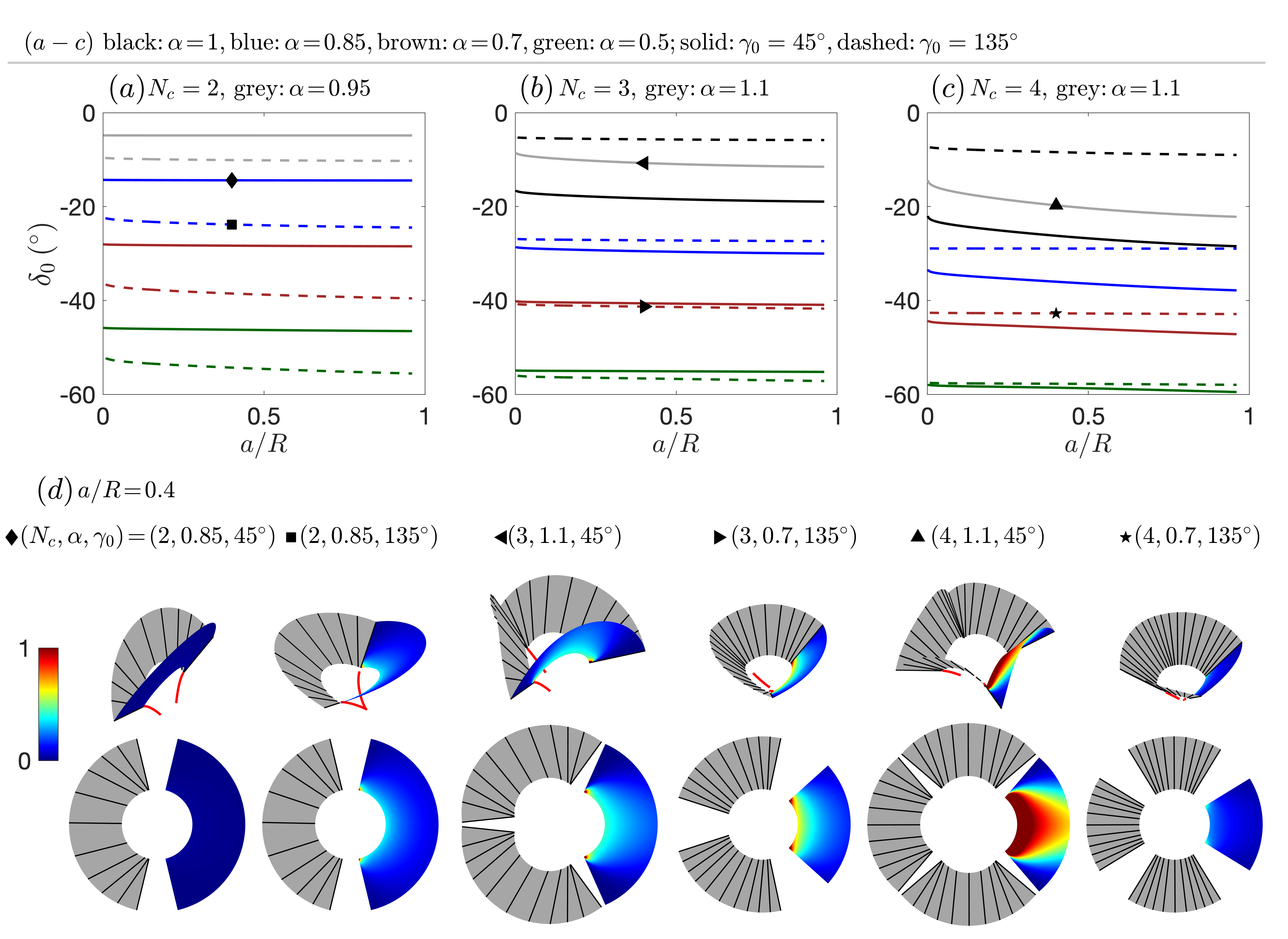}
	\caption{Solution curves of the folded state ($\delta_0$ versus $a/R$) with different ($N_c, \alpha, \gamma_0$). $K_cR/D$ is fixed to 20. (a) $N_c=2$. (b) $N_c=3$. (c) $N_c=4$. (d) Renderings that correspond to the symbols in $(a-c)$.} \label{fig:nfepsbifurphasefolded}
\end{figure}

Figures \ref{fig:nfepsbifurphasefolded}(a-c) present the solution curves of the folded state with $N_c=2$, $N_c=3$ and $N_c=4$, respectively. We employ $a/R$ as the continuation parameter and $\delta_0$ as the solution measure with different $(\alpha,\gamma_{0})$. Black curves are omitted in Figure \ref{fig:nfepsbifurphasefolded}(a) because the folded state with $(N_c,\alpha)=(2,1)$ remains flat and is trivial. The grey dashed curve with $\alpha=1.1$ is not included in Figures \ref{fig:nfepsbifurphasefolded}(b-c) because they cannot be obtained with the inextensible strip model due to the local contact between the edge of regression and the material surface.
The folded state is stable in the entire range $a/R \in [0.001, 0.96]$. Figure \ref{fig:nfepsbifurphasefolded}(d) displays several renderings corresponding to the symbols in Figures \ref{fig:nfepsbifurphasefolded}(a-c). The facets could be convex ($\blackdiamond$, $\blacksquare$, $\righttriangle$ and $\blackstar$) or concave ($\lefttriangle$ and $\triangleup$), depending on the choice of the geometric parameters $(\alpha,N_c,\gamma_{0})$. Convex facets tend to close the crease angle (i.e., $\gamma_{f0}<\gamma_{0}$), while concave facets tend to open the crease angle (i.e., $\gamma_{f0}>\gamma_{0}$).

\begin{figure}[h!]
	\centering
	\captionsetup[subfigure]{labelfont=normalfont,textfont=normalfont}
	\centering
	\includegraphics[width=0.95\textwidth]{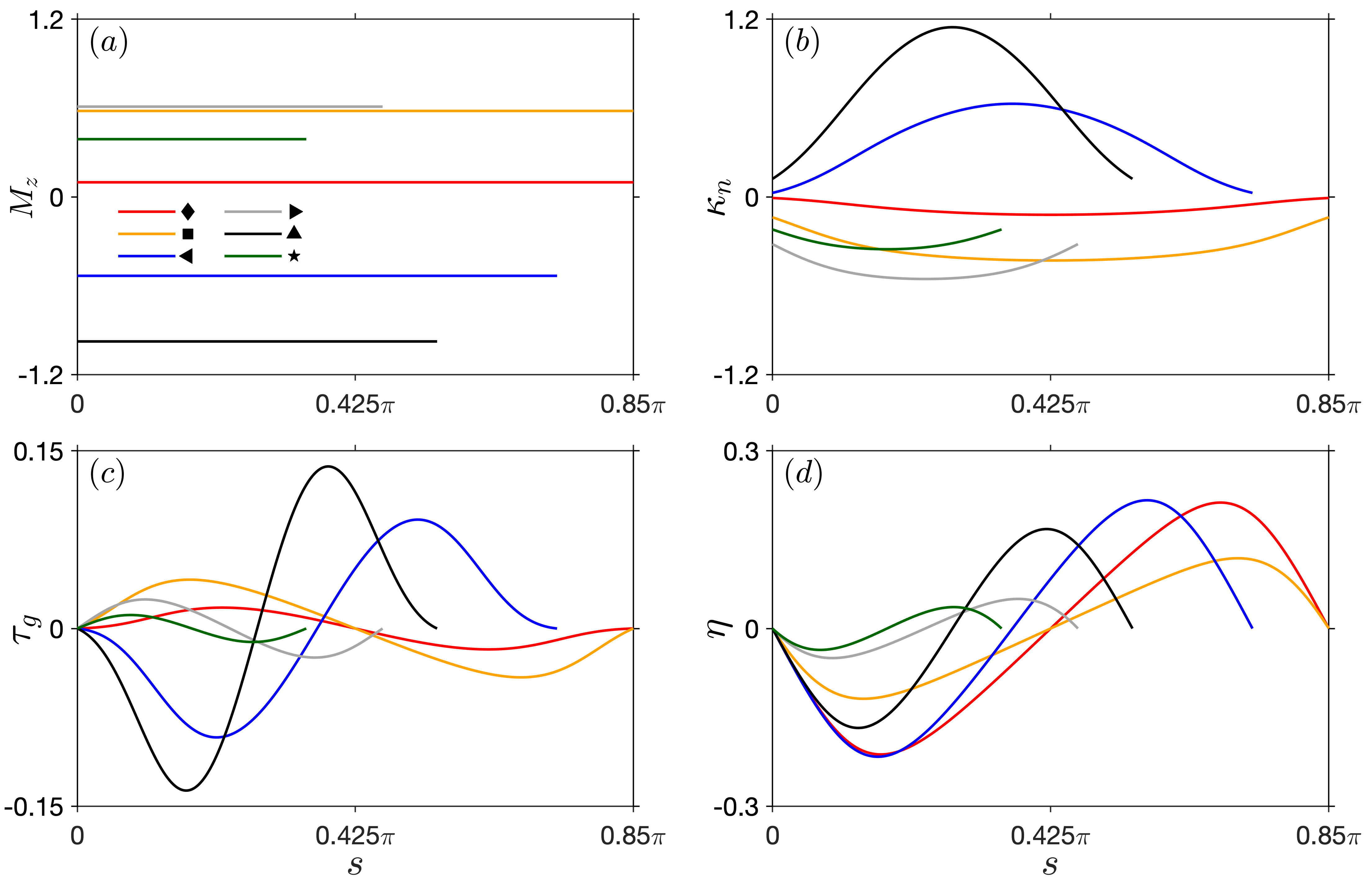}
	\caption{The Cartesian component of the contact moment and several geometric quantities, corresponding to the renderings in Figure \ref{fig:nfepsbifurphasefolded}(d). Because of the symmetry, only one facet is reported. (a) $M_z$. (b) Normal curvature $\kappa_n$. (c) Geodesic torsion $\tau_g$. (d) $\eta$.} \label{fig:nffoldedForcesCurvatures}
\end{figure}

\begin{figure}[h!]
	\centering
	\captionsetup[subfigure]{labelfont=normalfont,textfont=normalfont}
	\centering
	\includegraphics[width=0.9\textwidth]{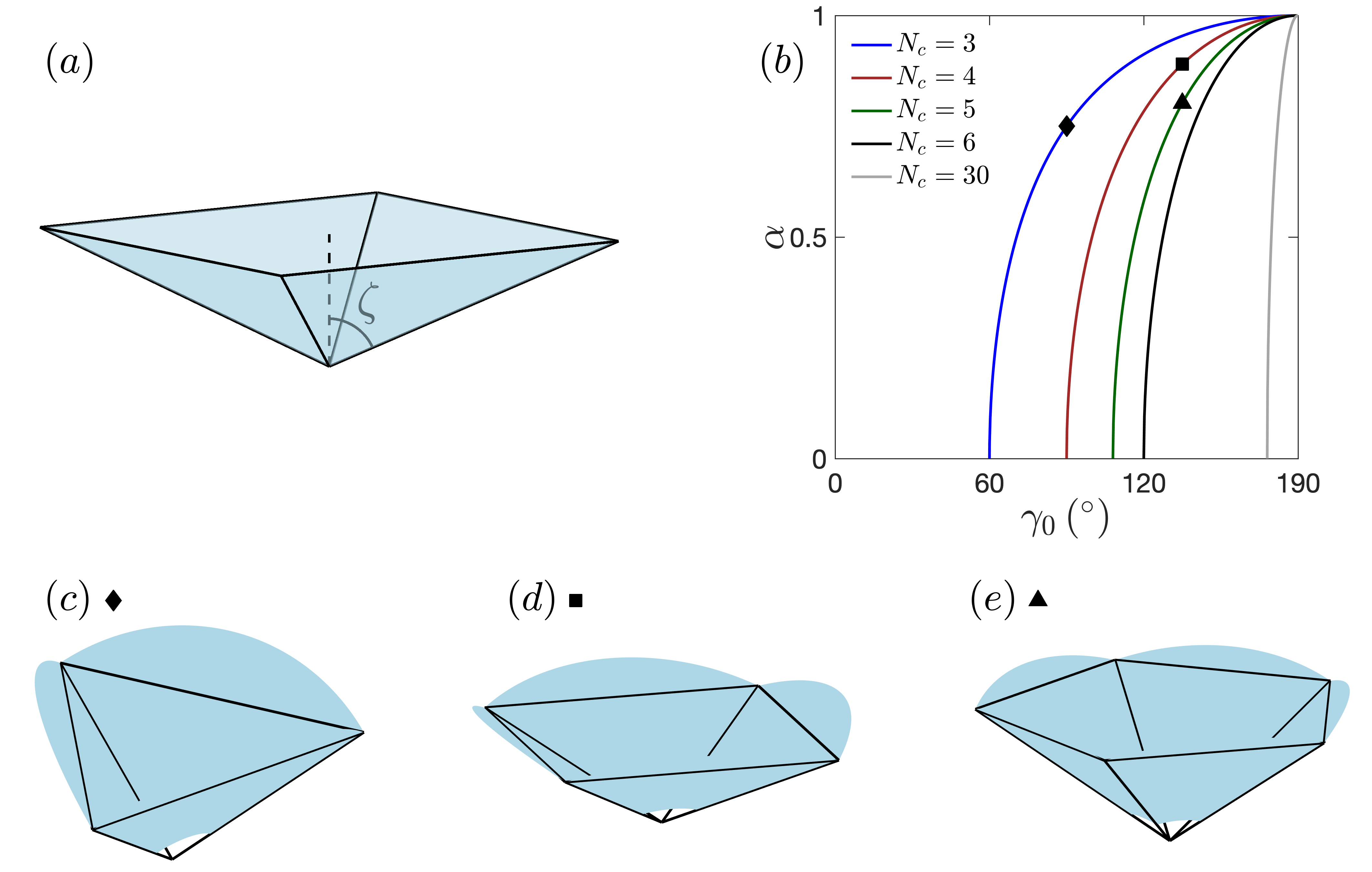}
	\caption{Energy-free folded states that lie on regular pyramids. (a) The vertex angle $\zeta$ is defined as the angle between the lateral edge (inclined solid lines) and the vertical axis (dashed line). (b) The relationship between $\alpha$ and $\gamma_{0}$ with different $N_c$ that leads to energy-free folded state. (c) An energy-free folded state with three creases (blue surface) which correspond to the symbol $\blackdiamond$ in (b). The pyramid is sketched by the black lines. (d) An example with four creases. (e) An example with five creases.} \label{fig:foldedEnergyFree}
\end{figure}

Figures \ref{fig:nffoldedForcesCurvatures}(a-d) present the Cartesian component of the contact moment $M_z$, the normal curvature $\kappa_n$, the geodesic torsion $\tau_g$, and $\eta$ of the renderings in Figure \ref{fig:nfepsbifurphasefolded}(d), respectively.
The configurations with convex facets have $M_z>0$ and $\kappa_n<0$ ($\blackdiamond$, $\blacksquare$, $\righttriangle$ and $\blackstar$), and the configurations with concave facets ($\lefttriangle$ and $\triangleup$) have $M_z<0$ and $\kappa_n>0$. 
In general, increasing $\gamma_{0}$ or decreasing $\alpha$ will lead to convex facets, while decreasing $\gamma_{0}$ or increasing $\alpha$ will result in concave facets. By carefully choosing $(\gamma_{0},\alpha,N_c)$, the facets of the folded state could remain flat with the crease being exactly the rest angle $\gamma_0$, which results in energy-free folded states. For $N_c \ge 3$, this requires

\begin{equation}\label{eq:energyfreefolded}
\begin{aligned}
\gamma_{0} = \pi - \cos^{-1} \left( \cos  \tfrac{2 \pi}{N_c} +2 \tan^2 \tfrac{\pi \alpha}{N_c} \cos^2 \tfrac{\pi}{N_c} \right)  \,,
\end{aligned}
\end{equation}

such that the folded state lies on the surface of a regular pyramid with a regular $N_c$-gonal base  and a ``vertex angle" $\zeta = \sin ^{-1} \left( \sin \tfrac{\pi \alpha}{N_c}  / \sin \tfrac{\pi}{N_c} \right)$, which is defined as the angle between the axis and the lateral edge of the pyramid (Figure \ref{fig:foldedEnergyFree}(a)). 
Notice that the hole size $a/R$ does not appear in Equation \eqref{eq:energyfreefolded}.
Because $\alpha \ge 0$, Equation \eqref{eq:energyfreefolded} further requires $\gamma_{0} \ge (\pi - \tfrac{2 \pi}{N_c})$, i.e., to obtain an energy free folded state, the rest angle $\gamma_{0}$ must be larger than the internal angle of the base polygon. Figure \ref{fig:foldedEnergyFree}(b) displays the relationship between the angle deficit $\alpha$ and the rest crease angle $\gamma_{0}$ in Equation \eqref{eq:energyfreefolded} with different $N_c$.
All the curves merge at the point $(\gamma_0,\alpha)=(180^{\circ}, 1)$, corresponding to a flat annulus. With $\alpha>1$, Equation \eqref{eq:energyfreefolded} does not have real solutions. With $\alpha$ slightly larger than unity, our numerical results show that the folded state always have a concave shape (e.g., the two renderings $\lefttriangle$ and $\triangleup$ in Figure \ref{fig:nfepsbifurphasefolded}(d)). In Figure \ref{fig:foldedEnergyFree}(b), a geometry from the left side of each curve corresponds to a concave shape, while a geometry from the right regime results in a convex shape. Figures \ref{fig:foldedEnergyFree}(c-e) display three energy-free renderings (blue surfaces) and their host pyramids (sketched by black lines), corresponding to the symbols in Figure \ref{fig:foldedEnergyFree}(b).  

\begin{figure}[htp!!!]
		\centering
		\captionsetup[subfigure]{labelfont=normalfont,textfont=normalfont}
		\centering
		\includegraphics[width=0.95\textwidth]{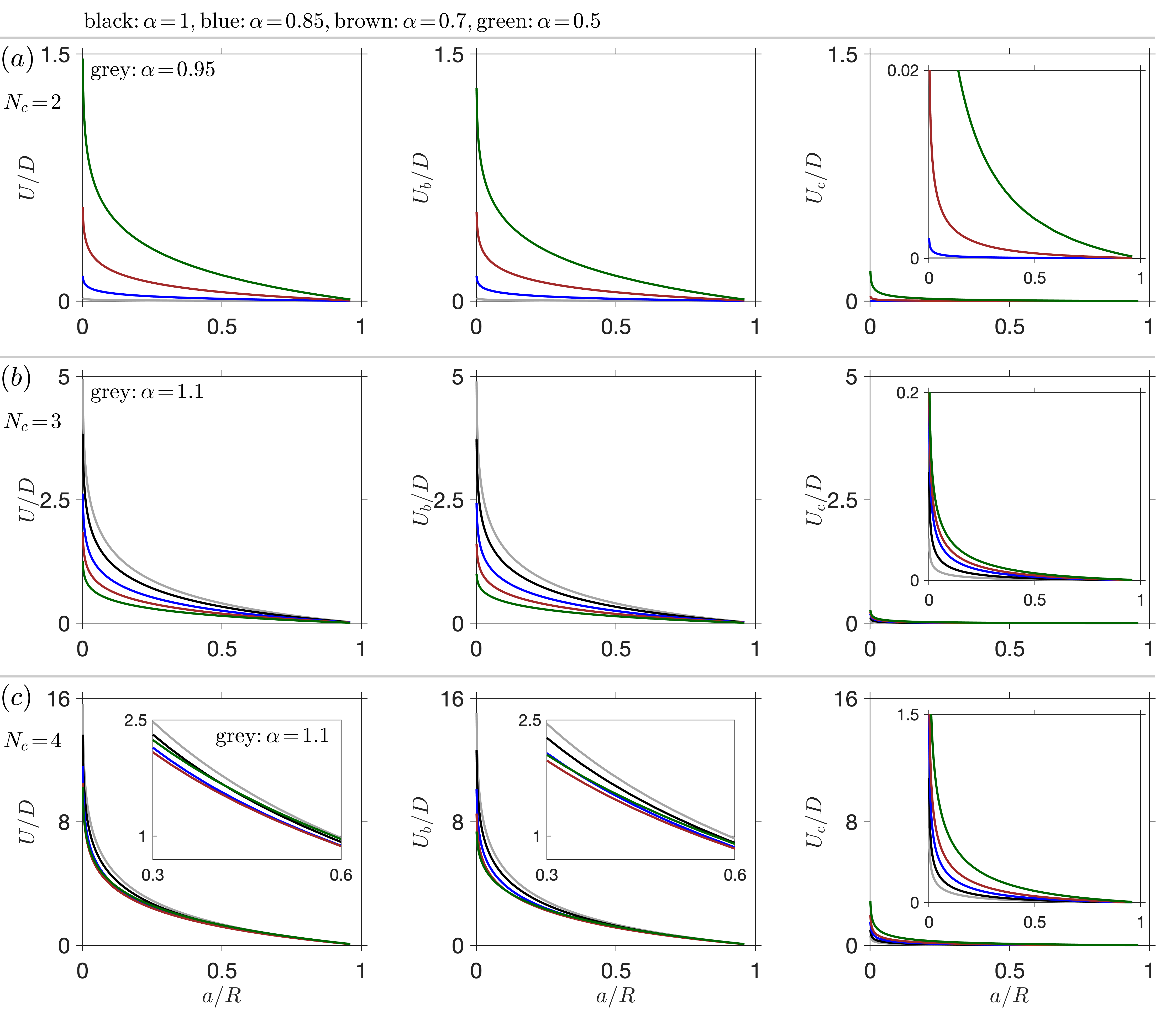}
		\caption{Normalized total energy $U/D$, bending energy $U_b/D$, and crease energy $U_c/D$. (a) corresponds to the solutions in Figure \ref{fig:nfepsbifurphasefolded}(a). (b) Figure \ref{fig:nfepsbifurphasefolded}(b). (c) Figure \ref{fig:nfepsbifurphasefolded}(c). }\label{fig:nfepsGamma45foldedEnergy}
\end{figure}

Figures \ref{fig:nfepsGamma45foldedEnergy}(a-c) report the normalized elastic energies of the solutions with $\gamma_{0}=45^{\circ}$ in Figures \ref{fig:nfepsbifurphasefolded}(a-c), respectively, including
the total energy $U/D$, the bending energy $U_b/D$, and the crease energy $U_c/D$. Compared with the elastic energies of the inverted state in Figure \ref{fig:alphaepsN2Gamma45Energy}(b) and Figures \ref{fig:nfepsN2Gamma45Energy}(b-c), the folded state contains much less elastic energy. In all the three panels of Figure \ref{fig:nfepsGamma45foldedEnergy}, the bending energy contributes much more to the total energy than the crease energy, and all the three energies decrease with the increase of the hole size $a/R$.

With $N_c=2$ in Figure \ref{fig:nfepsGamma45foldedEnergy}(a), decreasing $\alpha$ leads to the increase of all the three energies. With $N_c=2$ and $\alpha<1$, the folded state is always convex no matter what the rest crease angle is (see the two renderings $\blackdiamond$ and $\blacksquare$ in Figure \ref{fig:nfepsbifurphasefolded}(d)), which tends to close the crease angle (i.e., $\gamma_{f0} < \gamma_0$). Decreasing $\alpha$ generally makes the facets more convex, which increases the bending energy density and leads to the closing of the crease angle. The latter leads to higher crease energy. Our numerical results show that the bending energy also increases with the decrease of $\alpha$, implying that the intuitive decrease of the bending energy caused by decreasing $\alpha$ (which reduces the area of the facets) is exceeded by the increase of the bending energy density.

In Figure \ref{fig:nfepsGamma45foldedEnergy}(b) with $N_c=3$, decreasing $\alpha$ leads to the increase of the crease energy but the decrease of the bending energy and the total energy. From Figure \ref{fig:foldedEnergyFree}(b) we know that with $(N_c, \gamma_{0})=(3, 45^{\circ})$, the folded state is concave in the entire range $\alpha \in (0,1]$. Actually with $\alpha=1.1$, the structure is also concave, corresponding to $\lefttriangle$ in Figure \ref{fig:nfepsbifurphasefolded}(d). Here, decreasing $\alpha$ will reduce the facet area and flatten the facet, and thus decreases the bending energy. On the other hand, decreasing $\alpha$ tends to open the crease angle, which leads to an increase in the crease energy. Our results show that the bending energy is dominant here, and decreasing $\alpha$ leads to the decrease of the total energy.

In Figure \ref{fig:nfepsGamma45foldedEnergy}(c) with $(N_c, \gamma_{0})=(4, 45^{\circ})$, the transitions are slightly different. While the crease energy increases with the decrease of $\alpha$, the bending energy and the total energy does not change much as we vary $\alpha$.

\newpage

\section{Eccentricity - A family of stable inverted states }\label{se:eccentrichole}
	
\begin{figure}[h!]
	\centering
	\captionsetup[subfigure]{labelfont=normalfont,textfont=normalfont}
	\centering
	\includegraphics[width=0.85\textwidth]{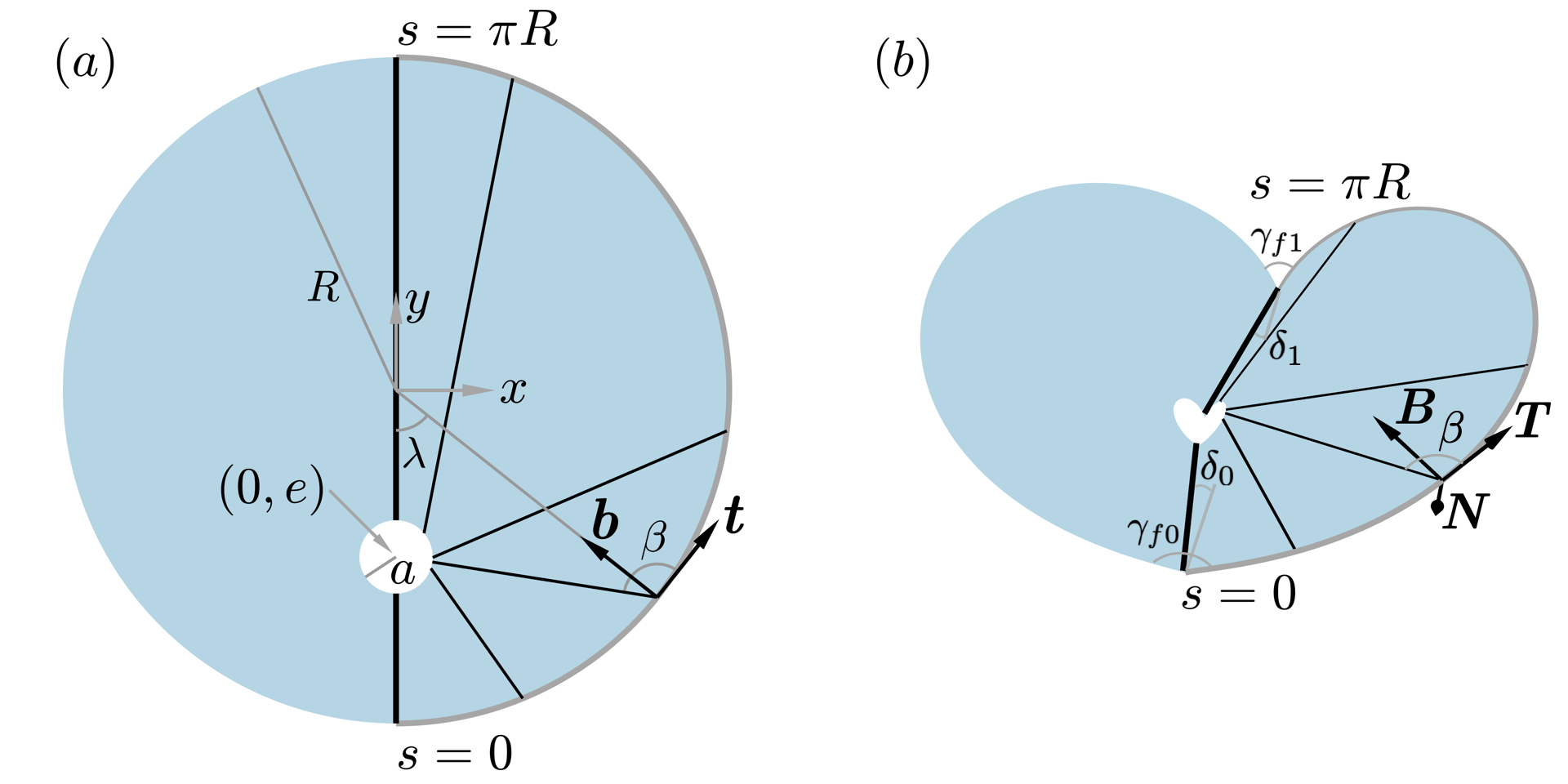}
	\caption{ (a) The flat configuration of a creased disk with a circular hole of radius $a$, located eccentrically at $(0,e)$. The thick black lines correspond to the crease and $(\bm{t},\bm{n},\bm{b})$ is a material frame attached to the outer circle, with $\bm{n}$ going into the plane. The generators are mapped from the inverted state in (b), where nonvanishing $|e/R|$ leads to different inclined angles $\delta_0$ and $\delta_1$ and different final crease angles $\gamma_{f0}$ and $\gamma_{f1}$ at the two ends.}\label{fig:KineEccen}
\end{figure}

In experiments, we observed that a creased disk can be inverted about almost anywhere along the crease to obtain a family of stable inverted states (Figure \ref{fig:exptmodel}(h)). In this section, we study the mechanics of this family of states  
by introducing a nonvanishing distance $e$ between the center of the hole and the center of disk, shown in Figure \ref{fig:KineEccen}(a). Figure \ref{fig:KineEccen}(b) shows the inverted state, which is obtained by first introducing a finite crease angle to the flat configuration in Figure \ref{fig:KineEccen}(a) and then inverting the crease.
We call $e/R$ the eccentricity of the hole. A nonvanishing $e/R$ breaks one of the two mirror symmetries in the inverted state, which now has a single mirror symmetry about the plane spanned by the two creases. Here we focus on the case with $\alpha=1$. The total energy can be written as

\begin{equation} \label{eq:totalenergyeccen}
	\begin{aligned}
	\frac{U}{D} = &\frac{K_{c} R}{D} \left(1-\frac{a}{R} +\frac{e}{R} \right) \left[1-\cos (\gamma_{f0} - \gamma_0)\right] \\
	& +  \frac{K_{c} R}{D} \left(1-\frac{a}{R} - \frac{e}{R}\right) \left[1-\cos (\gamma_{f1} - \gamma_0)\right] 
	+2 \int_0^{\pi R } YW ds  = \frac{U_c}{D} + \frac{U_b}{D}\,,
	\end{aligned}
\end{equation}
	
	where $\gamma_{f0}$ and $\gamma_{f1}$ represent the final crease angle at $s=0$ and $s=\pi R$, respectively. The first two terms represent the elastic energies in the two creases with different lengths, and the third term represents the bending energy of the facets. The Euler-Lagrange equations are the same as Equations \eqref{eq:stripgovern1}-\eqref{eq:stripgovern4}. The boundary conditions are slightly modified to account for the differences between the two creases. In addition, the algebraic constraint $\chi$ that implicitly determines $V$ is modified to
	
	\begin{equation}\label{eq:eccentricchi}
	\begin{aligned}
	\chi&= V^2+ (R^2 - 2VR -a^2) \sin^2 \beta + e \sin \beta [e \sin \beta - 2 R \cos \lambda \sin \beta +2 V \sin (\beta +\lambda)] \,,
	\end{aligned}
	\end{equation} 
	
	where $\lambda=s/R$, resulting in a ``nonautonomous" system.  $\lambda$ measures the angle between $-y$ and the radius (Figure \ref{fig:KineEccen}(a)). With $e=0$, Equation \eqref{eq:eccentricchi} degenerates to Equation \eqref{eq:circularkai}.  We follow standard techniques and transform the nonautonomous system into an autonomous system. Details of the transformation and the boundary conditions can be found in Appendix \ref{appse:BVPcontinuation}.

	Figure \ref{fig:EccenEpsAll} summarizes the solution curves ($\delta_0$ versus $e/R$), loci of the fold ($a/R$ versus $e/R$), and several renderings with different $(a/R, K_c R/D, \gamma_0)$. Small crease stiffness and large crease angle generally lead to a shallower inverted state with smaller $\delta_0$. In Figure \ref{fig:EccenEpsAll}(a), the curve with $a/R=0.03$ increases significantly with the decrease of $e/R$ when $e/R \rightarrow -1$. This is qualitatively different from the other curves. Upon a further examination of the solution, we find that the final crease angle $\gamma_{f0}$ corresponding to the upper left end of the curve is greater than $\pi$, which results in the increase of $\delta_0$ with the further decrease in $e/R$. 
	With a fixed $a/R$, the eccentricity $e/R$ is symmetrically bounded by two folds (one with $e/R<0$ and the other with $e/R>0$), where the inverted state (solid lines) loses stability through a fold which connects to the energy barrier (dashed lines). The inverted state and the energy barrier tend to form a closed loop. However, with small holes such as $a/R=0.03$ and 0.07, the curves do not close completely and terminate at the cross, where the bending energy blows up locally due to the local contact between the edge of regression and the material surface. Increasing $a/R$ tends to close the solution curves and shrink the closed loop, e.g., the loop with $a/R=0.15$ is smaller than the loop with $a/R=0.11$.
	This follows the typical feature of an isola center bifurcation, which is clearly seen in Figure \ref{fig:EccenEpsAll}(d) that shows the loci of the fold with different $(K_c R/D, \gamma_0)$. The area below the stability boundary corresponds to the bistable region where the inverted state exists. Increasing the eccentricity $|e/R|$ leads to the decrease of the critical hole size, and with a small hole, the eccentricity can be very large (i.e., $|e/R|$ could approach unity) without loss of the bistability. \emph{This matches with our experimental observation that a creased thin disk can be inverted almost anywhere along the crease (Figure \ref{fig:exptmodel}h)}.
	The renderings in Figure \ref{fig:EccenEpsAll}(e) correspond to the symbols in Figures \ref{fig:EccenEpsAll}(a-c), including several inverted states $\blackdiamond$, $\lefttriangle$ and $\triangleup$, and their energy barriers $\blacksquare$, $\righttriangle$ and $\blackstar$, respectively. More renderings of the inverted state with different $(a/R,e/R)$ are documented in Figure \ref{appfig:EccenConfigSet} of Appendix \ref{appse:morerenderings}.

	Figures \ref{fig:EccenEnergy}(a-b) report the normalized elastic energy of the solutions in Figures \ref{fig:EccenEpsAll}(a-b), respectively. The total energy of the inverted state generally decreases with the increase of the eccentricity $|e/R|$. With a weak crease and a large crease angle $(K_c R/D, \gamma_0)=(4,135^{\circ})$, the total energy is low and the inverted state is slightly deformed from the flat configuration. In addition, for each $a/R$, the bending energy is slightly larger than the crease energy. 
	With an intermediate crease stiffness and an intermediate crease angle $(K_c R/D, \gamma_0)=(20,90^{\circ})$, the total energy increases significantly, mainly from the contribution of the bending energy. The crease energy does not change too much, and its contribution is small compared with the bending energy.

	\begin{figure}[h!]
		\centering
		\captionsetup[subfigure]{labelfont=normalfont,textfont=normalfont}
		\centering
		\includegraphics[width=0.95\textwidth]{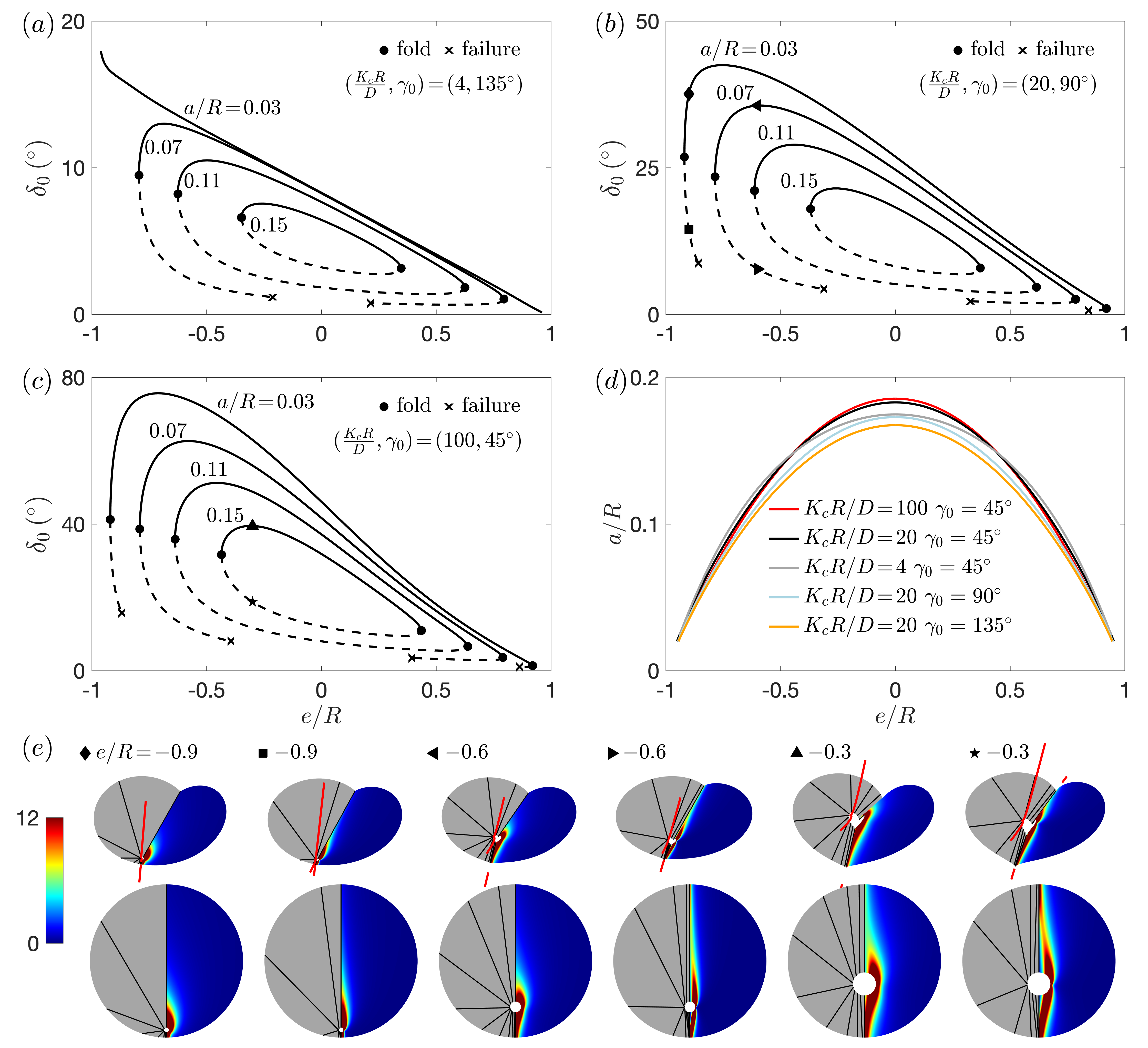}
		\caption{Solution curves ($\delta_0$ versus $e/R$) of the inverted state (solid lines) and the energy barrier (dashed lines), and loci of the fold ($a/R$ versus $e/R$) with different $(a/R,K_cR/D,\gamma_0)$. The angle deficit $\alpha$ is fixed to 1.
		 (a) $(K_cR/D, \gamma_0)=(4,135^{\circ})$. 
      (b) $(K_cR/D, \gamma_0)=(20,90^{\circ})$.
	(c) $(K_cR/D, \gamma_0)=(100,45^{\circ})$. 
	(d) Loci of the fold. (e) Renderings that correspond to the symbols in (a-c).} \label{fig:EccenEpsAll}
	\end{figure}

\begin{figure}[h!]
	\centering
	\captionsetup[subfigure]{labelfont=normalfont,textfont=normalfont}
	\centering
	\includegraphics[width=0.95\textwidth]{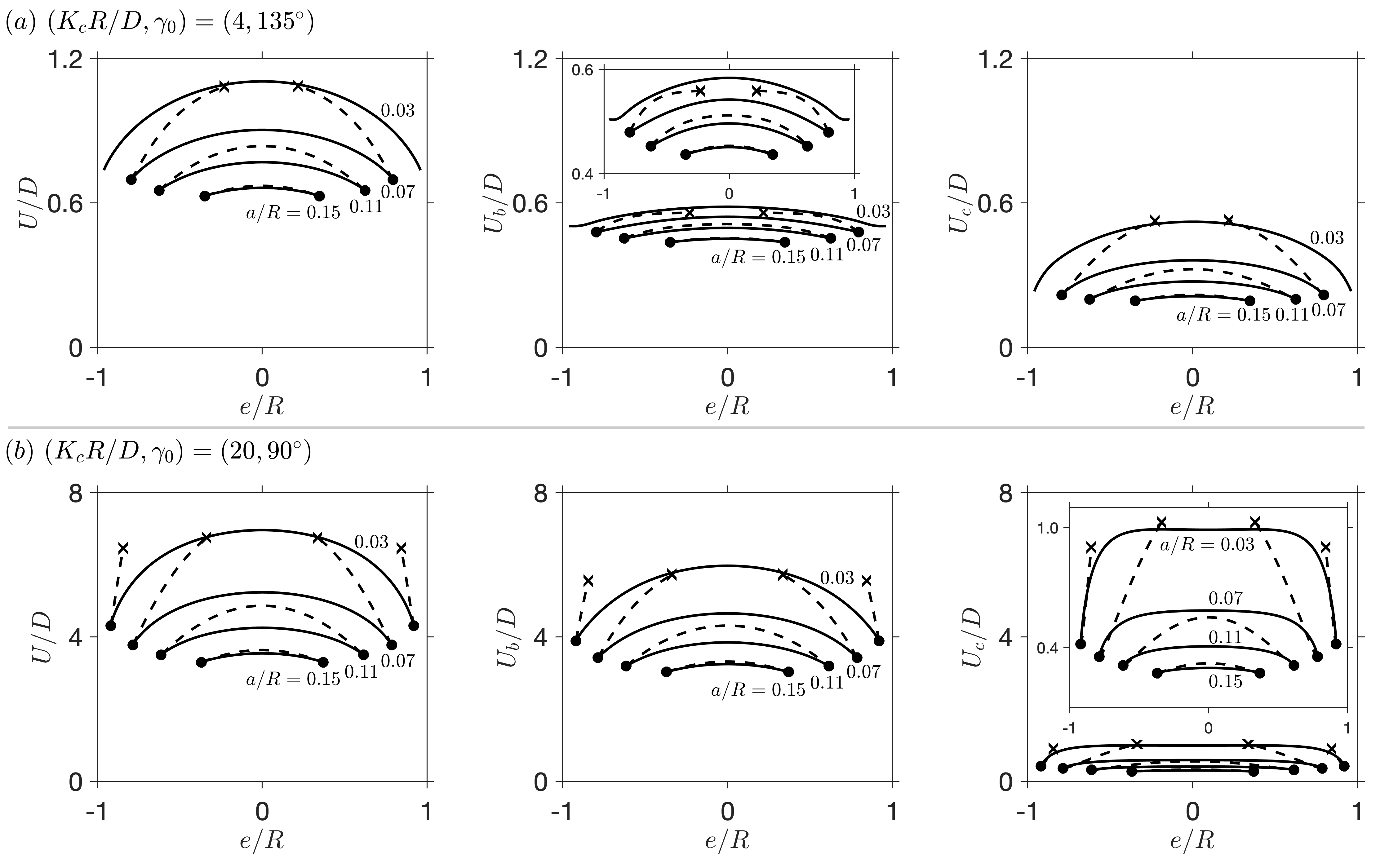}
	\caption{Normalized total energy $U/D$, bending energy $U_b/D$, and crease energy $U_c/D$. (a) and (b)
correspond to the solutions in Figures \ref{fig:EccenEpsAll}(a) and \ref{fig:EccenEpsAll}(b), respectively.
} \label{fig:EccenEnergy}
\end{figure}

\begin{figure}[h!]
		\centering
		\captionsetup[subfigure]{labelfont=normalfont,textfont=normalfont}
		\centering
		\includegraphics[width=0.95\textwidth]{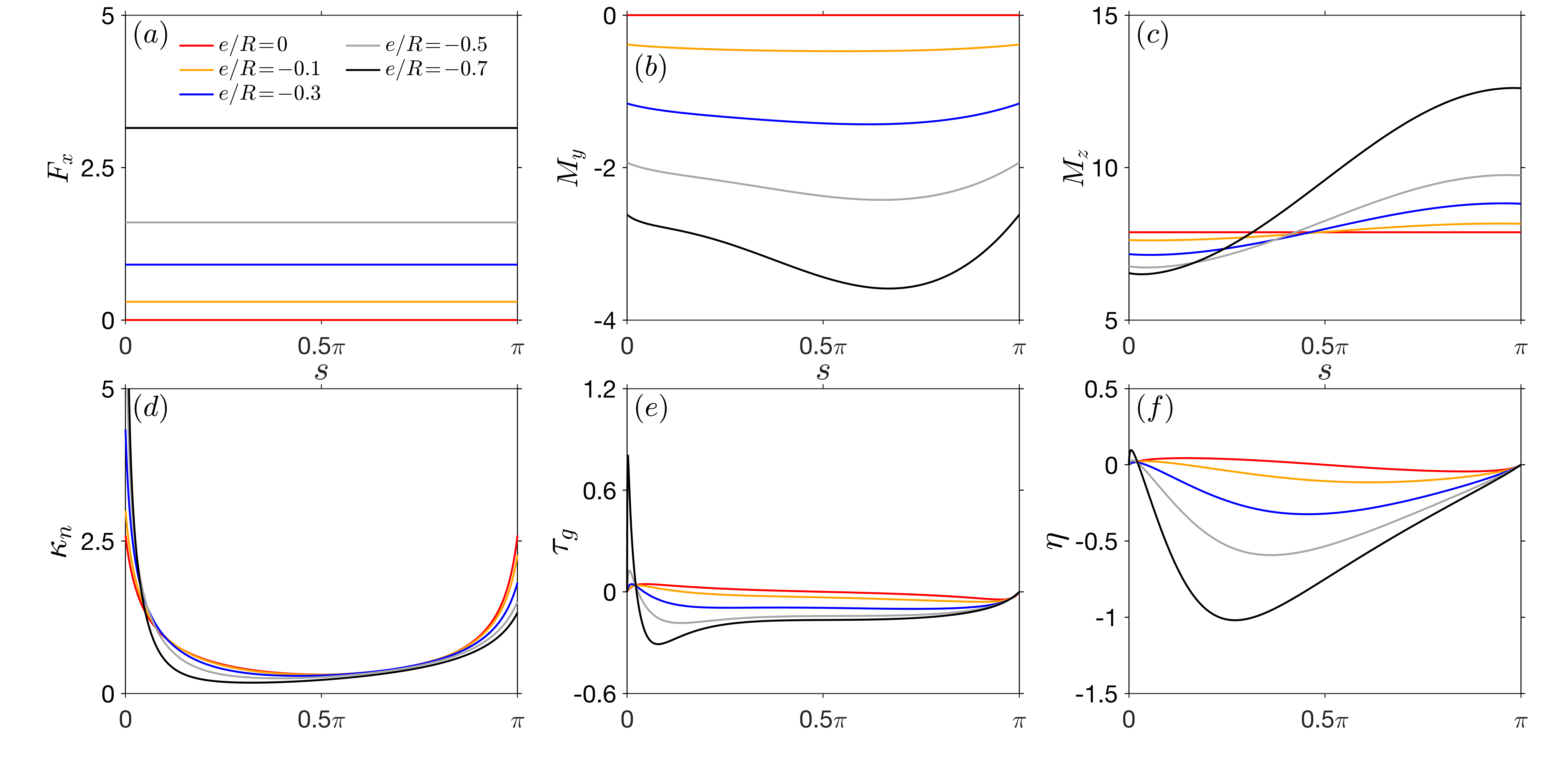}
		\caption{Distribution of the Cartesian components of contact force/moment and several geometric quantities with different $e/R$. $(K_cR/D, \gamma_0, a/R, \alpha)$ is fixed to $(20,90^{\circ},0.07,1)$. Because of the symmetry, only one facet of the structure is
			reported. (a) $F_x$. (b) $M_y$. (c) $M_z$. (d) Normal curvature $\kappa_n$. (e) Geodesic torsion $\tau_g$. (f) $\eta$. }\label{fig:EccenForcesCurvatures}
	\end{figure}

Nonvanishing eccentricity $e/R$ breaks the mirror symmetry of the inverted state about the $x-z$ plane, which now has a single mirror symmetry about the plane spanned by the two creases (i.e., the $y-z$ plane). This mirror symmetry forces the contact force to be a constant vector in the $x$ direction, and the contact moment to be nonconstant in the $y-z$ plane. Figures \ref{fig:EccenForcesCurvatures}(a-c) present respectively the Cartesian component of the contact force and moment $F_x$, $M_y$, and $M_z$, with $(K_cR/D, \gamma_0, a/R, \alpha)$ fixed to $(20,90^{\circ},0.07,1)$. With various eccentricities, $F_x$ is always found to be a positive constant, which matches with the symmetry analysis and further implies that with $e/R <0$, the extremity of the directrix $s=0$ is in tension while the other extremity $s=\pi R$ is under compression. In other words, a nonvanishing eccentricity will make the end of the directrix closer to the hole be in tension, while the farther end will be under compression. Our numerical results further show that for the inverted state, the crease angle of the shorter crease always opens more than the crease angle of the longer crease, which also qualitatively matches with our experimental observations (Figure \ref{fig:exptmodel}(h)).

With nonvanishing eccentricity, $M_y$ and $M_z$ vary along the arc length. The minimum $M_y$ corresponds to the highest point of the directrix $\bm{r}$ in the $z$ direction. Near the two ends $s=0$ and $s=\pi$, $M_z$ decreases a bit with the increase of $s$, corresponding to the fact that the two ends $s=0$ and $s=\pi$ are usually not the extremities in the $y$ direction. Instead, two material points slightly inside the two ends of the directrix have the largest and smallest $y$ coordinate, which can be seen in the projections of the outer and inner circumferences in Figure \ref{appfig:EccenEdgeProjection} (Appendix \ref{appse:3Dprofile2Dprojection}).   
Figures \ref{fig:EccenForcesCurvatures}(d-f) present respectively the distribution of the normal curvature $\kappa_n$, the geodesic torsion $\tau_g$, and $\eta$.

	\clearpage

\section{Summary and further discussion} \label{se:conclusiondiscussion}

Creases and cuts have been introduced to thin sheets to create novel structures called Origami and Kirigami, which can achieve morphable geometries and nontraditional mechanical properties \cite{sadik2021local,sadik2022local,Moshe19,yang2018multistable,castle2014making}. 
We found geometry determines the mechanics of creased thin disk and the influences of material properties are minimal. The novel mechanics phenomena studied in this work are general and not restricted to specific geometries or material properties. For example, it can be easily demonstrated through a piece of printing paper that the crease could be inverted about almost anywhere along the crease, independent of the crease angle and the geometry of the paper sheet. The influences of inserting or removing materials in the circumferential direction also appear to be general in creased thin sheets. Particularly, removing a circumferential sector could increase the critical hole size significantly, up to the size of the disk, in which case anisotropic rod model (which is free of singularity that might bother the inextensible strip) should work well. In a forthcoming work, we use anisotropic rod theory to further investigate bistable and looping behaviors of creased annular strips with a continuous description of creases \cite{yu2022continuous}].

In this work, we studied the mechanics of annular sheets and strips decorated with radial creases. Several geometric parameters that lead to novel mechanical phenomena are first identified through tabletop models. We then used an inextensible strip model to formulate a minimal facet as a two-point boundary value problem with the creases modeled as nonlinear hinges. Numerical continuation with AUTO 07P was conducted to obtain solution curves as certain geometric parameters vary. The numerical predictions match our experimental observations and further reveal unexpected nonlinear behaviors. We summarize our major findings and conclusions here:

	\begin{itemize}
	\item Our numerical results show that with $N_c \ge 2$ (i.e., with more than two evenly spaced creases), removing and inserting a small sector could significantly increase and decrease the critical hole size, respectively. For example, with the angle deficit $\alpha \le 0.7$ (i.e., cutting more than $30 \%$ of the annulus along the circumference), the critical hole size with $N_c=2$ could be as large as the disk; For $N_c=3$ and 4, this requires $\alpha \le 0.5$. On the other hand, $\alpha$ generally cannot exceed 1.1, otherwise the bistability will be destroyed even with an infinitesimal hole. In addition, increasing the hole size $a/R$ generally destroys the inverted branch through a fold.

	\item A thin disk with $N_c=1$ (i.e., a single crease) behaves differently from the ones with $N_c \ge 2$. First, the inverted state with a single crease can contain a hole as large as the disk without the requirement of cutting any sector. Second, inserting a sector could also destroy the inverted state and $\alpha$ generally cannot exceed 1.1, which is similar to the case with $N_c \ge 2$. However, instead of decreasing the largest critical hole size in the case with $N_c \ge2$, inserting a sector with $N_c =1$ could create a lower boundary for the hole size. In other words, with $N_c=1$ and $\alpha$ slightly larger than 1, decreasing $a/R$ could destroy the bistability, and at the same time, increasing $a/R$ will not lose the inverted state and the hole can still be as large as the disk.

   \item Several geometric parameters could conspire to create unexpected mechanical behaviors. For example, with $N_c=2$ and $\alpha \ge 1$, decreasing the rest crease angle $\gamma_0$ generally makes the inverted state more stable and could turn a monostable creased disk into a bistable one. On the contrary, with $N_c=2$ and $\alpha \le 0.99$, decreasing the crease angle $\gamma_0$ makes the inverted state less stable and could turn a bistable disk into a monostable one.

  \item  The folded state contains much less energy than the inverted state. With $N_c \ge 3$ (i.e., more than three creases), facets of the folded state are generally bent. An exception can be obtained by a careful choice of the geometric parameters such that the folded state lies on the lateral surface of a regular pyramid, resulting in energy-free folded states.
  
  \item Our results confirm that a creased disk can be inverted almost anywhere along the crease, resulting in a family of stable inverted states.

  \item The mechanics of the crease affect the nonlinear behaviors of the creased disk. With $N_c=2$ and a small hole, a crease following a sinusoidal angle-moment relationship could be flipped to create a pair of half flipped states and a flipped state, which do not exist with a crease following a linear angle-moment relationship. However in both cases, the system behaves similarly as we increase the hole size $a/R$ with different angle deficit $\alpha$. In other words, using different crease models does not affect the main conclusions and findings of this study.

\end{itemize}

Our findings demonstrate that with a simply creased disk, varying several geometric parameters could create extremely rich nonlinear behaviors. We have explored only a few novel phenomena in this system, which is worth further study. For example, the distribution of the creases could affect the mechanical behaviors of the creased disk. A tabletop model shows that with $N_c=2$ and an uneven distribution of creases, the critical hole size could be increased significantly. In addition, a creased thin sheet could be inverted simultaneously about several places along the creases, resulting in a system with several elastic singularities that could interact with each other. We reserve these topics for future study.

The inextensible strip model employed in this work is appropriate only for thin sheets, in which stretching of the surface is much more energetically expensive than bending. In experiments, thickness of the material is observed to be another factor that affects the mechanics of the creased disk. It is known that the competition between the mechanics of creases and the bending of facets in creased thin sheets is determined by the origami length \cite{lechenault2014mechanical}, which is proportional to the thickness of the material. With a thick creased disk, we found that the bistability may not exist even without a hole. It will be interesting to study the transitional behavior between thin and thick sheets in such systems.

Accurate modeling of the mechanics of creased thin sheets requires a precise description of the mechanics of the crease, which usually have complex relaxation phenomena and complicated mechanical responses under external loading \cite{thiria2011relaxation,jules2020plasticity,dharmadasa2020formation}. We have adopted only an elastic response for the crease in this study.  
In addition, we assumed a constant final crease angle along the crease. Our recent work with detailed FE modelings demonstrated that the final crease angle of the inverted state usually varies along the crease \cite{yu2021cutting}. The moment balance at the crease implemented in this work represents an overall balance and is not a pointwise local balance along the crease length. 
It appears that a non-constant final crease angle could be incorporated into the developable model to describe the inverted and the non-flat folded state, in which the crease is no longer a generator and will not remain straight. In other words, the crease will intersect with the nearby generators. This may require partitioning the deformed facet into several developable pieces with the generators bounded by different space curves \cite{badger2019normalized}, which is beyond the scope of this study. A recent study of the creased disk through FE modeling shows that near the crease, the lines of smallest principal curvature could intersect with the crease \cite{andrade2019foldable}.

\section{Acknowledgments}

The author is grateful to Jessica Flores for her diligent proofreading and constructive criticism of the manuscript. TY thanks James Hanna, Marcelo Dias, Ignacio Andrade-Silva, and Andy Borum for useful discussions. TY acknowledges partial support by U.S. National Science Foundation grant CMMI-2001262 to James Hanna and partial support by Princeton University Dean of Research Innovation Funding.

\newpage
	
	\appendix
	
	\section{Numerical implementation of the inextensible strip model} \label{appse:BVPcontinuation}
	Based on the inextensible strip model \cite{starostin2015equilibrium,dias2015wunderlich}, we formulate a minimal facet of the creased thin disk as a two-point boundary value problem and use continuation package AUTO 07P to conduct parametric studies \cite{doedel2007auto}. Euler angles $(\psi, \theta, \phi)$ are employed to describe the sequential rotations of the director frame $(\bm{T}, \bm{N}, \bm{B})$, following a $3-2-3$ convention. With $N_c$ evenly spaced creases $(N_c \ge 2)$, both the folded state and the inverted state have $N_c$-fold mirror symmetries. We take advantage of the symmetries and only solve $1/N_c$ of the structure bounded by two adjacent creases. With a single crease $N_c=1$, we solve half of the structure due to the single mirror symmetry.
	
\begin{figure}[h!]
		\centering
		\includegraphics[width=0.95\textwidth]{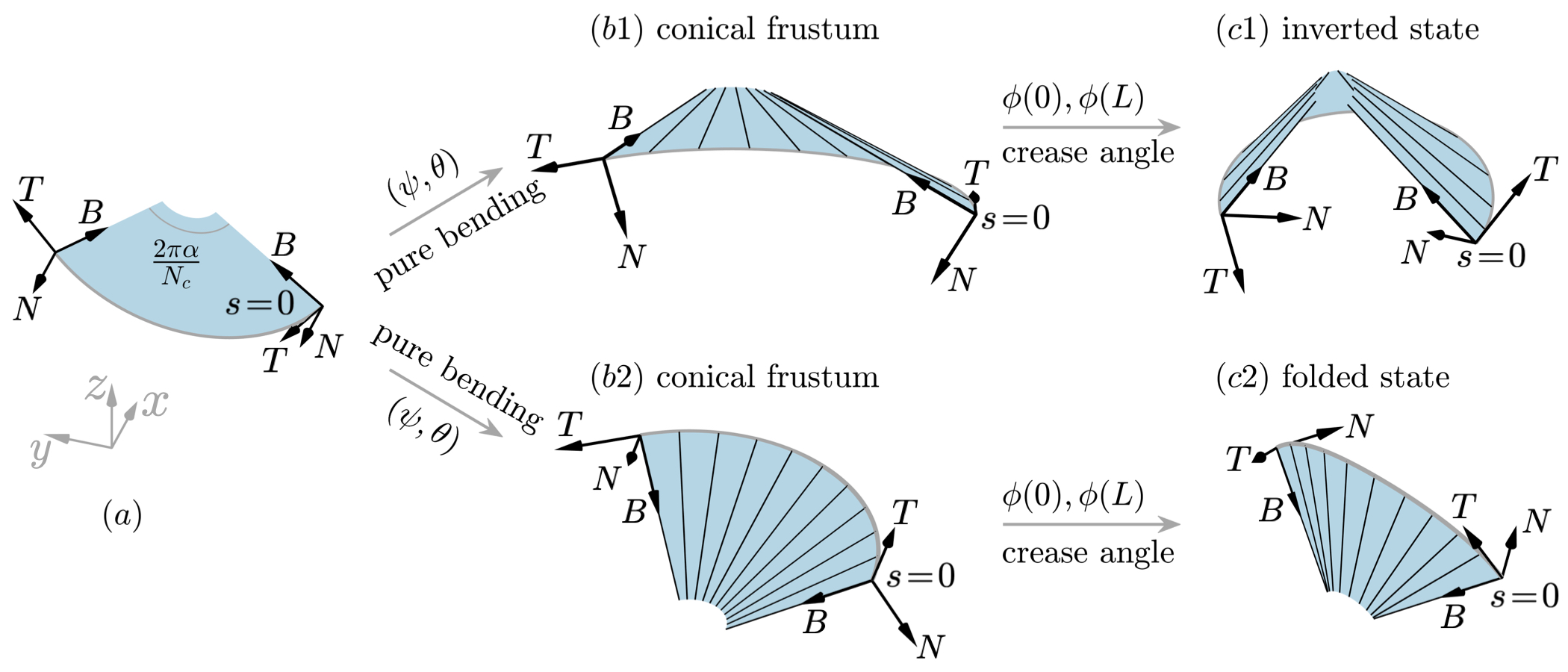}
		\caption{Euler angles $(\psi, \theta, \phi)$ are used to describe the rotations of the material frame attached to the outer circle of an annular sector, following a 3-2-3 rotation convention. The flat annular sector in (a) is deformed into two conical frustums in (b1) and (b2) by rotating the director frame about $\bm{B}(s)$ by $\psi (s)$ (a linear function of $s$), and then about $\bm{T}(s)$ by $\theta (s)$ (a constant). The crease angle is introduced to (b1) and (b2) by rotating the generator at the two ends, resulting in a minimal facet of the inverted state and the folded state in (c1) and (c2), respectively.}\label{appFig:Eulerrotation}
	\end{figure}

	Figure \ref{appFig:Eulerrotation} displays a series of deformations that first transform a flat annular sector with a central angle $2 \pi \alpha/N_c$ into conical frustums in panels $(b1)$ and $(b2)$, which are then deformed respectively into a minimal facet of the inverted and folded state by rotating 
	the generator at the two ends $s=0$ and $s=L \,\, (=2 \pi R \alpha /N_c)$ to match with the rest crease angle $\gamma_{0}$. Finite crease stiffness will be introduced later and so far the two creases are rigid.
	The two conical frustums in Figures \ref{appFig:Eulerrotation}(b1) and \ref{appFig:Eulerrotation}(b2) are mirror images of each other about the $x-y$ plane of a Cartesian coordinate system $x-y-z$. 
	The two ends $s=0$ and $s=L$ of the inverted and the folded state (Figures \ref{appFig:Eulerrotation}(c1) and \ref{appFig:Eulerrotation}(c2)) are symmetrically constrained in the $x-y$ plane to slide along the two rays $y=-\tan \tfrac{\pi}{N_c} x$ and $y=\tan \tfrac{\pi}{N_c} x$, respectively. The rotation axis of the complete structure is aligned with the $z$ axis.

	The Euler angles are further implemented through unit quaternions to avoid potential polar singularity. The relationship between the director frame and the Cartesian frame can be related through Euler angles and quaternions as, 
	
	\begin{equation}\label{appeq:323rotation} 
	\begin{aligned}
	\begin{bmatrix} \bm{-N} \\ \bm{T} \\ \bm{B} 
	\end{bmatrix}
	&=
	\begin{bmatrix} \cos \phi & \sin\phi & 0 \\ -\sin \phi & \cos \phi & 0 \\ 0 & 0 & 1 
	\end{bmatrix}
	\begin{bmatrix} \cos \theta & 0 & -\sin \theta \\ 0 & 1 & 0 \\ \sin \theta & 0 & \cos \theta 
	\end{bmatrix}
	\begin{bmatrix} \cos \psi & \sin\psi & 0 \\ -\sin \psi & \cos \psi & 0 \\ 0 & 0 & 1 
	\end{bmatrix}
	\begin{bmatrix} \bm{\hat{\bm{x}}} \\ \bm{\hat{\bm{y}}} \\ \bm{\hat{\bm{z}}} 
	\end{bmatrix} \\ 
	&=2 \begin{bmatrix} q_1^2+q_2^2-\frac{1}{2} & q_2 q_3 + q_1 q_4 & q_2 q_4 - q_1 q_3\\ q_2 q_3 - q_1 q_4 & q_1^2+q_3^2-\frac{1}{2} &q_3 q_4 +q_1 q_2 \\ q_2 q_4 +q_1 q_3  &  q_3 q_4 - q_1 q_2 & q_1^2+q_4^2-\tfrac{1}{2} 
	\end{bmatrix}
	\begin{bmatrix} \bm{\hat{\bm{x}}} \\ \bm{\hat{\bm{y}}} \\ \bm{\hat{\bm{z}}}
	\end{bmatrix}
	\end{aligned}
	\end{equation}
	
	For a $3-2-3$ rotation, quaternions can be written in terms of Euler angles as \cite{henderson1977euler}
	
	\begin{equation}\label{eq:angletoparameter} 
	q_1=\cos \tfrac{\theta}{2} \cos \tfrac{\phi + \psi}{2}\,,\; q_2=\sin \tfrac{\theta}{2} \sin \tfrac{\phi - \psi}{2}\,,\; q_3=\sin \tfrac{\theta}{2} \cos \tfrac{\phi - \psi}{2}\,,\; q_4=\cos \tfrac{\theta}{2} \sin \tfrac{\phi + \psi}{2} \, , 
	\end{equation}
	
	The derivatives of quaternion components can be written as
	\begin{equation}\label{eq:Dquaternion} 
	\begin{aligned}
	q'_{1}&=\tfrac{1}{2}(-q_4 \kappa_n +q_2 \kappa_g -q_3 \tau_g)\,,\; q'_{2}=\tfrac{1}{2}(-q_1 \kappa_g +q_3 \kappa_n -q_4 \tau_g)\,,\; \\ 
	q'_{3}&=\tfrac{1}{2}(-q_2 \kappa_n - q_4 \kappa_g + q_1 \tau_g)\,,\;\;\;\, q'_{4}=\tfrac{1}{2}(q_3 \kappa_g +q_1 \kappa_n +q_2 \tau_g)\, .
	\end{aligned}
	\end{equation}
	
	To obtain a system of first order ordinary differential equations (ODEs), we first differentiate the algebraic constitutive law in Equation \eqref{eq:stripgovern3} with respect to $s$ and combine with \eqref{eq:stripgovern4} to obtain a first order ODE for $\kappa_n$ and a second order ODE for $\eta$. The latter is transformed into two first order ODEs by introducing an intermediate variable $\Omega=\eta'$. In addition, we treat $V$ as an independent variable and differentiate the implicit relationship $\chi=0$ with respect to $s$, resulting in a first order ODE for $V$.
	Combining Equations \eqref{eq:stripgovern1}-\eqref{eq:stripgovern2}, \eqref{eq:Dquaternion}, $\bm{r}'=\bm{T}$, and the ODEs for $\kappa_n$, $\eta$, $\Omega$, and $V$, we have

	\begin{equation}\label{appeq:18ODE} 
	\begin{aligned}
	&\frac{d F_1}{d \bar{s}} -L(\kappa_n F_2 - \kappa_g F_3)=0 \,, \frac{d F_2}{d \bar{s}}+ L (\kappa_n F_1 -\kappa_n \eta F_3)=0 \,, \frac{d F_3}{d \bar{s}}+ L(\kappa_n \eta F_2 -\kappa_g F_1)=0 \,, \\
	&\frac{d M_1}{d \bar{s}}- L(\kappa_n M_2 - \kappa_g M_3)=0 \,, \frac{d M_2}{d \bar{s}}+ L (\kappa_n M_1-\kappa_n \eta M_3 -F_3)=0 \,, \frac{d M_3}{d \bar{s}}+ L (\kappa_n \eta M_2 -\kappa_g M_1+F_2)=0 \,, \\
	&\frac{d \eta}{d \bar{s}} =L \Omega \,, (AE-C^2) \frac{d  \Omega}{d \bar{s}} =L [(CB-AI) \Omega + AG -CJ] \,, (AE-C^2) \frac{d \kappa_n}{d \bar{s}} =L [(IC-BE) \Omega + JE -GC] \,, \\
	&\frac{d V}{d \bar{s}}=\left (-\frac{\chi_s}{\chi_V}-\frac{\chi_{\eta}}{\chi_{V}} \Omega \right) L \,, \\
	&\frac{d q_1}{d \bar{s}} =L [0.5(-q_4 \kappa_n +q_2 \kappa_g -q_3 \tau_g)+\mu q_1] \,,\; \frac{d q_2}{d \bar{s}}=L [0.5 (-q_1 \kappa_g +q_3 \kappa_n -q_4 \tau_g)+\mu q_2]\,,\;  \\
	&\frac{d q_3}{d \bar{s}}= L [0.5 (-q_2 \kappa_n - q_4 \kappa_g + q_1 \tau_g)+\mu q_3] \,,\;\, \frac{d q_4}{d \bar{s}}=L [0.5 (q_3 \kappa_g +q_1 \kappa_n +q_2 \tau_g)+\mu q_4] \, , \\
	&\frac{d x}{d \bar{s}}=2 L (q_2 q_3 -q_1 q_4) \,,\; \frac{d y}{d \bar{s}}=2 L (q_1^2 +q_3^2 -\tfrac{1}{2}) \,,\; \frac{d z}{d \bar{s}}=2 L (q_3 q_4 +q_1 q_2) \, ,\\
	&\frac{d s}{d \bar{s}}=L \,, \\
	\end{aligned}
	\end{equation} 
	
	where $L$ ($=2 \pi R \alpha /N_c$) corresponds to the length of the directrix $\bm{r}(s)$, $\frac{d()}{d \bar{s}}= L \frac{d ()}{d s} =L ()'$, and 
	\begin{equation}\label{appeq:ABCDcoefficients} 
	\begin{aligned}
	&A=Y_{\kappa_n \kappa_n} W \,, \\
	&B= Y_{\kappa_n \eta} W + Y_{\kappa_n} W_{\eta} + Y_{\kappa_n} W_{V} V_{\eta} \,, \\
	&C= Y_{\kappa_n \eta'} W + Y_{\kappa_n} W_{\eta'} \,,\\
	&I=Y_{\eta' \eta} W + Y_{\eta'} W_{\eta} +Y_{\eta'} W_{V} V_{\eta}+Y_{\eta}  W_{\eta'} +Y W_{\eta' \eta} +Y W_{\eta' V} V_{\eta} \,,\\
	&E=Y_{\eta' \eta'} W + 2 Y_{\eta'} W_{\eta'} +W_{\eta' \eta'}Y \,, \\
	&J= \eta' M_1 -F_2 + \kappa_g (M_1 - \eta M_3) - Y_{\kappa_n} W_V V_s \,,\\
	&G=Y_{\eta} W + Y W_{\eta} +Y W_{V} V_{\eta}- \kappa_n M_1  - Y_{\eta'} W_{V} V_{s} - Y W_{\eta' V} V_{s} \,,\\
	\end{aligned}
	\end{equation}

	where a subscript represents a partial derivative and a prime denotes an $s-$derivative. Through the introduction of $\bar{s}$ ($\in [0,1]$), we have normalized the length of the integral interval to unity, which is required by AUTO. In addition, the last ODE in Equation \eqref{appeq:18ODE} transforms the nonautonomous system into an autonomous system. 
	Varying $\alpha$ through $L$ allows us to remove or insert a sector.
	Following \cite{healey2006straightforward}, we have introduced a dummy parameter $\mu$ to enable a consistent prescription of boundary conditions for quaternions, through which the original pointwise constraint of the unit quaternions is required only at the two ends. $\mu$ is treated as a free parameter (i.e., a scalar unknown) in numerical continuation and its value should always be numerically zero  \cite{healey2006straightforward}.

	The $N_c$-fold mirror symmetries of the folded and the inverted state vanish the contact force identically and force the contact moment to be a constant vector in the $z$ direction. For the half flipped state with $N_c=2$ and the inverted state with $N_c=1$, they have a single mirror symmetry. 
	Forces in the plane of symmetry and the moment perpendicular to the plane of symmetry are set to zeros through the boundary conditions at the two ends of the minimal facet, which can be summarized as

	\begin{equation}\label{appeq:boundaryconditions} 
	\begin{aligned}
	&F_z (0)=0\,, F_b (0)=0\,, M_t (0)=0\,, s(0)=0\,, \\
	&\chi(V(0),0,\eta(0))=0 \,, \eta(0)=0\,, \eta (1)=0\,, \\
	&\frac{\kappa_n (0) (1+\eta ^2 (0))^2}{\eta' (0) + \kappa_g (0) (1+\eta^2 (0))} W(\eta' (0) , \eta (0)) - \eta(0) M_1 (0) - M_3 (0)=0\,,\\
	&q_1(0)=\cos \tfrac{\theta_0}{2} \cos \left[ \tfrac{1}{2} \left( \tfrac{\pi}{N_c}+\tfrac{1}{2} (\pi -\gamma_0) \right) \right]\,,
	q_2(0)=\sin \tfrac{\theta_0}{2} \sin \left[ \tfrac{1}{2} \left( \tfrac{\pi}{N_c}-\tfrac{1}{2} (\pi -\gamma_0) \right) \right]\,, \\
	&q_3(0)=\sin \tfrac{\theta_0}{2} \cos \left[ \tfrac{1}{2} \left( \tfrac{\pi}{N_c}-\tfrac{1}{2} (\pi -\gamma_0) \right) \right]\,,
	q_4(0)=-\cos \tfrac{\theta_0}{2} \sin \left[ \tfrac{1}{2} \left( \tfrac{\pi}{N_c}+\tfrac{1}{2} (\pi -\gamma_0) \right) \right]\,, \\
	&q_1 (1) =\cos \tfrac{\theta_1}{2} \cos \left[ \tfrac{1}{2} \left( \tfrac{\pi}{N_c}+\tfrac{1}{2} (\pi -\gamma_0) \right) \right]\,,
	q_2 (1)=\sin \tfrac{\theta_1}{2} \sin \left[ \tfrac{1}{2} \left( -\tfrac{\pi}{N_c}+\tfrac{1}{2} (\pi -\gamma_0) \right) \right]\,, \\
	&q_3 (1)=\sin \tfrac{\theta_1}{2} \cos \left[ \tfrac{1}{2} \left( -\tfrac{\pi}{N_c}+\tfrac{1}{2} (\pi -\gamma_0) \right) \right]\,,
	q_4 (1)=\cos \tfrac{\theta_1}{2} \sin \left[ \tfrac{1}{2} \left( \tfrac{\pi}{N_c}+\tfrac{1}{2} (\pi -\gamma_0) \right) \right]\,, \\
	&x(0) =-y(0) \cot \tfrac{\pi}{N_c} \,,z(0)=0 \,, \\
	&x (1) =y (1) \cot \tfrac{\pi}{N_c}\,, z (1)=0\,, y(0)+y (1)=0\,, \\ 
	\end{aligned}
	\end{equation}
	
	where $F_b(0)=\bm{F}(0) \cdot [-\cos (\tfrac{\pi}{N_c}) \hat{\bm{x}} +\sin (\tfrac{\pi}{N_c}) \hat{\bm{y}} ]$ and $F_z(0)=\bm{F}(0) \cdot \hat{\bm{z}}$ represents the contact force in the plane of symmetry, and $M_t(0)=\bm{M}(0) \cdot [ \sin (\tfrac{\pi}{N_c}) \hat{\bm{x}} +\cos (\tfrac{\pi}{N_c}) \hat{\bm{y}}] $ represents the moment perpendicular to the symmetry plane. Notice that the boundary conditions in Equation \eqref{appeq:boundaryconditions} admit the half flipped solutions with $N_c=2$, because we have only imposed one-fold mirror symmetry about the plane spanned by the two creases. $\theta_0$ and $\theta_1$ (both are negative in our definition) correspond respectively to the unknown second Euler angle at $s=0$ and $s=1$, and are treated as free parameters in numerical continuation. This is due to the fact that the two inclined angles $\delta_0$ and $\delta_1$ are unknown \emph{a priori}, and can be obtained respectively as $\delta_0=\tfrac{\pi}{2} + \theta_0$ and $\delta_1=\tfrac{\pi}{2} + \theta_1$. 
	Equation \eqref{appeq:boundaryconditions} contains 21 boundary conditions that are consistent with the number of unknowns, which include 18 state variables from Equation \eqref{appeq:18ODE}, and 3 free parameters $\mu$, $\theta_0$, and $\theta_1$. Equations \eqref{appeq:18ODE} and \eqref{appeq:boundaryconditions} lead to a well-posed two-point BVP.

	We use the conical frustum in Figures \ref{appFig:Eulerrotation}(b1) and \ref{appFig:Eulerrotation}(b2) as start solution for conducting numerical continuation on the inverted branch and folded branch, respectively. The start solution for the inverted state in Figure \ref{appFig:Eulerrotation}(b1) can be summarized as

	\begin{equation}\label{appeq:InitialGuessInverted} 
	\begin{aligned}
	F_1=0\,, F_2=0\,, F_3=0\,, M_1=0\,, M_2=\ln \frac{a}{R} \,, M_3=- \ln \frac{a}{R} \sqrt{\frac{1}{\alpha  ^2} - 1} \,, \\ 
	\kappa_n= \frac{1}{R} \sqrt{\frac{1}{\alpha  ^2} - 1}\,, \eta=0\,, \eta'=0\,,  \\
	q_1= \cos \left(\frac{\theta}{2} \right) \cos \left(\frac{\psi}{2} \right)\,,   
	q_2= -\sin \left(\frac{\theta}{2} \right) \sin \left(\frac{\psi}{2} \right)\,,
	q_3= \sin \left(\frac{\theta}{2} \right) \cos \left(\frac{\psi}{2} \right)\,,
	q_4= \cos \left(\frac{\theta}{2} \right) \sin \left(\frac{\psi}{2} \right)\,, \\
	x=\alpha R \cos \frac{\pi (1-2 \bar{s}) }{N_c}  \,, y=-\alpha R \sin \frac{\pi (1-2 \bar{s}) }{N_c} ,z=0, s=\frac{2 \pi R}{N_c} \alpha \bar{s}, V= R -a \,, \\
	\end{aligned}
	\end{equation}  
	
	with $\theta= -\sin^{-1} \alpha$, and $\psi= \frac{\pi}{N_c}(2\bar{s} - 1)$. The start solution for the folded state in Figure \ref{appFig:Eulerrotation}(b2) is different from Equation \ref{appeq:InitialGuessInverted} only in the sign of $M_3$ and $\kappa_n$ and the value of $\theta$, because the conical frustum in Figure \ref{appFig:Eulerrotation}(b1) is pointing upward, while the conical frustum in Figure \ref{appFig:Eulerrotation}(b2) is pointing downward. The different part can be rewritten as
	
	\begin{equation}\label{appeq:InitialGuessFolded} 
	\begin{aligned}
	M_3= \ln \frac{a}{R} \sqrt{\frac{1}{\alpha  ^2} - 1} \,, 
	\kappa_n= - \frac{1}{R} \sqrt{\frac{1}{\alpha  ^2} - 1}\,, 
	\theta= \sin^{-1} \alpha\ - \pi .
	\end{aligned}
	\end{equation}

	In numerical continuation, we always fix $R$ to 1 (i.e., $\kappa_g =-1$).
	Starting from a conical frustum (i.e., $\alpha <1$) with a small hole (e.g., $a/R=0.01$), we rotate the two end generators by decreasing $\gamma_{0}$ in Equation \eqref{appeq:boundaryconditions} to the target rest crease angle, which results in a configuration with rigid creases that will be used as start solution to introduce a finite crease stiffness.   
	
	To introduce flexible creases, the boundary conditions in Equation \eqref{appeq:boundaryconditions} need slight modifications. Particularly, the two final crease angles $\gamma_{f0}$ and $\gamma_{f1}$ (at $s=0$ and $s=2 \pi \alpha /N_c$, respectively) become unknowns. We replace $\gamma_0$ in $q_i(0)$ ($i=1,2,3,4$) with $\gamma_{f0}$ and $\gamma_0$ in $q_i(1)$ ($i=1,2,3,4$) with $\gamma_{f1}$, respectively. In addition, two additional boundary conditions representing the moment balance at the crease are added as following 
	
	\begin{equation}\label{appeq:momentBCsin} 
	\begin{aligned}
	&M_3 (0) =\frac{K_c R}{D} (1-\frac{a+e}{R}) \sin (\gamma_{f0} - \gamma_0) \,, \\
	&M_3 (1) = \frac{K_c R}{D} (1-\frac{a-e}{R}) \sin (\gamma_{f1} -\gamma_0) \,. \\
	\end{aligned}
	\end{equation}  
	
	 In numerical continuation, the scalar unknowns $\gamma_{f0}$ and $\gamma_{f1}$ are treated as free parameters. The consistency between the additional unknowns and the additional boundary conditions leads to a well-posed two-point BVP consisting of 23 unknowns and 23 boundary conditions.
	Equation \eqref{appeq:momentBCsin} implies that a large dimensional creases stiffness $K_{c} R/D$ will penalize $\gamma_{f0}$ and $\gamma_{f1}$ to be the rest crease angle $\gamma_0$.
	Starting with a rigid crease solution, we decrease $K_{c} R/D$ from a large number to the target finite crease stiffness. Equation \eqref{appeq:momentBCsin} also incorporates the eccentricity factor. Now, we are able to vary $a/R$, $\alpha$, $\gamma_{0}$ etc. to conduct parametric studies.
	
	The case with a single crease $N_c=1$ is similar to $N_c=2$. We solve half of the structure and impose only the crease boundary condition in Equation \eqref{appeq:momentBCsin} at $\bar{s}=0$. The $\bar{s}=1$ end is equivalent to a rigid crease with a rest crease angle $\pi$. 
	
	After obtaining the numerical results, an annular sector can be constructed as 
	\begin{equation}\label{appeq:3Dreconstruction} 
	\begin{aligned}
	\bm{X}(s,v)&=\bm{r}(s)+v[\bm{B}(s)+\eta(s)\bm{T}(s)] \,,\\
	&=(x+2v [\eta (q_2 q_3 - q_1 q_4)+q_2 q_4 +q_1 q_3]) \hat{\bm{x}} \\
	&  \hspace{0.12in} +(y+2v [\eta (q_1^2+q_3^2-\tfrac{1}{2})+q_3 q_4 - q_1 q_2]) \hat{\bm{y}} \\
	& \hspace{0.12in} +(z+2v [\eta (q_3 q_4 +q_1 q_2) +q_1^2+q_4^2-\tfrac{1}{2} ]) \hat{\bm{z}} \,,
	\end{aligned}
	\end{equation}

where $\eta=\tau_g / \kappa_n$, and $v \in [0,V] $. The complete structure is constructed by using symmetry properties. 
The edge of regression, on which adjacent generators intersect each other, can be defined as

\begin{equation}\label{appeq:3Dedgeregression} 
\begin{aligned}
\bm{c}(s)=\bm{r}(s)+ \frac{\sin \beta}{\beta'-\kappa_g}  \frac{\bm{B}(s)+\eta(s)\bm{T}(s)}{|\bm{B}(s)+\eta(s)\bm{T}(s)|} = \bm{r}(s)- \frac{\bm{B}(s)+\eta(s)\bm{T}(s)}{\eta'+\kappa_g(1+\eta^2)} \,.
\end{aligned}
\end{equation}

By differentiation, we have $\bm{c}'(s)=\tfrac{[\eta \kappa_g^2  (1+\eta^2) + \eta''+3 \kappa_g \eta \eta' ] }{[\eta'+\kappa_g (1+\eta^2)]^2} (\bm{B} +\eta \bm{T})$. The isolated points where $\eta''=-3 \kappa_g \eta \eta' - \eta \kappa_g ^2 (1+\eta^2)$ are called ``conical", because at such points, the edge of regression contains a cusp \cite{starostin2015equilibrium}. In addition,
the points where $\eta'=-\kappa_g (1+\eta^2)$ are called ``cylindrical".  At a cylindrical point, the mean curvature is constant along the local generator \cite{starostin2015equilibrium}.

The generators can be mapped onto the flat annular sector as
\begin{equation}\label{appeq:2Dreconstruction} 
\begin{aligned}
\bm{X}(s,v)&=\bm{r}(s)+v[\bm{b}(s)+\eta(s)\bm{t}(s)] \,, \\
& =(R \sin \lambda - v \sin \lambda  + v \eta \cos \lambda ) \hat{\bm{x}} + \left(  - R \cos \lambda + v \eta \sin \lambda + v \cos \lambda \right) \hat{\bm{y}} \,,
\end{aligned}
\end{equation}

where $\bm{b}(s)$ and $\bm{t}(s)$ are the binormal and the tangent of the outer circle, respectively. The edge of regression (which we did not include in the flat developments of the 3D renderings in this study) can be mapped onto the flat annular sector as 
\begin{equation}\label{appeq:2Dedgeregression} 
\begin{aligned}
\bm{c}(s)&=\bm{r}(s)- \frac{\bm{b}(s)+\eta(s)\bm{t}(s)}{\eta'+\kappa_g(1+\eta^2)} \,, \\
&= \left( R \sin \lambda+\frac{\sin \lambda  - \eta \cos \lambda}{\eta'+\kappa_g(1+\eta^2)} \right) \hat{\bm{x}} - \left ( R \cos \lambda + \frac{\eta \sin \lambda + \cos \lambda}{\eta'+\kappa_g(1+\eta^2)} \right) \hat{\bm{y}} \,.
\end{aligned}
\end{equation}

\section{Crease with a linear angle-moment relationship} \label{appse:linearcrease}

Here we give an example to show that with creases having a linear response, in certain parameter spaces a creased disk could behave differently from one with creases adopting a sinusoidal angle-moment relationship.
With linear creases, the crease boundary conditions in Equation \eqref{appeq:momentBCsin} are modified as

\begin{equation}\label{appeq:momentBClinear} 
\begin{aligned}
&M_3 (0) =\frac{K_c R}{D} (1-\frac{a+e}{R}) (\gamma_{f0} - \gamma_0) \,,\\
&M_3 (1) = \frac{K_c R}{D} (1-\frac{a-e}{R}) (\gamma_{f1} -\gamma_0) \,. \\
\end{aligned}
\end{equation}

\begin{figure}[h!]
	\centering
	\includegraphics[width=0.95\textwidth]{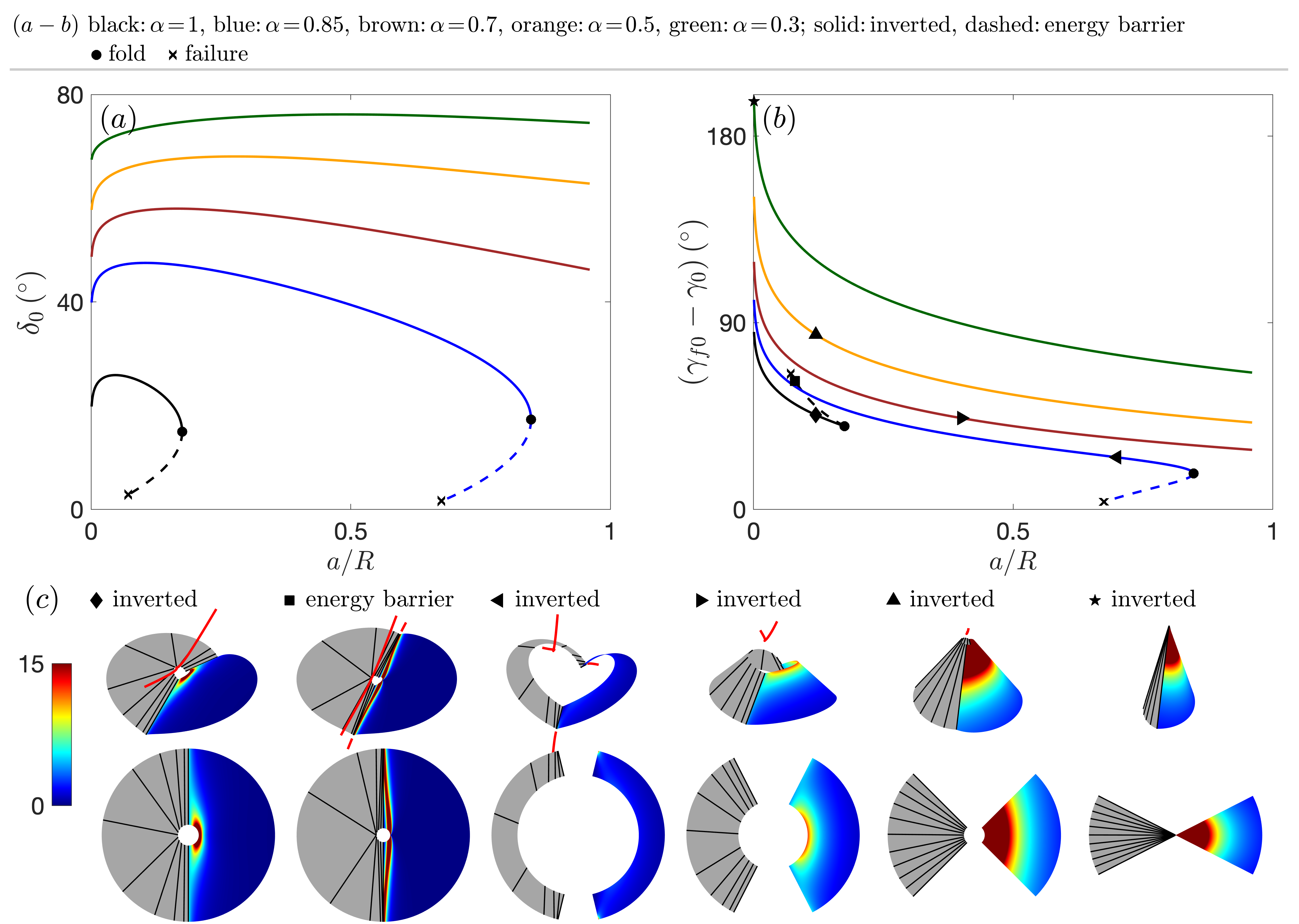}
	\caption{Solution curves with a linear crease and different $\alpha$. $(K_cR/D,N_c,\gamma_0)$ is fixed to $(4,2,45^{\circ})$. (a) $\delta_0$ versus $a/R$. (b) The same results presented in $(\gamma_{f0}-\gamma_0)$ versus $a/R$ plane. (c) Renderings that correspond to the symbols in $(b)$.}
	\label{appfig:alphaepsbiurLinear}
\end{figure}

The numerical results in Figure \ref{appfig:alphaepsbiurLinear}(a) adopt linear creases ( Equation \eqref{appeq:momentBClinear}) with $(K_cR/D,N_c,\gamma_0,e)$ fixed to $(4,2,45^{\circ},0)$, which contain the same parameter setting with the results in Figure \ref{fig:alphaepsbifurphase}(a). It is found there is not a lower boundary for $a/R$ with $\alpha \le 0.85$, which does exist in Figure \ref{fig:alphaepsbifurphase}(a). The reason is that a linear crease does not have a fictitious rest crease angle, e.g., $(\gamma_{f0}+\gamma_{0})$ in creases following a sinusoidal angle-moment relationship. Figure \ref{appfig:alphaepsbiurLinear}(b) reports the deviation of the crease from the rest angle $(\gamma_{f0} - \gamma_{0})$ at $s = 0$ for the solutions in Figure \ref{appfig:alphaepsbiurLinear}(a). With small $a/R$ the final crease angle opens significantly due to the large bending moment from the conical surface. Figure \ref{appfig:alphaepsbiurLinear}(c) displays several renderings corresponding to the symbols in Figure \ref{appfig:alphaepsbiurLinear}(b). With a small $\alpha=0.3$, the final crease angle $\gamma_{f0}$ could be much larger than $\pi$, e.g., the configuration indicated by a $\blackstar$. 

We want to emphasize that using a linear crease model does not change much the solution curves as we increase the hole size. For example, the hole size could have an upper boundary that destroys the bistability of the creased disk, decreasing $\alpha$ generally leads to the increase of the critical hole size, and with $\alpha \le 0.7$, the hole size could be as large as the disk without loss of bistability.

\newpage

\section{3D profile and corresponding 2D projections of the outer and inner circumferences} \label{appse:3Dprofile2Dprojection}

\begin{figure}[h!]
	\centering
	\includegraphics[width=0.9\textwidth]{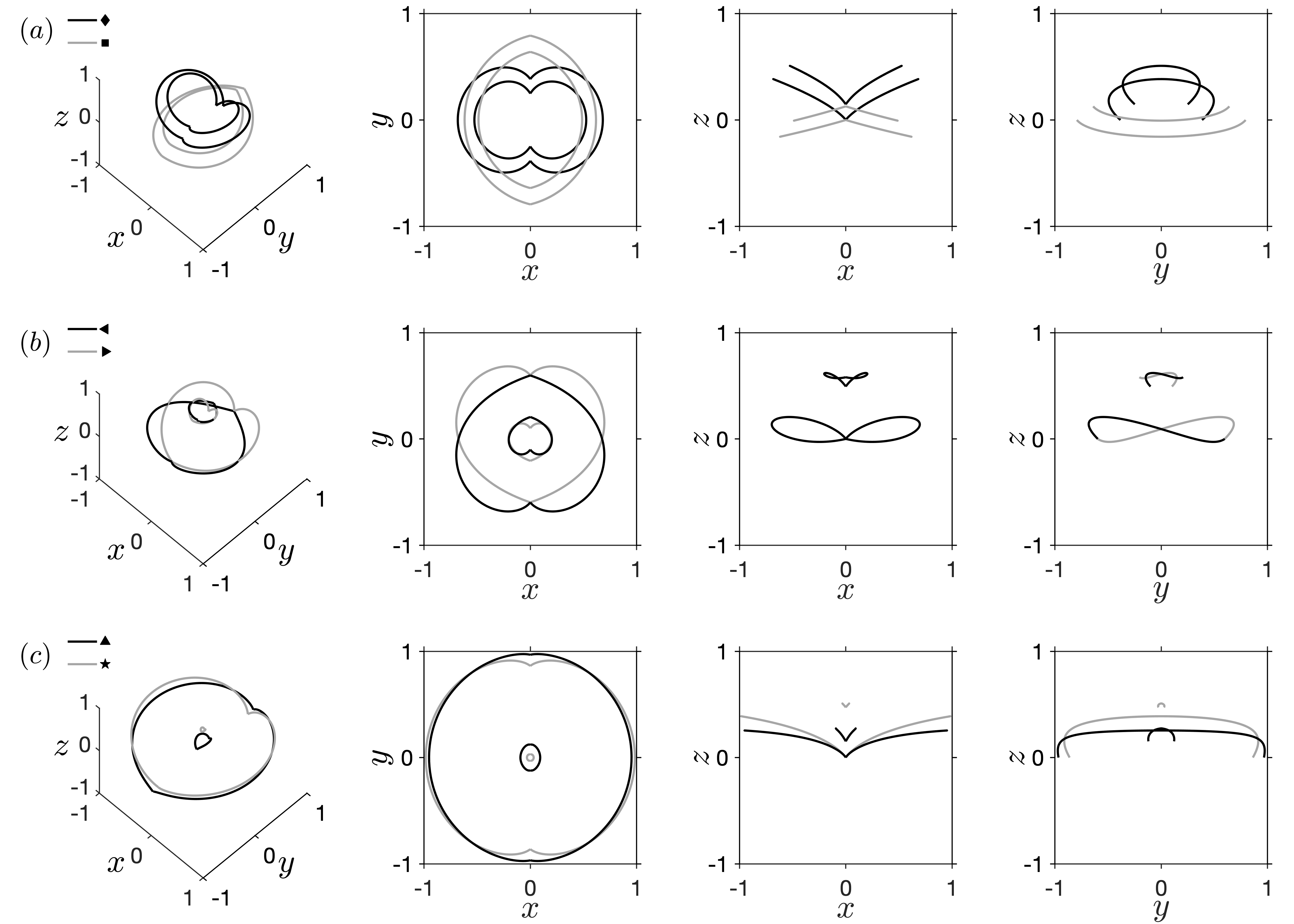}
	\caption{3D profile and corresponding 2D projections of the outer and inner circumferences of the renderings in Figure \ref{fig:alphaepsbifurphase}(e). (a) $\blackdiamond$ and $\blacksquare$. (b) $\lefttriangle$ and $\righttriangle$. (c) $\blacktriangle$ and $\blackstar$.}
	\label{appfig:alphaepsProjection}
\end{figure}

Here, we document the 3D profile and corresponding 2D projections of the outer and inner circumferences of some renderings presented in the main text.
Figure \ref{appfig:alphaepsProjection} displays the 3D profile and corresponding 2D projections of the renderings in Figure \ref{fig:alphaepsbifurphase}(e).

\newpage

\begin{figure}[h!]
	\centering
	\captionsetup[subfigure]{labelfont=normalfont,textfont=normalfont}
	\centering
	\includegraphics[width=0.9\textwidth]{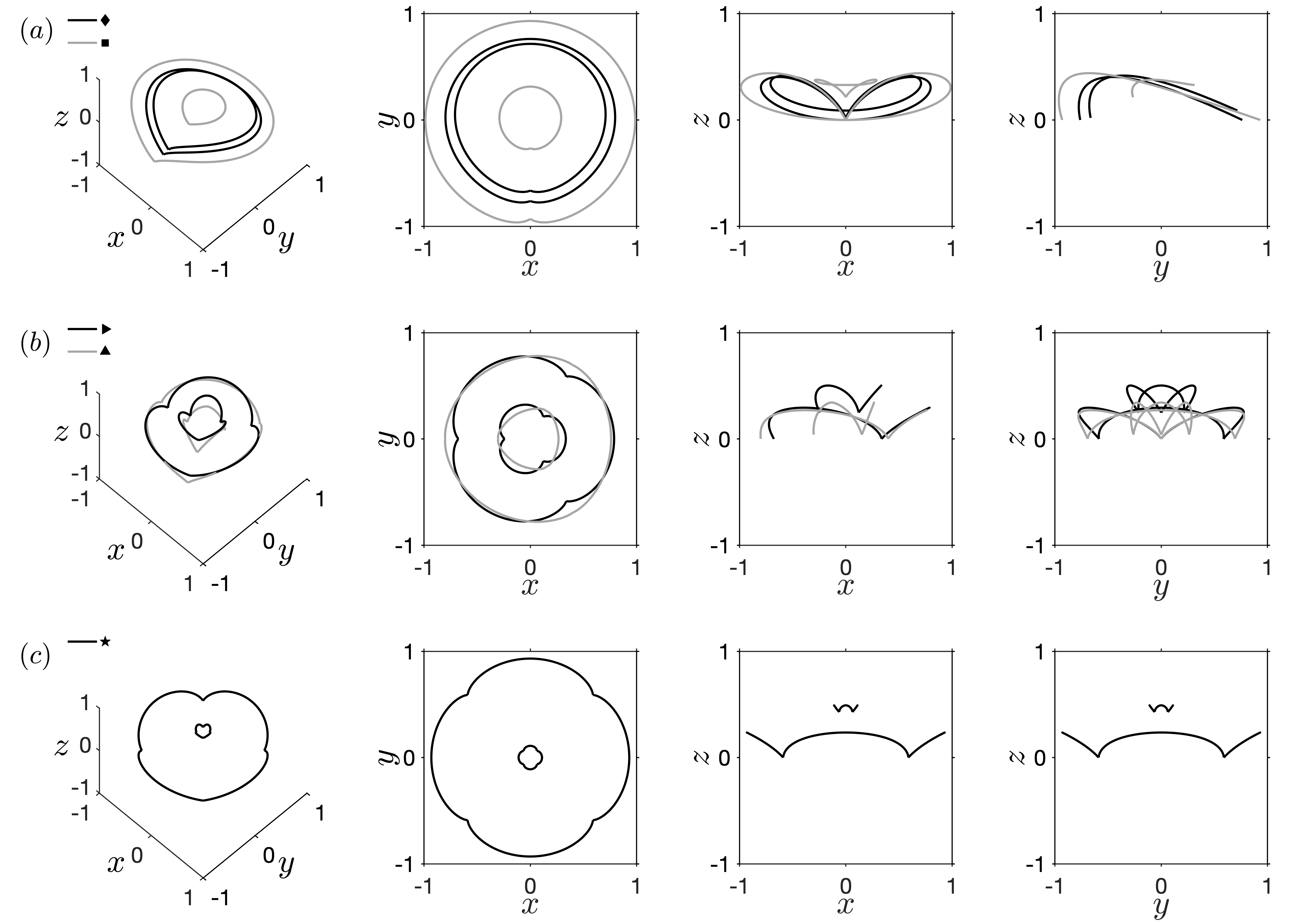}
	\caption{3D profile and corresponding 2D projections of the outer and inner circumferences of some of the renderings in Figure \ref{fig:nfepsbifurphaseinverted}(e). (a) $\blackdiamond$ and $\blacksquare$. (b) $\righttriangle$ and $\blacktriangle$. (c) $\blackstar$. }\label{appfig:InvertedStateProjection}
\end{figure}

Figure \ref{appfig:InvertedStateProjection} displays the 3D profile and corresponding 2D projections of the outer and inner circumferences of some renderings in Figure \ref{fig:nfepsbifurphaseinverted}(e).

\newpage

\begin{figure}[h!]
	\centering
	\captionsetup[subfigure]{labelfont=normalfont,textfont=normalfont}
	\centering
	\includegraphics[width=0.9\textwidth]{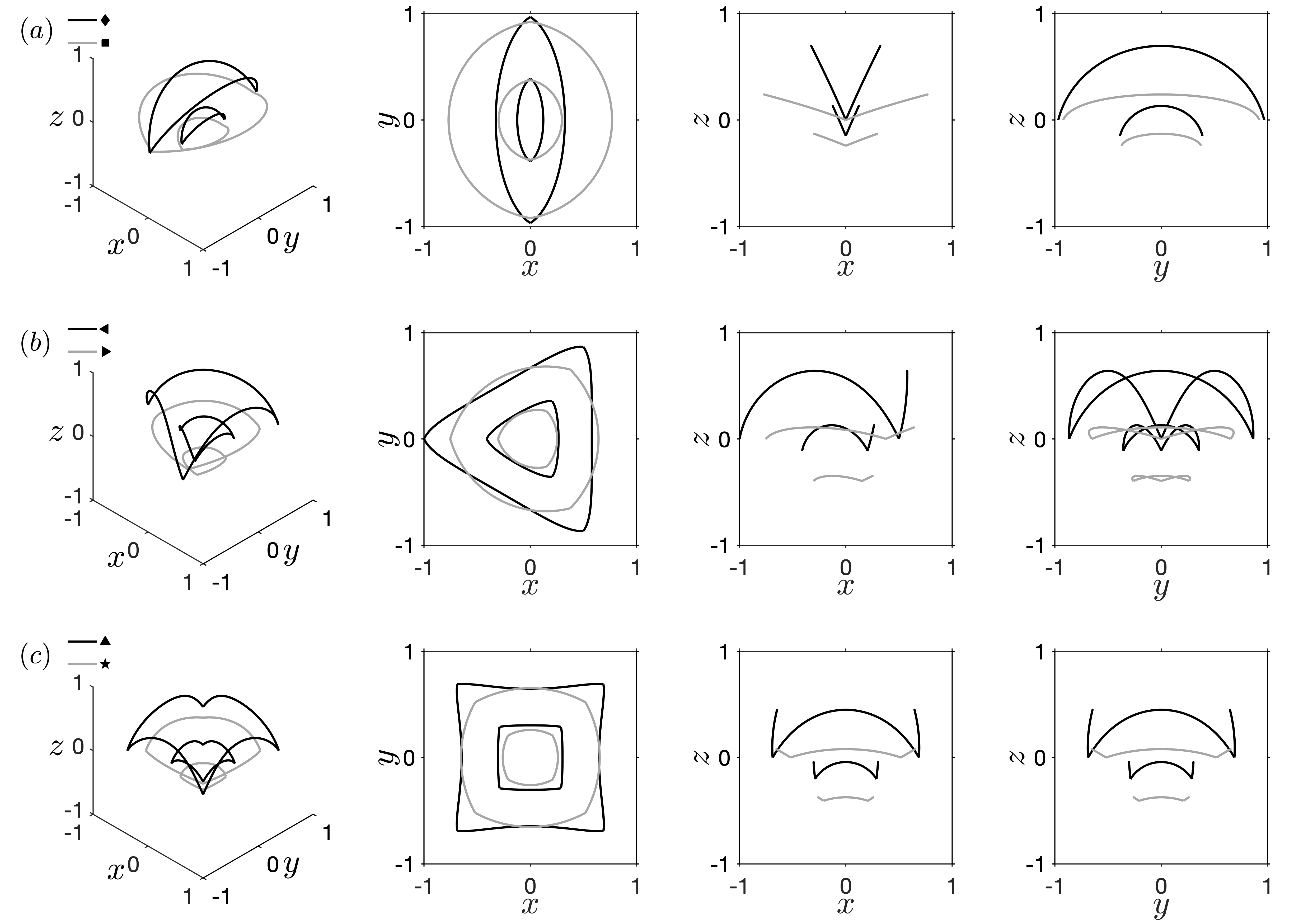}
	\caption{3D profile and corresponding 2D projections of the outer and inner circumferences of the renderings in Figure \ref{fig:nfepsbifurphasefolded}(e). (a) $\blackdiamond$ and $\blacksquare$. (b) $\lefttriangle$ and $\righttriangle$. (c) $\blacktriangle$ and $\blackstar$.} \label{appfig:FoldedStateProjection}
\end{figure}

Figure \ref{appfig:FoldedStateProjection} displays the 3D profile and corresponding 2D projections of the outer and inner circumferences of the renderings in Figure \ref{fig:nfepsbifurphasefolded}(e). 

\newpage

\begin{figure}[h!]
	\centering
	\captionsetup[subfigure]{labelfont=normalfont,textfont=normalfont}
	\centering
	\includegraphics[width=0.8\textwidth]{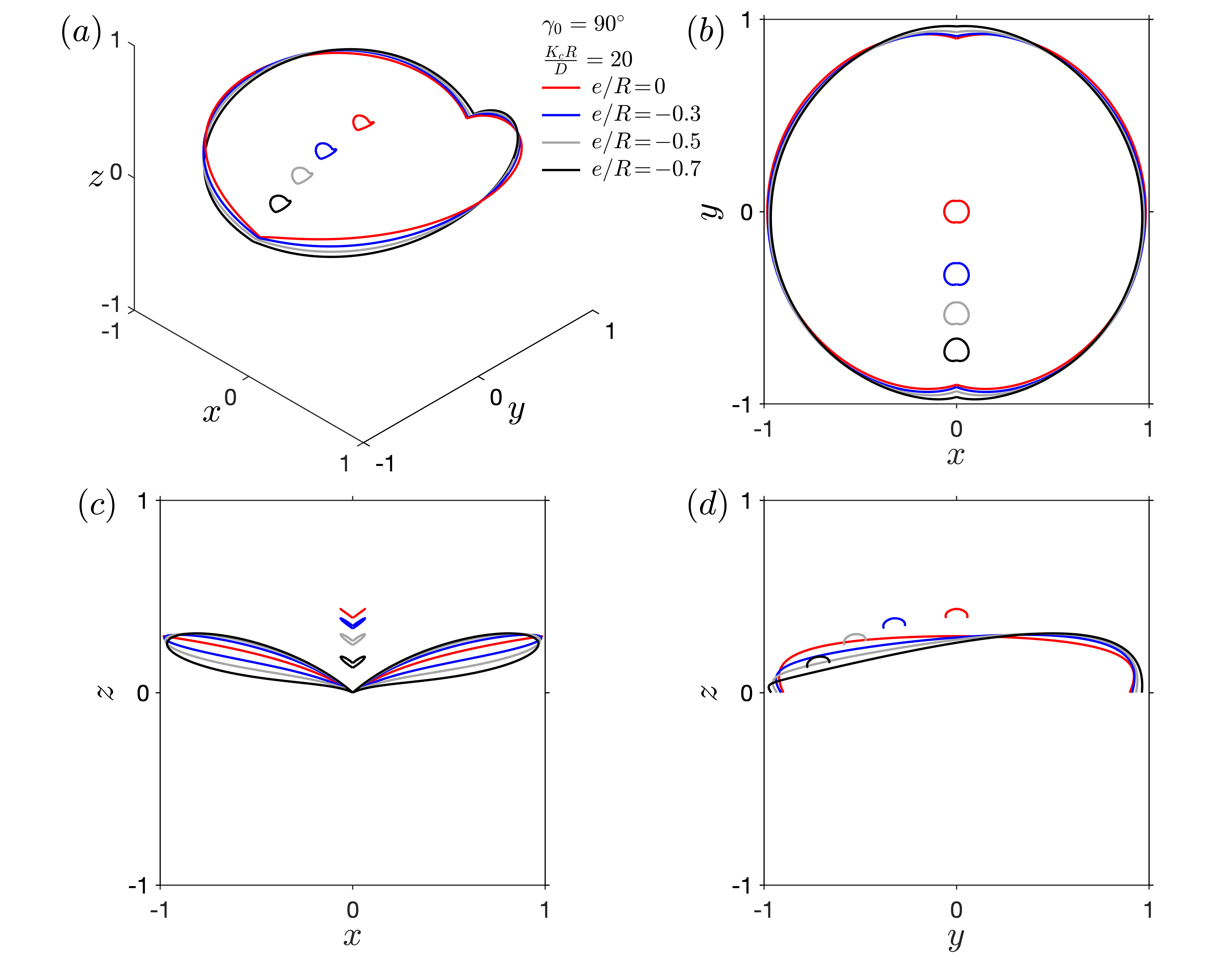}
	\caption{3D profile and corresponding 2D projections of the outer and inner circumferences of several inverted states with different eccentricities. Other parameters are fixed to $(K_cR/D, \gamma_0, a/R, \alpha) = (20,90^{\circ},0.07,1)$. } \label{appfig:EccenEdgeProjection}
\end{figure}

Figure \ref{appfig:EccenEdgeProjection} displays the 3D profile and corresponding 2D projections of the outer and inner circumferences of several inverted states with various eccentricities. Other geometric parameters are fixed to $(K_cR/D, \gamma_0, a/R, \alpha) = (20,90^{\circ},0.07,1)$.  

\clearpage

\section{Additional renderings} \label{appse:morerenderings}

\begin{figure}[h!]
	\centering
	\captionsetup[subfigure]{labelfont=normalfont,textfont=normalfont}
	\centering
	\includegraphics[width=\textwidth]{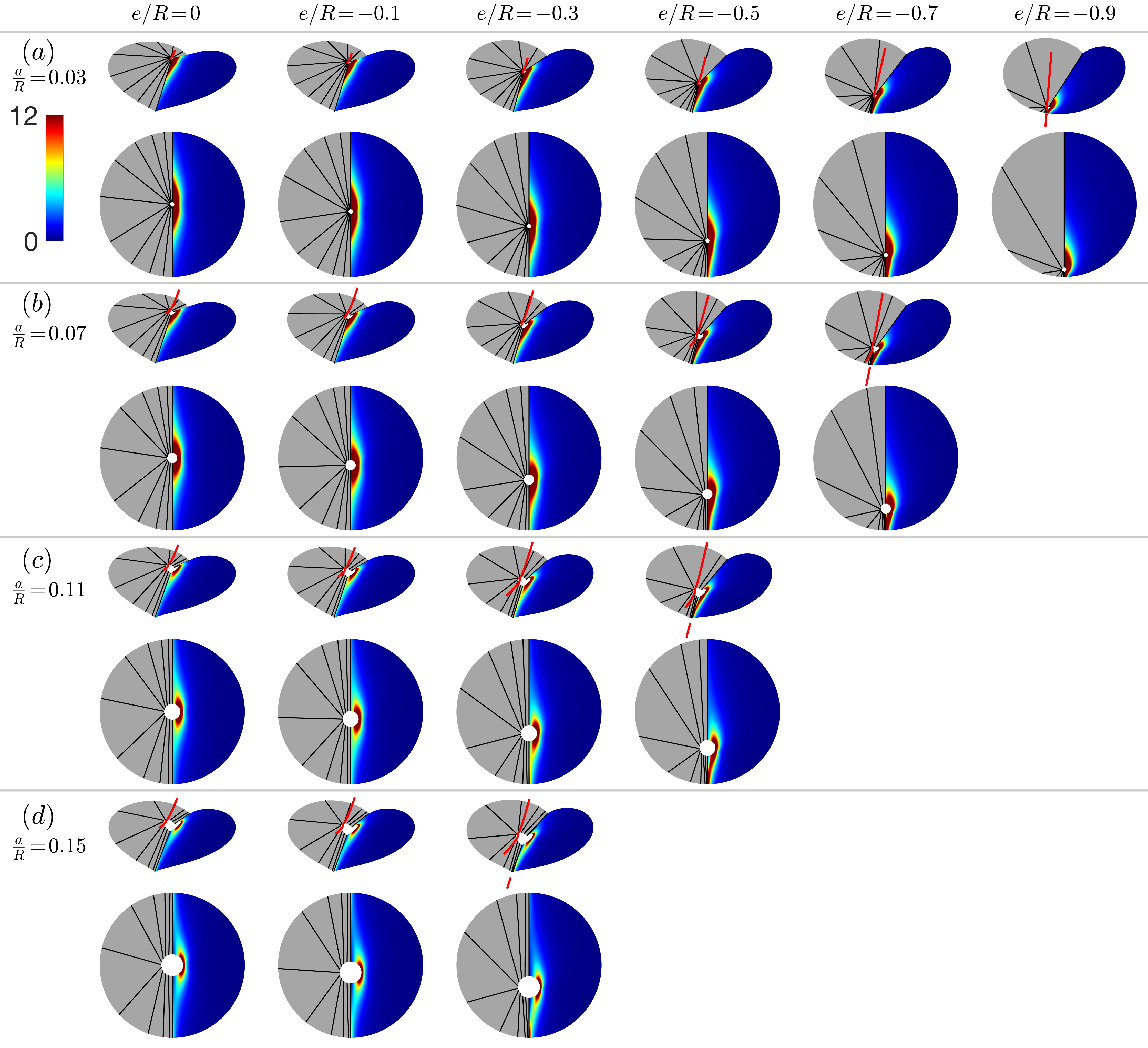}
	\caption{Renderings of the inverted state and their developments on the flat configurations with different $(a/R, e/R)$. $(K_c R / D, \gamma_0)$ is fixed to $(20, 90^{\circ})$. All the panels share the same color bar. (a) $a/R=0.03$. (b) $a/R=0.07$. (c) $a/R=0.11$. (d) $a/R=0.15$. }\label{appfig:EccenConfigSet}
\end{figure}

Here, we document additional renderings obtained from numerical continuation of the inextensible strip model for the interest of the reader. Figure \ref{appfig:EccenConfigSet} displays renderings of the inverted state and their flat developments with different $(a/R, e/R)$. $(K_c R / D, \gamma_0)$ is fixed to $(20, 90^{\circ})$.

\newpage

\begin{figure}[h!]
	\centering
	\captionsetup[subfigure]{labelfont=normalfont,textfont=normalfont}
	\centering
	\includegraphics[width=\textwidth]{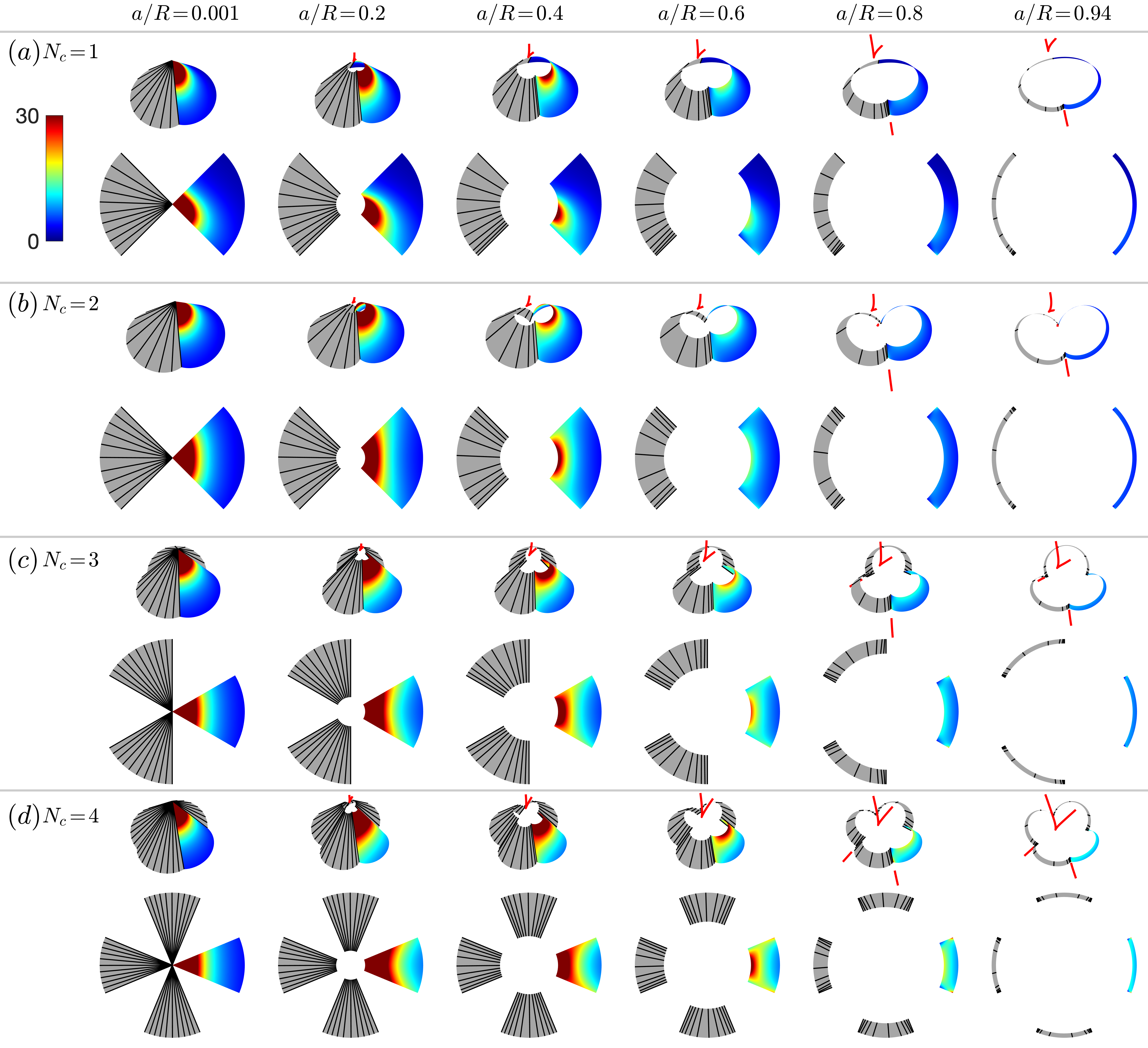}
	\caption{Renderings of the inverted state and their developments on the flat configurations with different combination $(a/R,N_c)$. $(\alpha,K_c R / D,\gamma_0)$ is fixed to $(0.5,20,45^{\circ})$. All the panels share the same color bar. (a) $N_c=1$. (b) $N_c=2$. (c) $N_c=3$. (d) $N_c=4$. }\label{appfig:nfepsConfigSet}
\end{figure}

Figure \ref{appfig:nfepsConfigSet} displays renderings of the inverted state and their flat developments with different $(a/R,N_c)$. $(\alpha,K_c R / D,\gamma_0)$ is fixed to $(0.5,20,45^{\circ})$.

\newpage

\begin{figure}[h!]
	\centering
	\captionsetup[subfigure]{labelfont=normalfont,textfont=normalfont}
	\centering
	\includegraphics[width=\textwidth]{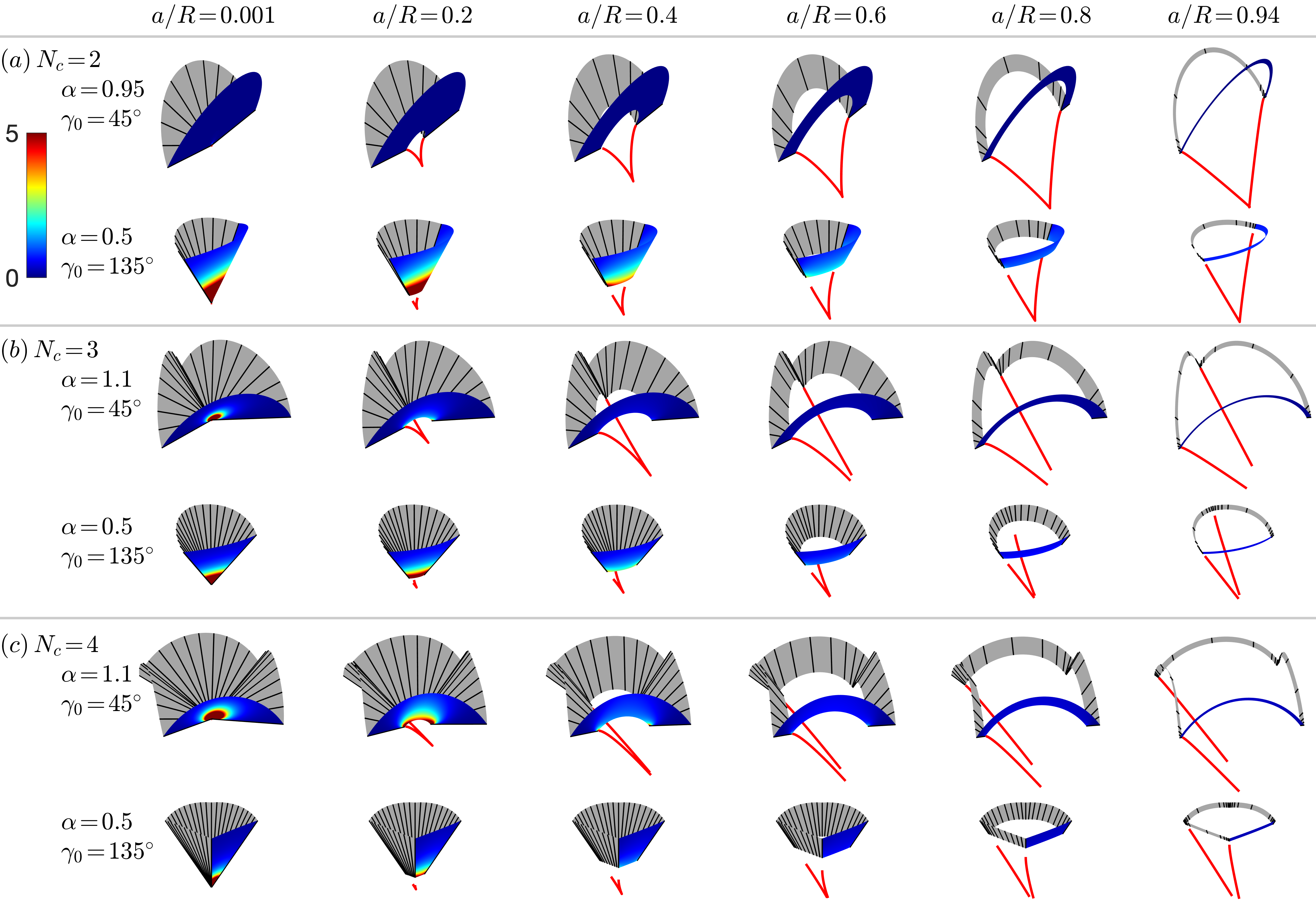}
	\caption{3D renderings of the folded state with different $(N_c,\alpha,\gamma_0,a/R)$. $K_c R / D$ is fixed to 20. All the panels share the same color bar. (a) $N_c=2$. (b) $N_c=3$. (c) $N_c=4$. }\label{appfig:nfepsConfigSetfolded}
\end{figure}

Figure \ref{appfig:nfepsConfigSetfolded} displays renderings of the folded state with different $(N_c,\alpha,\gamma_0,a/R)$. $K_c R / D=20$ is fixed to 20.

	\clearpage
	\bibliographystyle{unsrt}
	\bibliography{creasedannularribbon}
	
\end{document}